%% file: text6.tex
\documentclass[useAMS,usenatbib]{mn2e}
\usepackage{graphicx}
\def\apj{ApJ}
\def\apjs{ApJS}

\def\aap{A\&A}

\def\aj{AJ}
\def\mnras{MNRAS}

\def\pasa{PASA}

\def\pasp{PASP}
\def\nat{Nature}

\def\rmxaa{Revista Mexicana de Astronomia y Astrofisica}

\title[Star-forming galaxies with hard radiation]
{Large Binocular Telescope observations of new six compact star-forming 
galaxies with [Ne\,{\sc v}] $\lambda$3426\AA\ emission}
\author[Y. I. Izotov et al.]{Y. I.\ Izotov$^{1}$,
T. X.\ Thuan$^{2}$ and N. G.\ Guseva$^{1}$\\
                $^{1}$Bogolyubov Institute for Theoretical Physics,
                     National Academy of Sciences of Ukraine,
                     14-b Metrolohichna str., Kyiv, 03143, Ukraine,\\
                     yizotov@bitp.kiev.ua, nguseva@bitp.kiev.ua\\
                $^{2}$Astronomy Department, University of Virginia, 
                     P.O. Box 400325, Charlottesville, VA 22904-4325, USA,\\
                     txt@virginia.edu\\
}
\begin{document}


\pagerange{\pageref{firstpage}--\pageref{lastpage}} \pubyear{2012}

\maketitle

\label{firstpage}

\begin{abstract}
We report the discovery of [Ne\,{\sc v}]\,$\lambda$3426 emission, in addition
to He\,{\sc ii}\,$\lambda$4686 emission, in six compact star-forming galaxies.
These observations considerably increase the sample
of eight such galaxies discovered earlier by our group. For four of the new 
galaxies, the optical observations are supplemented by near-infrared spectra.
All galaxies, but one, have H\,{\sc ii} regions that are dense, with 
electron number densities of $\sim$\,300\,--\,700\,cm$^{-3}$. They are all  
characterised 
by high H$\beta$ equivalent widths EW(H$\beta$)\,$\sim$\,190\,--\,520\AA\
and high O$_{32}$\,=\,[O\,{\sc iii}]\,$\lambda$5007/[O\,{\sc ii}]\,$\lambda$3727 ratios
of 10\,--\,30, indicating young starburst ages and the presence of high ionization radiation. All
are low-metallicity objects with 12\,+\,logO/H\,=\,7.46\,--\,7.88. The spectra
of all galaxies show a low-intensity broad component of the H$\alpha$ line and
five out of six objects show Wolf-Rayet features. 
 Comparison with photoionization models shows that pure stellar ionization 
 radiation from massive stars is not hard enough to produce such strong
 [Ne\,{\sc v}] and He\,{\sc ii} emission in our galaxies. The 
 [Ne\,{\sc v}]\,$\lambda$3426/He\,{\sc ii}\,$\lambda$4686 flux ratio of
 $\sim$\,1.2 in J1222$+$3602 is consistent with some contribution of 
active galactic nucleus ionizing radiation. However, in the remaining 
five galaxies, this ratio is considerably lower, 
$\la$\,0.4. The most plausible models are likely to be   
non-uniform in density, where He\,{\sc ii} and [Ne\,{\sc v}] lines are 
emitted in low-density channels made by outflows and illuminated by harder 
ionizing radiation from radiative shocks propagating through these channels,
whereas [O\,{\sc iii}] emission originates in denser regions exposed to 
softer stellar ionizing sources.

\end{abstract}


\section{Introduction}\label{sec:INT}

The  presence  of  hard  radiation in dwarf star-forming galaxies (SFGs) has long been known
thanks to the detection of a relatively strong nebular 
He\,{\sc ii}\,$\lambda$4686 emission line in some of them.  This line can serve as a 
probe of extreme-UV ionizing radiation with photon energies higher than 4 ryd 
(54 eV), corresponding to wavelengths of $\leq$ 228\AA\ \citep{PA86}.

The flux  of this line has been shown to increase  with  
decreasing  metallicity  of the ionized gas 
\citep*[e.g. ][]{G91,GIT00,IT04,TI05}. 
Presently, the lowest-metallicity galaxies with known 
He\,{\sc ii}\,$\lambda$4686 emission are J1631$+$4426 with 
12\,$+$\,logO/H = 6.90 from the wide-field optical imaging data obtained with 
the Subaru/Hyper Suprime-Cam (HSC) \citep{K21}, and two galaxies selected
in the Sloan Digital Sky Survey (SDSS) and later observed with the Large 
Binocular Telescope (LBT), J0811$+$4730 with 
12\,$+$\,logO/H = 6.98 \citep*{I18a} and J1234$+$3901 with 
12\,$+$\,logO/H = 7.04 \citep*{I19}. 

Other high ionization emission lines of heavy elements are also seen in the 
spectra of some dwarf SFGs. Thus, the 
[Fe\,{\sc v}]\,$\lambda$4227 emission line was first detected by \citet{Fr01} 
and \citet*{I01} in the blue compact dwarf (BCD) galaxies Tol\,1214$-$277 and 
SBS\,0335$-$052E, respectively. The 
presence of this line also requires ionizing radiation with photon energies 
in excess of 4 ryd. Later, this line was detected in many more SFGs
\citep*[e.g. ][]{TI05,I17} with fluxes several times smaller than that 
of the He\,{\sc ii}\,$\lambda$4686 emission line. As of today, the 
lowest-metallicity galaxy with known [Fe\,{\sc v}]\,$\lambda$4227 emission is the 
SFG J1234$+$3901 with 12\,$+$\,logO/H = 7.04 \citep{I19}.

Furthermore, \citet{I01} found [Fe\,{\sc vi}] -- [Fe\,{\sc vii}] emission
in the BCD SBS\,0335$-$052E, implying that it contains hard radiation with 
photon energies above ionization potentials of 5.5 ryd (i.e. 75 eV) for 
Fe$^{5+}$, corresponding to wavelengths of radiation 
below 165\AA, and above 7.3 ryd (or 100 eV) for Fe$^{6+}$,
corresponding to wavelengths of radiation below 90\AA. 

Another very high-ionization line is the [Ne\,{\sc v}]\,$\lambda$3426 emission 
line which, along with the He\,{\sc ii}\,$\lambda$4686 emission 
line, is the subject of this paper. It was first discovered by 
\citet{I04a} in the BCD Tol\,1214$-$277. Later, \citet{TI05} and \citet*{I12b}
found this emission line in seven more dwarf SFGs. This line requires
the presence of extreme-UV radiation with photon energies above 7.1 ryd,
(or $\geq$ 97 eV), corresponding to wavelengths of radiation 
$\lambda$~$\leq$~128\AA. It is often seen in high-excitation Seyfert~2 
galaxies, where 
it can be considerably stronger than the He\,{\sc ii} $\lambda$4686 emission 
line. This situation is contrary to the one in SFGs where [Ne\,{\sc v}] is 
invariably weaker than He\,{\sc ii}. The higher strength of [Ne\,{\sc v}] in 
Seyfert 2 (Sy2) galaxies is caused partly by the high metallicity of the 
interstellar medium (ISM) of Sy2 galaxies and, possibly, by the harder 
non-thermal radiation produced by an active galactic nucleus (AGN), compared to 
that of SFGs. 

The origin of the extreme-UV radiation in dwarf SFGs remains unclear. 
Several mechanisms for producing hard ionizing radiation have been proposed, 
such as AGNs \citep{IT08}, intermediate-mass black holes 
($\ga$\,100\,M$_\odot$), formed from supermassive ($>$\,300\,M$_\odot$)
stars \citep{K21}, Wolf-Rayet (WR) stars \citep{S96}, high-mass X-ray 
binaries \citep*{G91,SFI19}, and fast radiative shocks \citep{DS96,I04a,I12b}. 
However, a large data base is needed to confront models with observations. In 
fact, [Ne\,{\sc v}]\,$\lambda$3426 emission has been detected only in eight SFGs 
\citep{I04a,TI05,I12b}, while He\,{\sc ii}\,$\lambda$4686 emission line has been 
detected in many more (several thousands) SFGs 
\citep{GIT00,TI05,SB12}. The detection of [Ne\,{\sc v}]\,$\lambda$3426 emission
line requires a large telescope equipped with spectrographs operating 
efficiently in the near-UV range. 

The BCD SBS\,0335$-$052E, with an oxygen abundance 12\,+\,logO/H = 7.30, is 
the lowest-metallicity SFG with detected [Ne\,{\sc v}] emission.
\citet{I01,I06b} and \citet{TI05} have discussed the origin of
He\,{\sc ii} and [Ne\,{\sc v}] emission in that object. Those authors found that the BCD shows 
a He\,{\sc ii} emission which is spatially distinct from the hydrogen H$\alpha$ and 
H$\beta$ emission, and which extends over a broader area \citep{I06b}. They thus concluded that the main source of hard 
radiation is not stars, but more likely radiative shocks.
Furthermore, the presence of [Ne\,{\sc v}] emission in SBS\,0335$-$052E strengthens
that conclusion. Observations of other SFGs with [Ne\,{\sc v}] emission
\citep{I04a,I12b} also support the radiative shock mechanism.

Recently \citet{K18} have reconsidered the origin of the
He\,{\sc ii}\,$\lambda$4686 emission in SBS\,0335$-$052E.
They analysed several sources of ionizing radiation and ruled out 
  significant contributions from X-ray sources, shocks and WR stars.
To study the role of stellar radiation in producing He\,{\sc ii} emission,
\citet{K18} adopted the Binary Population and Spectral Synthesis ({\sc bpass})
v2.1 stellar population synthesis models by \citet{E17}. These differ from other
models, e.g. {\sc starburst99} \citep{L99,L14}, as they take into account binary
stellar evolution. They found that the He\,{\sc ii} emission of
SBS 0335$-$052E can only be produced by either single, rotating metal-free 
stars or a binary population with $Z$ $\sim$ 10$^{-5}$ and a ’top-heavy’
Initial Mass Function (IMF). 
However, such a discrepancy between the predicted metallicity (0 or 10$^{-5}$)
and the observed ISM metallicity (about 1/25 solar) appears to be too large 
to be reasonable, even for the most metal-deficient galaxies (about 1/50 solar).

\input{tab1_1.tex}

\input{tab2_1.tex}

In this paper, we present LBT\footnote{The LBT 
is an international collaboration among institutions in the United States, 
Italy and Germany. LBT Corporation partners are: The University of Arizona on 
behalf of the Arizona university system; Istituto Nazionale di Astrofisica, 
Italy; LBT Beteiligungsgesellschaft, Germany, representing the Max-Planck 
Society, 
the Astrophysical Institute Potsdam, and Heidelberg University; The Ohio State 
University, and The Research Corporation, on behalf of The University of Notre 
Dame, University of Minnesota and University of Virginia.} spectroscopic
observations of six compact SFGs. These observations are part of a long-term
LBT spectroscopic
program to study compact SFGs with high-excitation H\,{\sc ii} regions. Our 
sample consists of six SFGs with detected [Ne\,{\sc v}]\,$\lambda$3426 emission.
Five of them, excluding HS\,1851$+$6933, are SDSS objects. 
Coordinates, redshifts and other characteristics of the sample objects, obtained from the 
photometric and spectroscopic data, are shown in Table\,\ref{tab1}. All of 
these SFGs are characterised by a high O$_{32}$ = 
[O\,{\sc iii}]\,$\lambda$5007/[O\,{\sc ii}]\,$\lambda$3727 flux ratio, in the range
10 -- 30, indicating a high ionization parameter.

Some objects in Table\,\ref{tab1} have been discussed earlier in various contexts. 
Thus, \citet*{I07} and \citet{IT08} selected J1222$+$3602 from the SDSS as a possible
low-metallicity AGN candidate because of its very broad hydrogen low intensity emission.
The study of its properties does imply that J1222$+$3602 is likely an AGN, but 
its SDSS spectrum did not show [Ne\,{\sc v}]\,$\lambda$3426\AA\ emission. 
\citet{TI05} analysed the SDSS spectrum of J0240$-$0828 and noted the 
presence of [Fe\,{\sc v}]\,$\lambda$4227\AA\ emission. 
As for W\,1702$+$18, it has been selected by \citet{Gr11} as a 
galaxy with an extremely red {\sl WISE} $W1$\,--\,$W2$ colour, an indication
of hot dust emission. Here, $W1$ and $W2$ are apparent magnitudes in the Vega
system at 3.4\,$\mu$m and 4.6\,$\mu$m, respectively. Those authors also presented optical 
spectroscopy and derived the chemical composition of the ISM in this galaxy.
Lastly, the galaxy J1205$+$4551 was selected 
by \citet{I17} to be observed with the LBT because it has a high O$_{32}$ ratio (23).

Here we use LBT observations to study the hard ionizing
radiation in the objects listed in Table\,\ref{tab1}, extending the work of  \citet{I17}. 
The LBT observations and data reduction are described in 
Sect.\,\ref{sec:observations}. We derive element abundances of the sample SFGs in 
Sect.\,\ref{sec:abundances}. How the properties of galaxies depend on the hard ionizing radiation intensity  
is considered in Sect.\,\ref{sec:properties}. In Sect.\,\ref{sec:highion} we 
discuss the 
origin of hard ionizing radiation responsible for the [Ne\,{\sc v}]\,$\lambda$3426 and
He\,{\sc ii}\,$\lambda$4686 emission lines. We summarize our main results in 
Sect.\,\ref{sec:conclusions}.

\section{LBT Observations and data reduction}\label{sec:observations}

\subsection{MODS observations}

We have obtained LBT long-slit spectrophotometric observations in the
optical range of the six sample galaxies
during the period 2013 -- 2019, using the MODS1 (Multi-Object Double Spectrograph)
instrument\footnote{This paper used data obtained with the MODS 
spectrographs built with
funding from NSF grant AST-9987045 and the NSF Telescope System
Instrumentation Program (TSIP), with additional funds from the Ohio
Board of Regents and the Ohio State University Office of Research.} in the monocular mode.

MODS1 spectra were obtained with exposures 
ranging from 1800\,s to 3600\,s, in the wavelength 
range 3200 -- 10000\AA\ with 1.0 and 1.2 arcsec wide slits (Table\,\ref{tab2}). 
The seeing changed from 0.8 to 1.8 arcsec during different nights.
The airmass varied from 1.01 to 1.33, meaning that the effect of atmospheric 
refraction is small for all objects \citep{F82}. 
MODS1 spectra of several spectrophotometric standard stars were obtained during the 
same nights with a 5 arcsec wide slit for flux calibration and correction 
for telluric absorption lines in the red part.
Calibration frames of biases, flats and argon comparison lamps
were obtained during the daytime, before or after the observations.

\begin{figure*}
\centering
\includegraphics[angle=-90,width=0.90\linewidth]{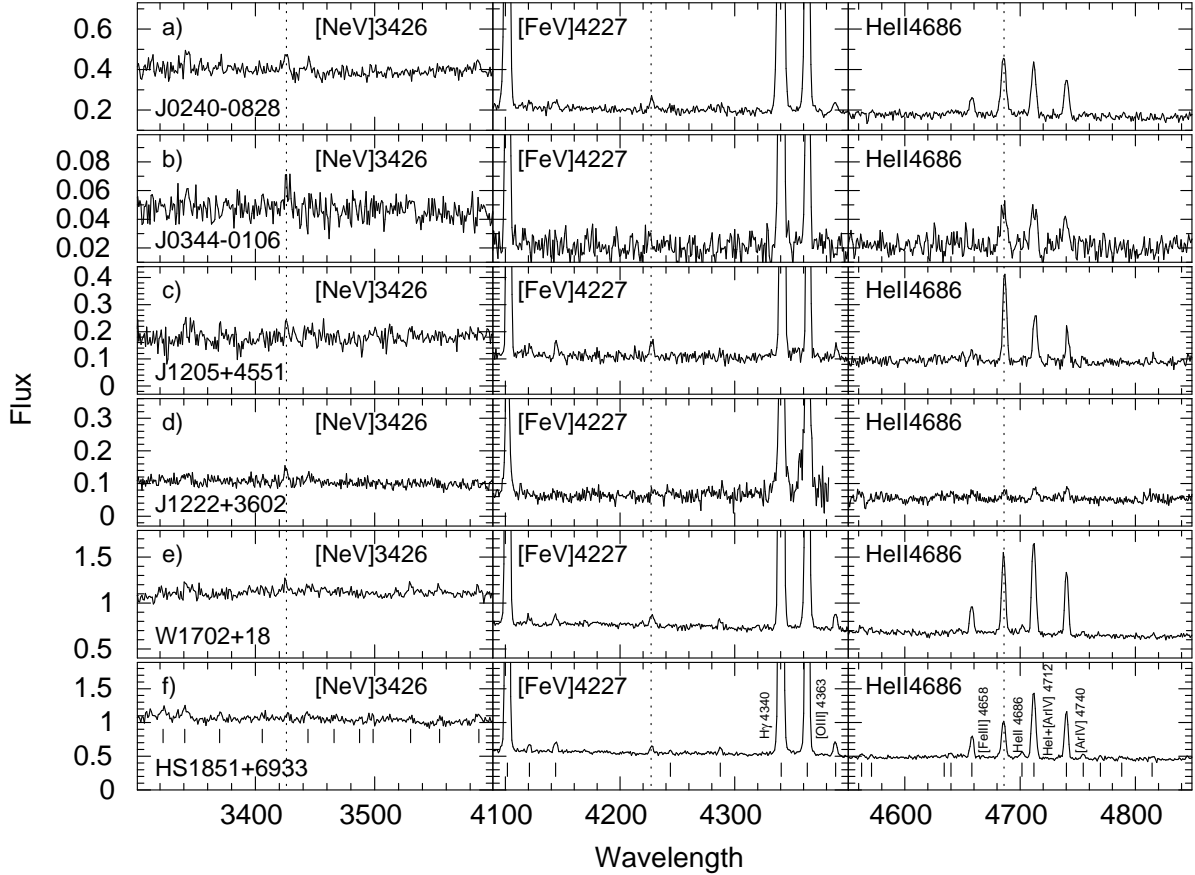}
\caption{Segments of non-corrected for extinction restframe LBT spectra 
with the [Ne\,{\sc v}]\,$\lambda$3426 emission line (left panels), 
the [Fe\,{\sc v}]\,$\lambda$4227 emission line (middle panels) and the
He\,{\sc ii}\,$\lambda$4686 emission line (right panels). Their locations
are indicated by vertical dotted lines. Locations of other weak and 
strong emission lines in the UV range of the HS\,1851+6933 spectrum 
(Table\,\ref{taba1}) are indicated in
{\bf f)} by short vertical lines. The brightest lines are labelled in {\bf f)}.}
\label{fig1}
\end{figure*}

Bias subtraction, flat field correction, wavelength and flux calibration have
been done with the MODS Basic CCD Reduction package {\sc modsCCDRed}\footnote{http://www.astronomy.ohio-state.edu/MODS/Manuals/ MODSCCDRed.pdf} and 
{\sc iraf}\footnote{{\sc iraf} is distributed by the 
National Optical Astronomy Observatories, which are operated by the Association
of Universities for Research in Astronomy, Inc., under cooperative agreement 
with the National Science Foundation.}. Selected segments of one-dimensional spectra 
of the galaxies listed in Table\,\ref{tab1}, extracted in a 1.2 arcsec aperture along the spatial axis, 
are shown in Fig.\,\ref{figa1}. Strong emission lines can be seen in these 
spectra, suggesting active star formation. Some of these lines are
labelled in the MODS spectrum of HS\,1851$+$6933 
(Fig.\,\ref{figa1}n -- \ref{figa1}o).
In particular, a strong [O\,{\sc iii}]\,$\lambda$4363 emission line is 
detected with a high signal-to-noise ratio in all galaxies, allowing a 
reliable abundance determination of their ISM. Other striking features 
are the [Fe\,{\sc v}]\,$\lambda$4227 emission line, detected in MODS1 spectra of 
nearly all galaxies, with the exception of J1222$+$3602, and the 
He\,{\sc ii}\,$\lambda$4686 emission line seen in all galaxies (Fig.\,\ref{fig1}, 
Table\,\ref{taba1}). Finally,  the [Ne\,{\sc v}]\,$\lambda$3426 emission line is 
also detected in the totality of galaxies. We also note the strong
broad component of the H$\alpha$ emission line in the spectrum of J1222$+$3602 
and the considerably weaker broad H$\beta$ component, indicating considerable 
collisional excitation of the former line (Fig.\,\ref{figa1}h). As mentioned 
before, this galaxy has been studied by \citet{IT08} who concluded that it is 
likely a low-metallicity AGN.

\subsection{LUCI observations}

In addition, near-infrared spectra in the 9000 -- 13000\,\AA\ wavelength range
during the 2013 -- 2019 period have been obtained with the LUCI1 and LUCI2
(LBT Utility Camera in the Infrared) spectrographs\footnote{LUCI1 and LUCI2 are 
mounted at the front Bent Gregorian f/15 focal stations of the LBT, and were 
designed and built by the LUCI consortium, which consists of Landessternwarte 
Heidelberg, MPE, MPIA, and Astronomisches Institut der Universität Bochum.} 
of four out of the six sample SFGs. 

These observations were aimed at detecting the He\,{\sc i}\,$\lambda$10831\AA\
emission line, the flux of which (relative to the H$\beta$ flux) is very sensitive to the 
electron number density in the H\,{\sc ii} region. The two galaxies with the highest 
redshifts in our sample, J0344$-$0106 and J1222$+$3602, were not observed because the
He\,{\sc i}\,$\lambda$10831\AA\ in these galaxies falls in the region of strong 
telluric absorption. For subtraction of the sky background, we obtained a series
of equal numbers of subexposures at two positions A and B along the slit.
The duration of each subexposure was 240\,s. 
Background subtraction has been done with the following three steps. First,
frames at positions A and B were separately co-added. Then the co-added frame at
position B was subtracted from the co-added A frame, resulting in a 
background-subtracted frame with a positive spectrum at position A and a 
negative spectrum at position B. Finally, the negative spectrum is shifted to
the position of the positive spectrum and subtracted from it, producing a final
spectrum. Each observation of a galaxy was supplemented by the observation of 
an A0 star with known brightness at two positions A and B along the slit
and at a similar
airmass for flux calibration and correction of the spectrum for telluric
absorption. Furthermore, a sequence of frames with the Ar comparison lamp and 
dark current with equal exposures were obtained during the daytime. Then
the co-added frame with dark current was subtracted from the co-added frames 
with the spectra of the Ar comparison lamp.

We note that, in the near-infrared range, very strong 
He\,{\sc i}\,$\lambda$10831\AA\ emission is observed in all our objects (compared to the 
nearby hydrogen P$\gamma$\,$\lambda$10941\AA\ emission line) (Fig.\,\ref{figa1}) 
indicating a high electron number density. This line and some other 
strong emission lines are labelled in the LUCI spectrum of HS\,1851$+$6933
(Fig.\,\ref{figa1}p).

\subsection{Extinction-corrected emission-line fluxes}

The observed emission-line fluxes and their errors were measured using the 
{\sc iraf splot} routine.
The fluxes were measured in the flux-calibrated spectra, whereas their
  errors were measured in the non-flux-calibrated spectra using Poisson
  statistics. The errors were scaled to flux units by using the ratio of the
  flux to the line count in the non-flux-calibrated spectrum. Both the fluxes
  and errors were corrected for extinction, 
derived from the observed decrement of the hydrogen Balmer emission lines
after their correction for underlying stellar absorption, following 
\citet*{ITL94}. The equivalent widths of the underlying stellar Balmer 
absorption lines are assumed to be the same for each line.
They were derived iteratively and simultaneously with the extinction
coefficient so as to reproduce the observed hydrogen Balmer decrements.
The extinction-corrected fluxes are shown 
in Table \ref{taba1}. Visible and near-infrared observations have been
obtained during different epochs, with different weather conditions and slit
widths. Therefore, in order to link the near-infrared observations to the 
optical ones, the NIR fluxes have been scaled to
obtain the theoretical recombination $I$(P$\gamma$\,$\lambda$10941)/$I$(H$\beta$)
flux ratio, after correction for extinction, and adopting the electron 
temperature $T_{\rm e}$(O\,{\sc iii}). The Table also includes, for each galaxy, 
the extinction 
coefficient $C$(H$\beta$), the rest-frame equivalent width EW(H$\beta$),
the equivalent width of the Balmer absorption lines, and the observed
H$\beta$ flux $F$(H$\beta$). 

\begin{figure*}
\centering
\includegraphics[angle=-90,width=0.99\linewidth]{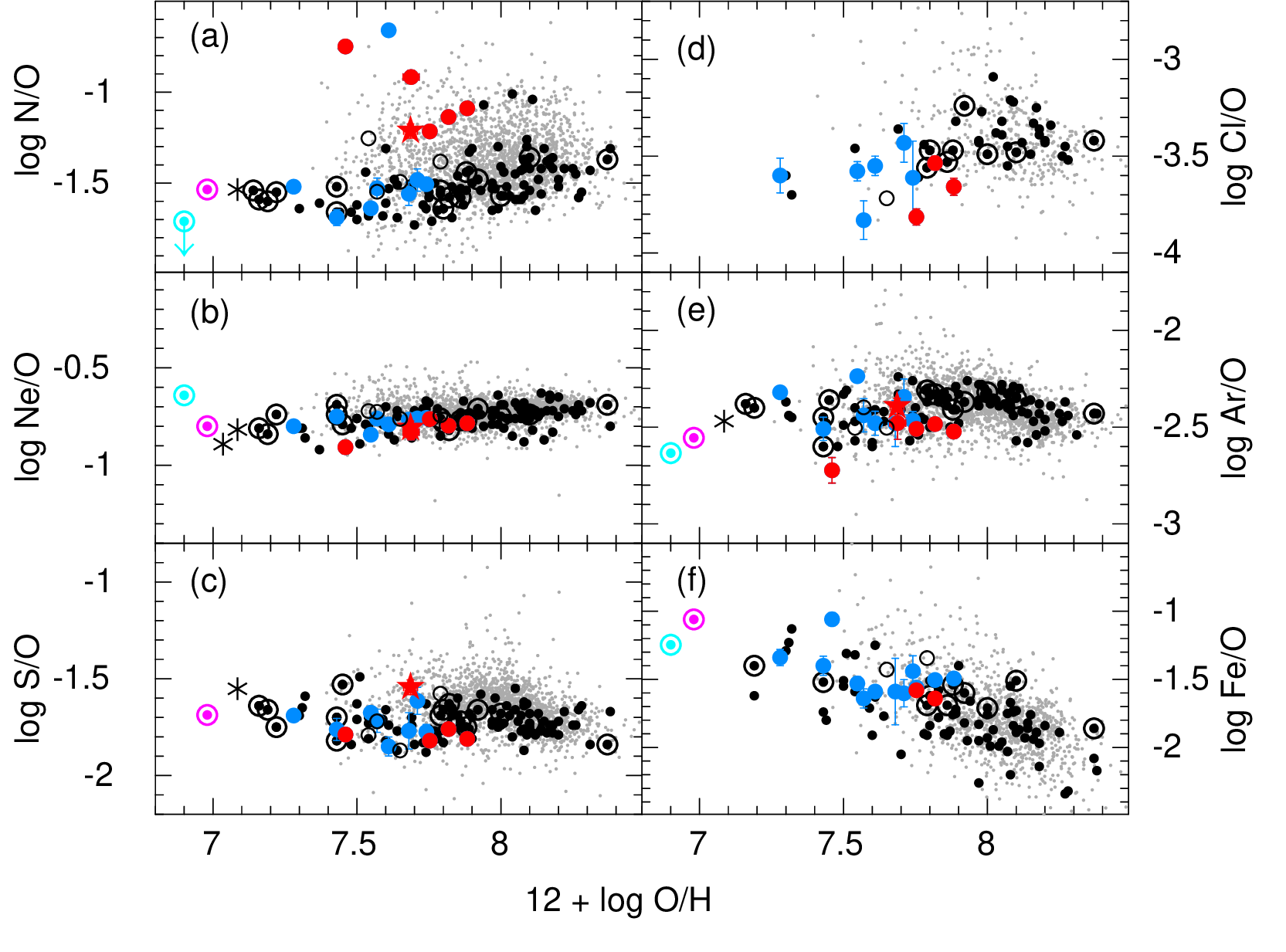}
\caption{Dependences of different elemental abundance ratios on the oxygen 
abundance 12\,+\,log O/H. The new galaxies in this paper with [Ne\,{\sc v}] 
emission, excluding J1222+3602, are shown by red filled circles. J1222+3602, a 
dwarf AGN, is shown by a red filled star. Previously known galaxies with 
[Ne\,{\sc v}] emission \citep{I04a,TI05,I12b} are shown by blue filled circles. 
For all these SFGs (blue and red symbols), we show 1$\sigma$
error bars. However, in most cases, the error bars are so small compared to the size of the circles that they are not easily seen.
For comparison, we show two most metal-deficient SFGs, J1631$+$4426 
\citep[cyan symbol, ][]{K21} and J0811$+$4730 \citep[magenta symbol, ][]{I18a},
with 12\,+\,logO/H $<$ 7.0, two other most metal-deficient SFGs,
J1234$+$3901 and J2229$+$2725, with 
12\,+\,logO/H in the range 7.0 -- 7.1 \citep[asterisks, ][]{I19,I21},
SFGs from the SDSS DR16 with [O\,{\sc iii}]\,$\lambda$4363 fluxes measured 
with an accuracy better than 4$\sigma$ (grey dots), and the HeBCD sample from 
\citet{IT04} and \citet{I04b}, used for primordial helium abundance 
determination (black filled circles). Encircled black filled circles are those
galaxies from the HeBCD sample, which have been observed in the wavelength
range covering [Ne\,{\sc v}]\,$\lambda$3426 emission line, but with no
detection of this line \citep{TI05}. Four compact SFGs, observed with the LBT, 
with extreme O$_{32}$ = 23 -- 43 and non-detected [Ne\,{\sc v}]\,$\lambda$3426 
emission line are shown by open circles \citep*{I17}.}
\label{fig2}
\end{figure*}

\section{Element abundances}\label{sec:abundances}

Following \citet{I06a}, we use the direct $T_{\rm e}$-method to derive the 
electron temperatures and element abundances.
The temperature $T_{\rm e}$(O\,{\sc iii}) is obtained from the 
[O\,{\sc iii}]\,$\lambda$4363/($\lambda$4959\,+\,$\lambda$5007) emission-line flux 
ratios. It is used to derive the abundances of O$^{3+}$, O$^{2+}$ and Ne$^{2+}$.
The abundances of N$^{+}$, O$^{+}$, S$^{+}$ and Fe$^{2+}$ are derived with the 
electron temperature $T_{\rm e}$(O\,{\sc ii}), obtained from the relation
between $T_{\rm e}$(O\,{\sc ii}) and $T_{\rm e}$(O\,{\sc iii}) \citep{I06a}, 
whereas the abundances of S$^{2+}$, Cl$^{2+}$ and Ar$^{2+}$ are obtained with the 
electron temperature $T_{\rm e}$(S\,{\sc iii}) \citep{I06a}.
The electron number densities $N_{\rm e}$(S\,{\sc ii}) are calculated from the 
[S\,{\sc ii}]\,$\lambda$6717/$\lambda$6731 emission line flux ratios.

The electron number density can also be derived from the fluxes of He\,{\sc i} 
emission lines \citep*{I14}. In general, emission in He\,{\sc i} lines
comes from recombination processes. However, it is modified by two 
non-recombination mechanisms, collisional and fluorescent excitation. The 
relative role of these two
processes is different depending on the He\,{\sc i} emission line. In particular,
the He\,{\sc i}\,$\lambda$6678 emission line flux is least sensitive to both
the collisional and fluorescent excitations, whereas the flux of 
He\,{\sc i}\,$\lambda$10831 emission line is not sensitive to fluorescent 
excitation, but is extremely sensitive to the electron number density.
Thus, the He\,{\sc i}\,$\lambda$10831/$\lambda$6678 flux ratio is a good measure
of the electron number density in a He$^+$ zone. To derive 
$N_{\rm e}$(He\,{\sc i}), we adopt the \citet*{I13} approximations of 
the \citet{P13} He\,{\sc i} emissivities.

The total oxygen abundance O/H is derived as the sum of the O$^{+}$/H$^+$, 
O$^{2+}$/H$^+$ and O$^{3+}$/H$^+$ ionic abundances using the relations of 
\citet{I06a}. For the ionic and total abundances of nitrogen, neon, sulfur, 
chlorine, argon and iron, we also use the relations of \citet{I06a}.

The electron temperatures, electron number densities, ionic abundances,
ionization correction factors, ionic and total N, O, Ne, S, Cl, Ar and Fe 
abundances are presented in Table\,\ref{taba2}. All observed galaxies have
high $T_{\rm e}$(O\,{\sc iii}), in the range 16000 -- 20600K, and atypically high
$N_{\rm e}$(S\,{\sc ii}), in the range $\sim$ 300 -- 700 cm$^{-3}$ for five 
galaxies. Only in one galaxy, J0344$-$0106, $N_{\rm e}$(S\,{\sc ii}) is as low as 
28 cm$^{-3}$. However, this galaxy is the faintest in the sample, with an 
apparent SDSS $g$-band magnitude of 21.52 mag (Table\,\ref{tab1}). Therefore, its
spectrum is noisier compared to other galaxies, and the determination of 
its $N_{\rm e}$(S\,{\sc ii}) is somewhat uncertain. The electron number densities 
$N_{\rm e}$(He\,{\sc i}) derived for the four galaxies with available 
near-infrared observations are also high, in agreement with those obtained from 
the [S\,{\sc ii}] emission-line flux ratios.

\begin{figure*}
\centering
\includegraphics[angle=-90,width=0.48\linewidth]{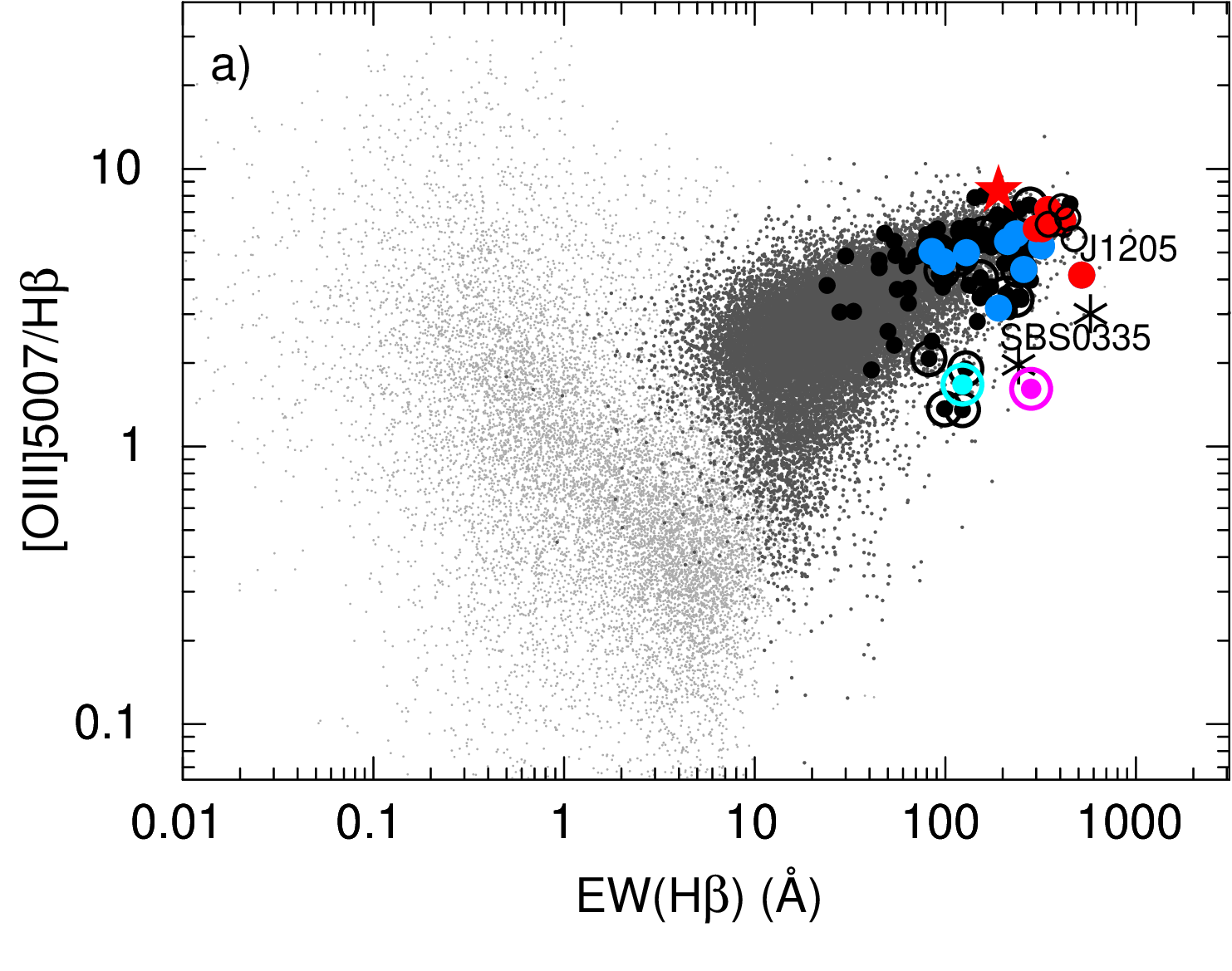}
\includegraphics[angle=-90,width=0.48\linewidth]{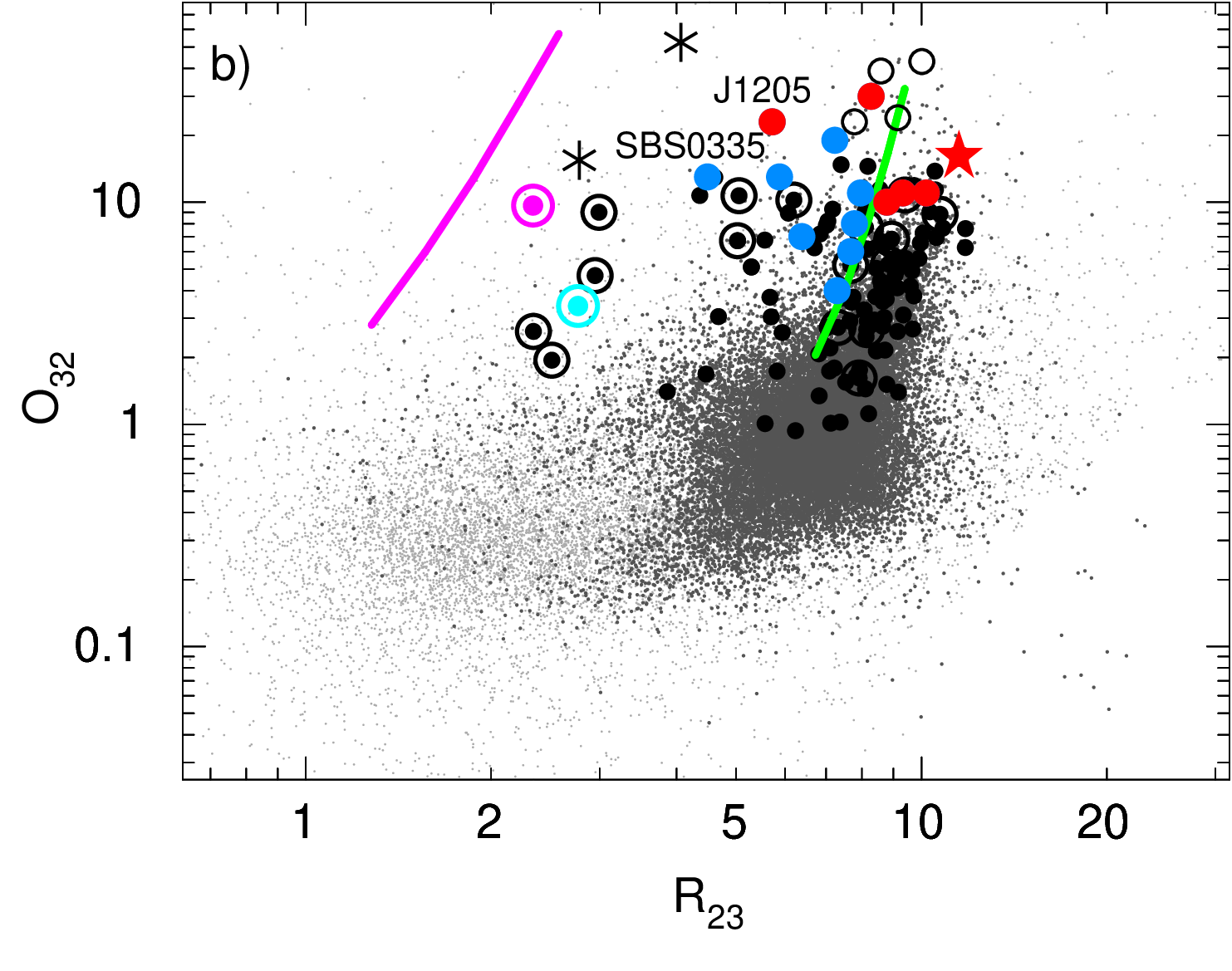}
\includegraphics[angle=-90,width=0.48\linewidth]{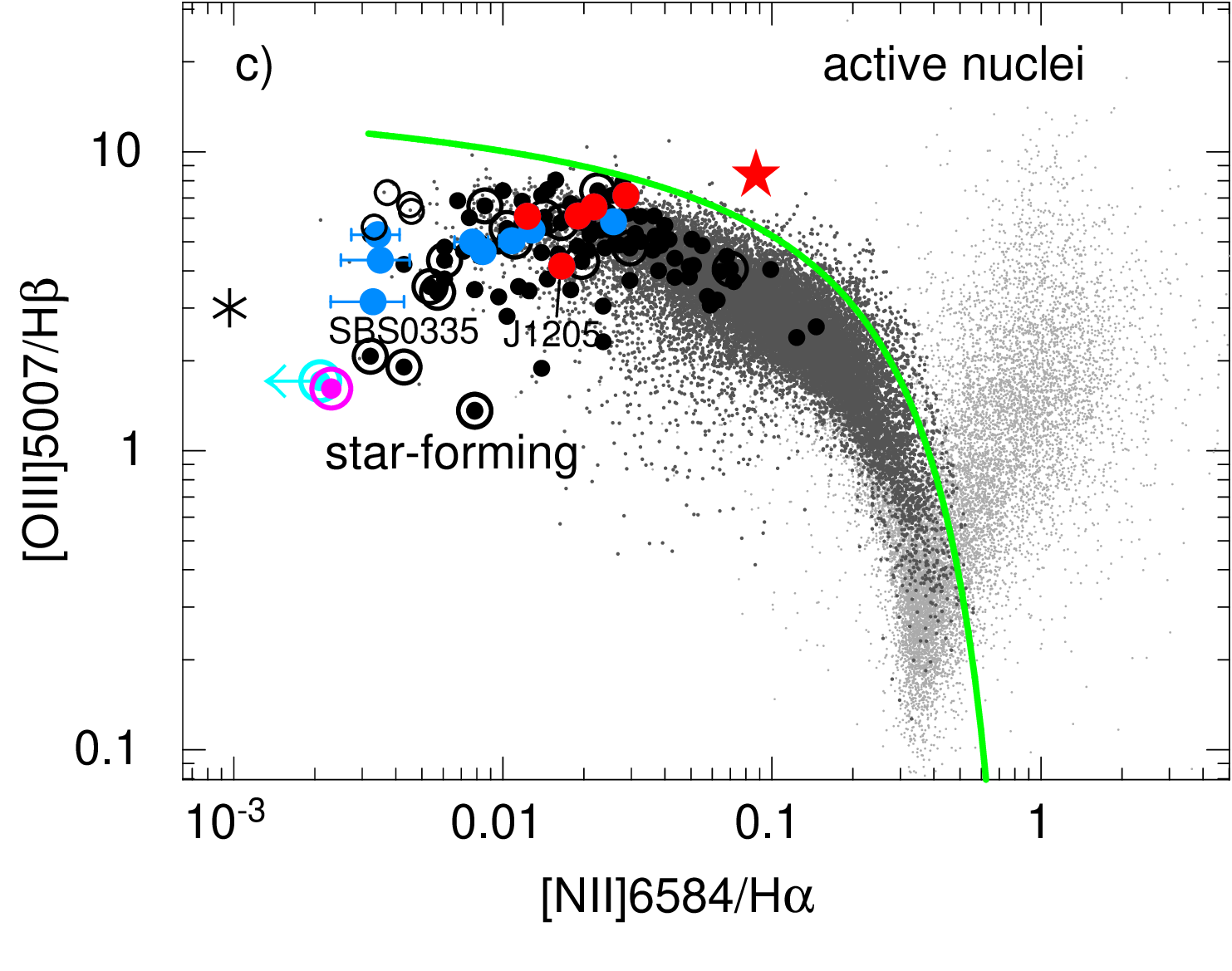}
\includegraphics[angle=-90,width=0.48\linewidth]{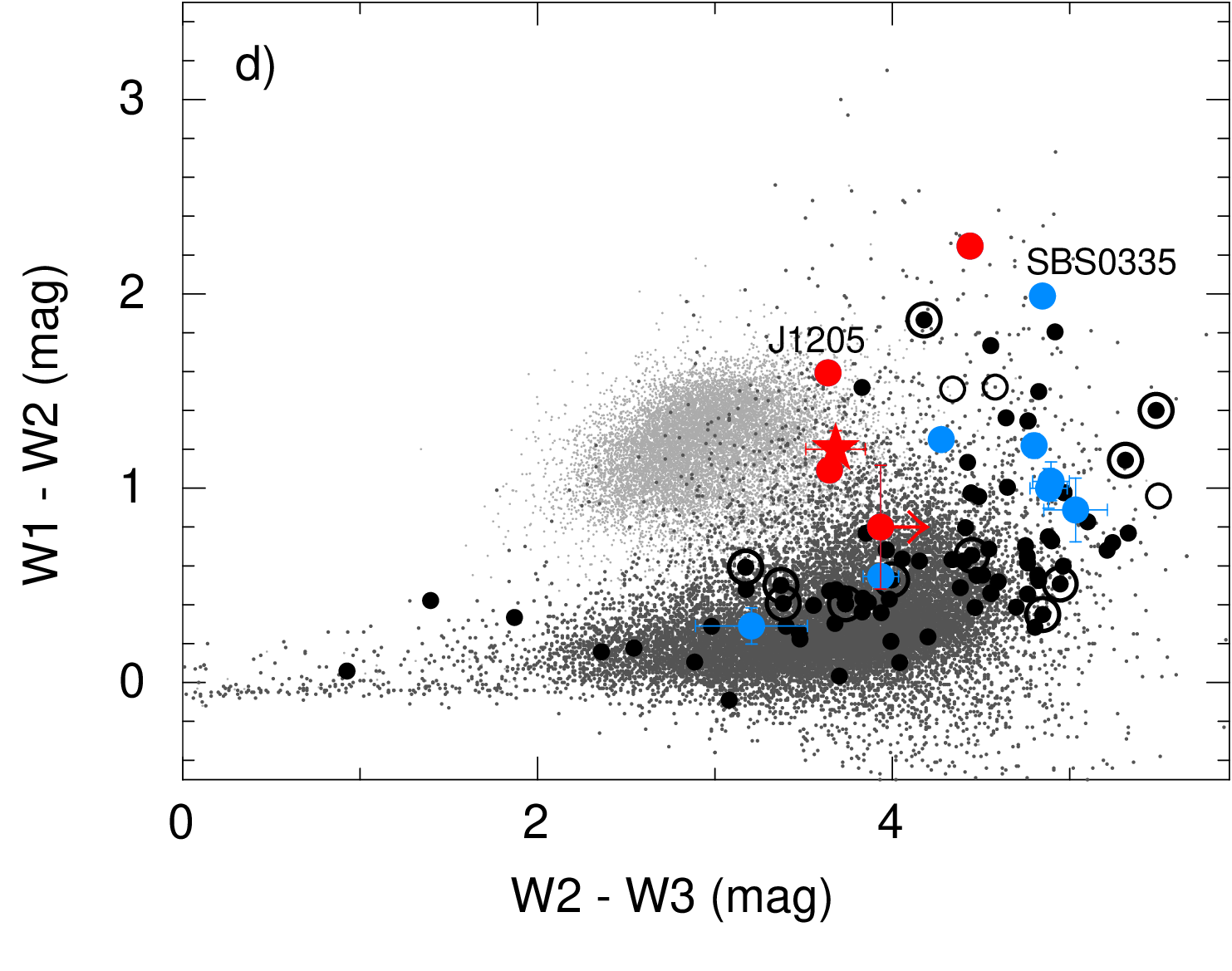}
\caption{{\bf a)} [O\,{\sc iii}]\,$\lambda$5007/H$\beta$ flux ratio vs 
equivalent width EW(H$\beta$). The new galaxies in this paper with [Ne\,{\sc v}] emission, 
excluding J1222$+$3602, are shown by red filled circles. J1222$+$3602 is shown 
by a red filled star. Previously known galaxies with [Ne\,{\sc v}] emission
\citep{I04a,TI05,I12b} are shown
by blue filled circles. For comparison, a sample of SDSS SFGs is 
represented by dark-grey dots, SDSS DR7 galaxies including AGN are shown by 
light-grey dots. Meaning of other symbols is the same as in Fig.\,\ref{fig2}. {\bf b)} O$_{32}$ = 
[O\,{\sc iii}]\,$\lambda$5007/[O\,{\sc ii}]\,$\lambda$3727 flux ratio vs. R$_{23}$ =
([O\,{\sc ii}]\,$\lambda$3727 + [O\,{\sc iii}]\,$\lambda$4959 +
[O\,{\sc iii}]\,$\lambda$5007)/H$\beta$ flux ratio. The dark-grey and 
light-grey dots
are the same as in {\bf a)}. Loci of values modelled with {\sc cloudy} code
for various ionization parameters in the range of log $U$ between $-$1.3 and 
$-$3.0 and adopting constant 12\,+\,logO/H = 7.0 and 8.0
are shown by magenta and green lines, respectively.
{\bf c)} The Baldwin-Phillips-Terlevich (BPT) diagram \citep*{BPT81}. The  
dark-grey and light-grey dots are the same as in {\bf a)}. The green solid line 
by \citet{K03} separates AGN and SFGs. {\bf d)} The {\sl WISE} ($W1$ -- $W2$) - 
($W2$ -- $W3$) colour-colour diagram, where  $W1$, $W2$, $W3$ are apparent
magnitudes in the Vega system at 3.4 $\mu$m, 4.6 $\mu$m and 12 $\mu$m, 
respectively. SDSS SFGs are represented by dark-grey dots and SDSS QSOs
\citep{S10} by
light-grey dots. HS\,1851$+$6933 is not shown because of the lack of {\sl WISE} 
data. The meaning of other symbols in {\bf b)} - {\bf d)} is the same as 
in {\bf a)}. Two very low-metallicity galaxies with detected [Ne\,{\sc v}]
emission, SBS\,0335$-$052E (blue-filled circle) and J1205+4551 (red-filled 
circle), are labelled in all panels.}
\label{fig3}
\end{figure*}

How do our abundance measurements compare with previous determinations? We derive a nebular oxygen abundance 12\,+\,logO/H in W\,1702+18 of 7.75$\pm$0.02, somewhat
higher than the value of 7.63$\pm$0.06 obtained by \citet{Gr11}. The oxygen abundance
in J1222+3602 of 7.69$\pm$0.02 is somewhat lower than the value of 7.88$\pm$0.05 obtained
by \citet{IT08}. Our determination of 12\,+\,logO/H = 7.88$\pm$0.02 for J0240$-$0828 is in 
good agreement with the value of 7.89$\pm$0.02 obtained by \citet{TI05}.

The Ne/O, S/O, Ar/O and Fe/O abundance ratios for these galaxies 
(Table \ref{taba2}, Fig.\,\ref{fig2}) follow the relations obtained for other
low-metallicity SFGs (grey dots, black filled circles, encircled black 
filled circles, open circles, asterisks), and the 
two most metal-deficient SFGs known with 12\,+\,log\,O/H\,$<$\,7.0, 
J1631$+$4426 \citep[cyan symbol, ][]{K21}
and J0811$+$4730 \citep[magenta symbol, ][]{I18a}, which are shown for
comparison. Some of the comparison galaxies, which are shown by asterisks,
open circles and encircled symbols (excluding the cyan encircled circle), have
been observed in the wavelength range covering [Ne\,{\sc v}]\,$\lambda$3426 
emission line, but with no detection. We note that, although [Ne\,{\sc v}] 
emission is not detected in J0811$+$4730 and has not been observed in J1631$+$4426, 
a relatively strong He\,{\sc ii}\,$\lambda$4686
emission line, with an intensity of $\sim$ 2 per cent that of the H$\beta$ 
emission line, is seen in both galaxies. 
Fig.\,\ref{fig2} shows that J0811$+$4730 (magenta symbol) follows well sequences 
for other SFGs, whereas J1631$+$4426 (cyan symbol) deviates somewhat more from these
sequences.

On the other hand, 
the Cl/O abundance ratios for galaxies with [Ne\,{\sc v}] emission are somewhat
lower than those of other SFGs. We also note that the N/O abundance ratios
for the new galaxies with detected [Ne\,{\sc v}] emission discussed in this paper
are systematically higher (by 0.4 -- 0.9 dex) than the average
values for the low-metallicity SFGs from the HeBCD sample used for the 
determination of primordial He abundance \citep{IT04,I04b}, in the same range 
of oxygen abundances (black filled circles). However, 
previous galaxies with detected [Ne\,{\sc v}] emission from 
\citet{I04a}, \citet{TI05} and \citet{I12b} have, with the exception of one 
galaxy, N/O ratios that are similar to those of the HeBCD sample.

The rapid decrease of the Fe/O abundance ratio with increasing oxygen 
abundance in Fig.\,\ref{fig2}f is remarkable. \citet{I06a} interpret this 
behaviour as a depletion of iron, and in a lesser extent of oxygen due to the 
formation of dust grains, which is more effective at higher metallicities. This 
is supported by the slight increase of the Ne/O abundance ratio with increasing 
oxygen abundance due to depletion of oxygen, whereas the noble element neon is 
not depleted. Contrary to that point of view, \citet{K21} have analysed the data
for the extremely metal-deficient galaxies J1631$+$4426 and
J0811$+$4730, and suggested that their very high Fe/O ratios are due to the 
increased Fe production by very massive stars, with masses $>$ 300 M$_\odot$, 
which may have existed in these galaxies.
However, that suggestion may not work because similarly high Fe/O 
are found in some compact SFGs with much higher metallicities 
(Fig.\,\ref{fig2}f).

In conclusion, the overall abundance patterns of the galaxies with detected 
[Ne\,{\sc v}] emission show no anomalies compared to the bulk of compact SFGs.
An exception is for the N/O ratio. The N/O ratio in about half of the galaxies with  
a detected [Ne\,{\sc v}]\,$\lambda$3426 emission line is similar to that in 
galaxies without [Ne\,{\sc v}] emission. On the other hand, the N/O abundance
ratio in the remaining half is by a factor of $\sim$\,3\,--\,10 higher 
(Fig.\,\ref{fig2}a). \citet{I06b} assumed that the N/O enhancement of the
interstellar medium can be due to stellar winds from Wolf-Rayet stars
enriched in nitrogen. They argued that chemical evolution models of the 
well-mixed uniform medium cannot reproduce high N/O ratios. On the other hand,
the considerable N/O enhancement can be achieved adopting a medium with dense 
nitrogen-enriched clumps around Wolf-Rayet stars embedded in the low-density 
interstellar medium.
 
\section{Integrated properties of galaxies with [Ne\,{\sc v}] emission}
\label{sec:properties}

Since the \citet{I12b} study, we have found six more SFGs with detected 
[Ne\,{\sc v}] emission, increasing the number of such SFGs to fourteen
and expanding further the range of oxygen abundances to higher values, up to
12\,+\,logO/H\,$\sim$\,7.9. We have also extended that range to the lowest 
values by including one of the most metal-deficient galaxies known, 
J0811$+$4730, with 12$+$logO/H\,=\,6.98.
Using the LBT observations of that galaxy by \citet{I18a}, 
we have estimated the 1$\sigma$ upper limit of its 
[Ne\,{\sc v}]\,$\lambda$3426/He\,{\sc ii}\,$\lambda$4686 flux ratio to be 0.15. 
We have also included the extremely metal-deficient SFG J1631$+$4426 observed 
by \citet{K21}. These authors provided intensities of emission lines including 
He~{\sc ii}\,$\lambda$4686, but not [Ne~{\sc v}]\,$\lambda$3426.

The growing number of galaxies with detected 
[Ne\,{\sc v}] emission allows us to study better their properties 
and investigate possible mechanisms for generating hard ionizing radiation.
For comparison, we use a sample of SFGs, some of which were observed in
the wavelength range which includes the [Ne\,{\sc v}]\,$\lambda$3426 emission line. This
sample was described in Sect.\,\ref{sec:abundances}. We wish to study 
similarities and differences in the integrated properties of SFGs with 
detected and non-detected [Ne\,{\sc v}]\,$\lambda$3426 emission.

\subsection{Diagnostic diagrams}

Fig.\,\ref{fig3}a shows that all galaxies with [Ne\,{\sc v}] emission are high-excitation objects that are located at the extreme end of the relation between [O\,{\sc iii}]\,$\lambda$5007/H$\beta$ 
and EW(H$\beta$) (Fig.\,\ref{fig3}a). This relation is a sort of evolutionary diagram, 
in the sense that the starburst age increases with both decreasing  
[O\,{\sc iii}]\,$\lambda$5007/H$\beta$ and EW(H$\beta$). Most of the  
galaxies with [Ne\,{\sc v}] emission 
follow well the sequence of SDSS galaxies at its upper end. 
However, there are three outliers. 
The first outlying galaxy with detected [Ne\,{\sc v}] 
emission, J1222$+$3602 (red star), likely a dwarf AGN, is offset
to a higher [O\,{\sc iii}]\,$\lambda$5007/H$\beta$ ratio, probably 
because of the presence of a non-thermal source. 
Two other galaxies, SBS\,0335$-$052E (blue filled circle)
and J1205$+$4551 (red filled circle), are offset from the
sequence of compact SDSS SFGs (dark-grey dots) to lower 
[O\,{\sc iii}]\,$\lambda$5007/H$\beta$ ratios because of their low 
metallicities, with 12\,+\,logO/H\,$\sim$\,7.3\,--\,7.4. 

For a fixed EW(H$\beta$),
there are a few other lowest-metallicity galaxies, but with non-detected
[Ne\,{\sc v}] emission, that are more offset from the
sequence of compact SDSS SFGs, to lower values 
of [O\,{\sc iii}]\,$\lambda$5007/H$\beta$. These galaxies are I\,Zw\,18NW and
I\,Zw\,18SE (the two outermost black encircled filled circles), J1234+3901 and 
J2229+2725 (asterisks with lower and higher EW(H$\beta$), respectively), 
and J1631+4426 (cyan encircled filled circle) with
lower metallicities. The most deviant galaxy is J0811$+$4730 (magenta encircled 
filled circle) because of its extremely low metallicity.

There is a considerable number of other SFGs from the HeBCD sample (black
symbols) which occupy the same region in Fig.\,\ref{fig3}a as the SDSS SFGs 
and the most intense [Ne\,{\sc v}] emitters, indicating 
similar excitation properties in their H\,{\sc ii} regions and metallicities.
Some of these galaxies were observed in the wavelength range covering 
the [Ne\,{\sc v}]\,$\lambda$3426 emission line (black encircled filled circles).
However, no [Ne\,{\sc v}] emission was detected, possibly due to the 
weakness of the line and/or an insufficient signal-to-noise ratio in the 
spectrum.

Fig.\,\ref{fig3}b shows that the galaxies with [Ne\,{\sc v}] emission are 
characterised by high O$_{32}$ ratios. Three of them are among the objects with 
the highest known O$_{32}$  ratios, $>$\,20. Most are offset from the relation 
defined by SDSS galaxies (black and grey dots), due to lower $R_{23}$ (=
([O\,{\sc ii}]\,$\lambda$3727 + [O\,{\sc iii}]\,$\lambda$4959 +
[O\,{\sc iii}]\,$\lambda$5007)/H$\beta$)) values, because of lower 
12\,$+$\,logO/H, the most extreme offset being for SBS\,0335$-$052E, with the 
lowest value 12\,$+$\,logO/H\,=\,7.3 \citep{TI05} in the [Ne~{\sc v}] sample. 
For comparison, by magenta and 
green lines are shown loci of values modelled with the {\sc cloudy} code for 
two oxygen abundances 12\,+\,logO/H = 7.0 and 8.0, respectively, and varying
log of the ionization parameter log $U$ in the range of --1.3 - --3.0.
This offset is in accord with the analysis of 
\citet{I18a} for the lowest-metallicity SFGs. In Fig.\,\ref{fig3}b is shown
for comparison the location of some lowest-metallicity galaxies with
metallicities considerably lower than those in the galaxies with
detected [Ne\,{\sc v}] emission, including the 
most deviant galaxy J0811$+$4730 (magenta encircled filled circle). This fact
indicates that J0811$+$4730 has a lower metallicity than J1631$+$4426
(cyan encircled filled circle), contrary to the claim by \citet{K21}.

It is seen in Fig.\,\ref{fig3}b that the galaxies with detected (blue and red 
symbols) and non-detected (open circles and some black encircled filled circles)
[Ne\,{\sc v}]\,$\lambda$3426 emission line are overlapping, again indicating
their similar properties, although the HeBCD galaxies (black encircled 
filled circles), which were selected for determination of the primordial 
He abundance, have systematically lower oxygen abundances.

A commonly used diagnostic diagram to distinguish between different sources of 
ionizing radiation in galaxies is the 
[O\,{\sc iii}]\,$\lambda$5007/H$\beta$ -- [N\,{\sc ii}]\,$\lambda$6584/H$\alpha$
diagram of Baldwin, Phillips \& Terlevich (BPT) \citep*{BPT81}, shown in Fig.\,\ref{fig3}c. We find that thirteen out of 
fourteen galaxies with [Ne\,{\sc v}]\,$\lambda$3426 emission are located in the
region of star-forming galaxies, indicating that hard ionizing radiation is 
produced by the hottest stars or is a result of stellar evolution.
The location of these galaxies in the diagram is similar to that of
many galaxies with non-detected [Ne\,{\sc v}]\,$\lambda$3426 emission (black open 
circles and black encircled filled circles in Fig.\,\ref{fig3}c). There is 
only one outlying galaxy, J1222$+$3602, where the source producing hard ionizing
radiation is an AGN, according to the BPT diagram.

Another possible diagnostic diagram for distinguishing whether the ionizing 
radiation comes from stars or an AGN is the {\sl WISE} colour -- colour diagram
($W1$ -- $W2$) -- ($W2$ -- $W3$), where $W1$, $W2$, $W3$ are mid-infrared {\sl WISE} 
magnitudes at 3.4, 4.6, and 12 $\mu$m. In principle, emission of galaxies in 
the $W3$ band is produced by warm dust, whereas in $W1$ and $W2$ bands it is of 
stellar and nebular origin. However, AGN can also heat dust to high
temperatures of several hundred degree resulting in a high $W1$~--~$W2$ colour
index. Thus, the location of a galaxy in the {\sl WISE} colour diagram is
determined by the properties of dust emission. In particular, the presence of
hot dust with a temperature of 
several hundred K will result in an increase of emission at 4.6 $\mu$m and 
consequently with a large (red) $W1$~--~$W2$ colour index.
In Fig.\,\ref{fig3}d are 
shown SDSS SFG (dark-grey dots) and AGN (light-grey dots) points that are 
clearly separated, indicating the presence of hot dust in AGN. 
There is however an overlapping of AGN with a small fraction of SFGs 
characterized by $W1$ -- $W2$ $\ga$ 0.5 mag and $W2$ -- $W3$ $\ga$ 3 mag. Galaxies with 
[Ne\,{\sc v}]\,$\lambda$3426 emission are shown with large coloured symbols. 
Three of them, J0240$-$0828, J1205$+$4551 and J1222$+$3602, are located at the 
edge of the AGN region, where, however, many other SFGs are detected. Thus, the 
{\sl WISE} colour -- colour diagram in Fig.\,\ref{fig3}d is not sufficient by 
itself to distinguish between stellar and
nonthermal ionizing radiation. Furthermore, two galaxies in Fig.\,\ref{fig3}d
with [Ne\,{\sc v}]\,$\lambda$3426 emission, W\,1702$+$18 and SBS\,0335$-$052E, have
$W1$\,--\,$W2$\,$>$\,2.0 mag and $W2$\,--\,$W3$\,$>$\,4.0 mag, considerably higher 
than the corresponding colour indices for AGN \citep{Gr11}. Comparing with the
galaxies from the comparison sample, we note that there is no evident 
difference in distributions of SFGs with detected and non-detected 
[Ne\,{\sc v}]\,$\lambda$3426 emission in Fig.\,\ref{fig3}d.

We conclude from the above diagnostic diagrams that hard radiation produced by 
an AGN is 
likely present only in one galaxy, J1222$+$3602, whereas there is no evidence 
for such emission in the other thirteen galaxies. Overall, 
[Ne\,{\sc v}]\,$\lambda$3426 emission is produced in high-excitation H\,{\sc ii}
regions ionized by hot massive stars in most of the galaxies shown in 
Fig\,\ref{fig3} by blue and red symbols. However, this emission is not seen in 
some other SFGs with similar excitation conditions. Therefore, diagnostic 
diagrams are not sufficient by themselves for the selection of SFGs with 
[Ne\,{\sc v}]\,$\lambda$3426 emission.

\begin{figure}
\includegraphics[angle=-90,width=0.99\linewidth]{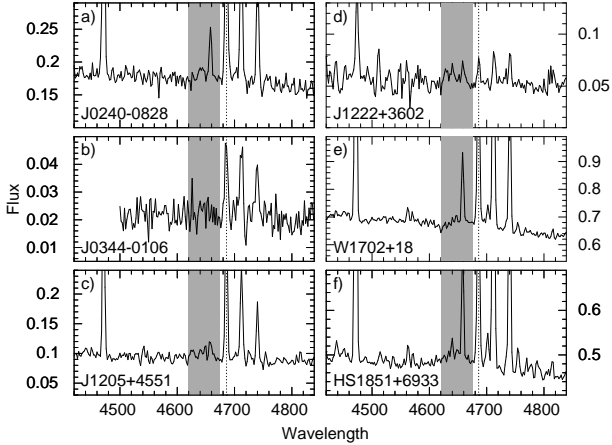}
\caption{Segments of non-corrected for extinction rest-frame LBT spectra 
of new compact SFGs with a detected [Ne\,{\sc v}]\,$\lambda$3426 emission line. 
The location of the He\,{\sc ii}\,$\lambda$4686 emission line is indicated by a 
vertical dotted line. The location of broad Wolf-Rayet lines is shown by shaded 
regions.}
\label{fig4}
\end{figure}

\subsection{Evidence for stellar winds}\label{wr}

Wolf-Rayet (WR) stellar populations with ages $<$ 4 -- 5 Myr have been suggested
\citep[e.g., ][]{SS99,SB12} to provide hard ionizing radiation with 
photon energy $\geq$ 4 ryd, and to be responsible for He\,{\sc ii} emission.
WR broad emission at $\sim$\,$\lambda$4650 (blue bump), which is a blend of 
stellar N\,{\sc iii}\,$\lambda$4634,4640, C\,{\sc iv}\,$\lambda$4658 and 
He\,{\sc ii}\,$\lambda$4686 emission lines, and a broad C\,{\sc iv}\,$\lambda$5808
emission line (red bump) are frequently detected in spectra of star-forming 
galaxies with sufficiently high signal-to-noise ratio in the continuum 
\citep[e.g., ][]{GIT00}. However, the WR broad emission is rarely seen in 
nebular He\,{\sc ii}-emitting galaxies. Furthermore, \citet{GIT00} found that 
the nebular He\,{\sc ii}\,$\lambda$4686/H$\beta$ ratio increases with decreasing
metallicity, whereas the WR population decreases with decreasing metallicity.
However, there are some exceptions, the most striking case being that of the SFG
I\,Zw\,18 with 12\,+\,logO/H $\sim$ 7.2. Both a strong nebular 
He\,{\sc ii}\,$\lambda$4686 emission line and broad blue and red WR bumps are 
seen in the spectrum with a high signal-to-noise ratio in the continuum
of its northwest component \citep{I97,L97}.

We now check for the presence of WR features in the spectra of the new SFGs with
[Ne\,{\sc v}]\,$\lambda$3426 emission. No broad C~{\sc iv}\,$\lambda$5808
emission, the ``red bump'', a signature of the early carbon WR stars, has
been detected in any of the galaxies. Additionally, no appreciable broad
He~{\sc ii}\,$\lambda$4686 emission in the ``blue bump'' is seen.
Segments of spectra that include regions of the
WR blue bump are shown in Fig.\,\ref{fig4} (shaded regions). It is seen that WR 
emission is present in the spectra of five out of the six studied galaxies,
indicating the presence of late nitrogen WR stars. The 
exception is J0344$-$0106, where the signal-to-noise ratio in the continuum is 
too low to draw definite conclusions.
The fluxes of the WR features were measured with the {\sc iraf splot}
  routine and are shown in Table\,\ref{taba1}, after removal of the nebular
  [Fe~{\sc iii}]\,$\lambda$4658 emission line in the spectra of all galaxies,
  and of the   narrow N~{\sc iii}\,$\lambda$4640 emission line in the spectrum
  of HS\,1851$+$6933.
We note that a WR population is present even in the galaxy J1205$+$4551, with
the lowest oxygen abundance 12\,+\,logO/H = 7.46 in our sample of galaxies.
Taking at face value the
very high equivalent width of 519\AA\ of its H$\beta$ emission line, this 
indicates a very young starburst in J1205$+$4551, and that its broad WR emission
is produced by the most massive stars, with masses of $\sim$\,100\,M$_\odot$.

Two galaxies, J0240$-$0828 and Tol\,1214$-$277, out of the fourteen galaxies with
detected [Ne\,{\sc v}]\,$\lambda$3426 emission, have been observed with the {\sl HST}
in the far-UV range. A broad stellar N\,{\sc v}\,$\lambda$1240 line with 
a P\,Cygni profile has been detected in the spectra of both galaxies \citep{TI97,J17}.
This implies that stellar winds, including those from WR stars, are at least 
present in some galaxies with [Ne\,{\sc v}]\,$\lambda$3426 emission.

However, WR emission is detected in only half of the fourteen
galaxies with [Ne\,{\sc v}] emission discussed here, including galaxies in
\citet{TI05} and \citet{I12b}. Furthermore, \citet{I12b} modelled the intensity 
of nebular He\,{\sc ii}\,$\lambda$4686 emission line and found that it is 
several times lower than the one observed even in the WR stage. Thus, we do not 
consider WR stars to be the main source of ionizing radiation in our sample. 

\subsection{Evidence for fast gas motions}

Another possible source of hard ionizing radiation producing 
He\,{\sc ii}\,$\lambda$4686 and [Ne\,{\sc v}]\,$\lambda$3426 emission
is radiative shocks produced
in the star-forming region by explosive processes like supernovae (SNe). 
\citet{I04a,I12b} have argued that, to explain the presence of both 
[Ne\,{\sc v}] and He\,{\sc ii} emission, the propagation of shocks at a 
velocity of 300\,--\,500\,km\,s$^{-1}$ would be necessary. Observationally, these 
shocks should manifest themselves as the broad components of strong emission 
lines, such as the H$\alpha$ emission line, formed in expanding envelopes and
produced by multiple SN events.

\begin{figure}
\includegraphics[angle=-90,width=0.99\linewidth]{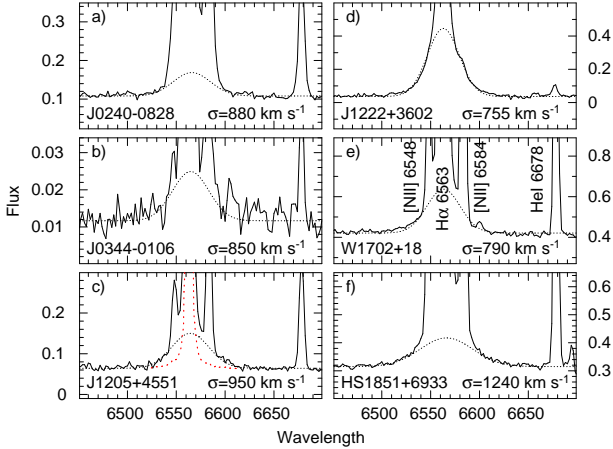}
\caption{Parts of the rest-frame LBT spectra which include the H$\alpha$ 
emission line of new compact SFGs with a detected
[Ne\,{\sc v}]\,$\lambda$3426 emission line. Emission lines are labelled in 
panel {\bf e)}. The comparison arc emission line is shown for comparison in
{\bf c)} by the red dotted line. Its maximum intensity is scaled to be equal to
the maximum intensity of the H$\alpha$ emission line.
Note the weak broad components of the H$\alpha$\,$\lambda$6563 
emission line in all spectra. They are considerably broader than the
arc line. Their fits by single Gaussians with labelled 
velocity dispersions $\sigma$ are shown by black dotted lines.}
\label{fig5}
\end{figure}

We note that \citet{I04a,I12b} and \citet{TI05} observed galaxies with 
[Ne\,{\sc v}] emission only in the blue range. On the other hand, the galaxies 
observed with the LBT/MODS presented here, have been observed 
both in the blue and red ranges. This allows us to search for low-intensity 
broad components of the H$\alpha$ emission line. In Fig.\,\ref{fig5}, we show by
a black dotted line the fit to the broad component of the H$\alpha$ line
for all galaxies in our sample.

Only the wings of the weak broad component are seen in the spectra, its
  central part being hidden by the 
  very strong narrow H$\alpha$ emission line and strong [N~{\sc ii}] emission
  lines. Therefore, we derive the parameters of the broad
  component by using a single Gaussian to fit its wings.
  As an example, in Fig.~\ref{fig5}c we show the profile
  of a comparison arc line (red dotted line), scaling its maximum profile
  intensity to the H$\alpha$ maximum intensity. It is seen that the arc line is
  considerably narrower than the H$\alpha$ line and shows much weaker broad
  wings, indicating that the stronger observed broad emission from the galaxies
  is not due to an instrumental effect, but real.
We find that the velocity dispersions range from 
$\sim$ 750 km s$^{-1}$ to $\sim$ 1240 km s$^{-1}$. 
For all galaxies, excluding the dwarf AGN J1222$+$3602, the flux ratio of 
broad to narrow H$\alpha$ components is a few per cent. On the other hand, 
this flux ratio is very high in J1222$+$3602. \citet{IT08} found it to be $\sim$\,72 per cent,
whereas the ratio derived from the LBT spectrum is $\sim$ 64 per cent.
The two observations having been made respectively in March 2005 and May 2015,
the difference may be due to some small flux  variability of the AGN.

No broad components of other emission lines have been detected in any 
  galaxy, with the exception of J1222$+$3602. A broad H$\beta$ component is detected
  in J1222$+$3602 with an intensity $\sim$ 10 times lower than that of
  the broad H$\alpha$ component and a velocity dispersion of
  $\sim$ 500 km s$^{-1}$ or 2/3 of the broad H$\alpha$ velocity dispersion.
  A broad [O~{\sc iii}] $\lambda$5007 component is also detected in the
  spectrum of J1222$+$3602, with a 
  ([O~{\sc iii}] $\lambda$5007/H$\beta$)$_{broad}$ flux ratio of $\sim$ 8.7
  and a velocity dispersion similar to that of the H$\beta$
  velocity dispersion.
  No other broad emission has been detected in this galaxy.

\section{Modelling of H\,{\sc ii} regions with hard ionizing 
radiation}\label{sec:highion}

In this section, we consider some mechanisms which can produce hard
ionizing radiation in SFGs, and verify whether they are able to reproduce 
the observed intensities of the He\,{\sc ii}\,$\lambda$4686 and [Ne\,{\sc v}]\,$\lambda$3426
emission lines. For this purpose, we have calculated models of photoionized H\,{\sc ii} 
regions with the publicly available software code {\sc cloudy} \citep{F13},
in combination with ionizing radiation derived from {\sc bpass} stellar population 
models alone \citep{E17}, or from a combination of stellar radiation of 
{\sc starburst99} population synthesis 
models \citep{L99}, with ionizing radiation produced by shocks \citep{A08} 
or by non-thermal radiation from an AGN. In these models, the
lines of  high-ionization species He\,{\sc ii} and [Ne\,{\sc v}] are due to the
ionization by
hard stellar or non-stellar ionizing radiation, whereas the
lines of low-ionization species  H\,{\sc i}, He\,{\sc i}, [O\,{\sc ii}], [O\,{\sc iii}] etc are produced
by softer stellar ionizing radiation. We aim to find models which satisfactorily
reproduce the observed ratios of both the high- and low-ionization emission 
lines.

\subsection{Stellar radiation} \label{stellar}

Hot massive stars are the main source of soft ionizing radiation in star-forming
galaxies producing strong emission lines of H\,{\sc i}, He\,{\sc i}, [O\,{\sc ii}],
[O\,{\sc iii}] and some other emission lines, mainly of singly and doubly
ionized species. However, many studies have suggested that stellar emission can 
also be an important or
even dominant source of hard ionizing radiation in low-metallicity galaxies, 
producing He\,{\sc ii} emission \citep[e.g. ][]{SS99,SB12,K18,K21}.
On the other hand, only few studies on the role of stellar radiation in the 
origin of [Ne\,{\sc v}] emission have yet been carried out \citep{I04a,I12b}. 
In general, the origin of high-ionization species is not satisfactorily
solved and the role of stellar radiation in producing both  
He\,{\sc ii} and [Ne\,{\sc v}] emission remains unclear. 
In particular, \citet{TI05} have found from the {\sc starburst99}
models that stellar radiation, including that of WR stars, is 
too soft for producing the observed He\,{\sc ii}\,$\lambda$4686 emission in 
star-forming galaxies.

The situation is even worse for producing He\,{\sc ii}\,$\lambda$4686 emission 
with fluxes above 2 per cent that of the H$\beta$ emission line and detectable
[Ne\,{\sc v}]\,$\lambda$3426 emission in low-metallicity SFGs 
\citep[Table\,\ref{taba1}, ][]{I04a,TI05,I12b}. 
Unlike He\,{\sc ii} emission, ionizing radiation with
energy above 7 Ryd is needed to produce [Ne\,{\sc v}]\,$\lambda$3426 emission.
None of the {\sc starburst99} population synthesis models with stellar 
ionizing radiation are able to reproduce this emission at the detectable flux
level of $\sim$\,0.1\,--\,1 per cent that of the H$\beta$ emission line. 

Alternatively, {\sc bpass} population synthesis models can, possibly,  
account for the origin of hard ionizing radiation. These models include 
the evolution of close binaries producing hard radiation in accretion discs, 
capable
of generating ions with high degrees of ionization. Below, we consider the 
{\sc bpass} stellar models including radiation from close binaries, and 
show that, contrary to expectations, these models are unable to reproduce the 
observed emission-line flux ratios
and their equivalent widths in spectra of SFGs with [Ne\,{\sc v}] emission.

\begin{figure*}
\hbox{
\includegraphics[angle=-90,width=0.325\linewidth]{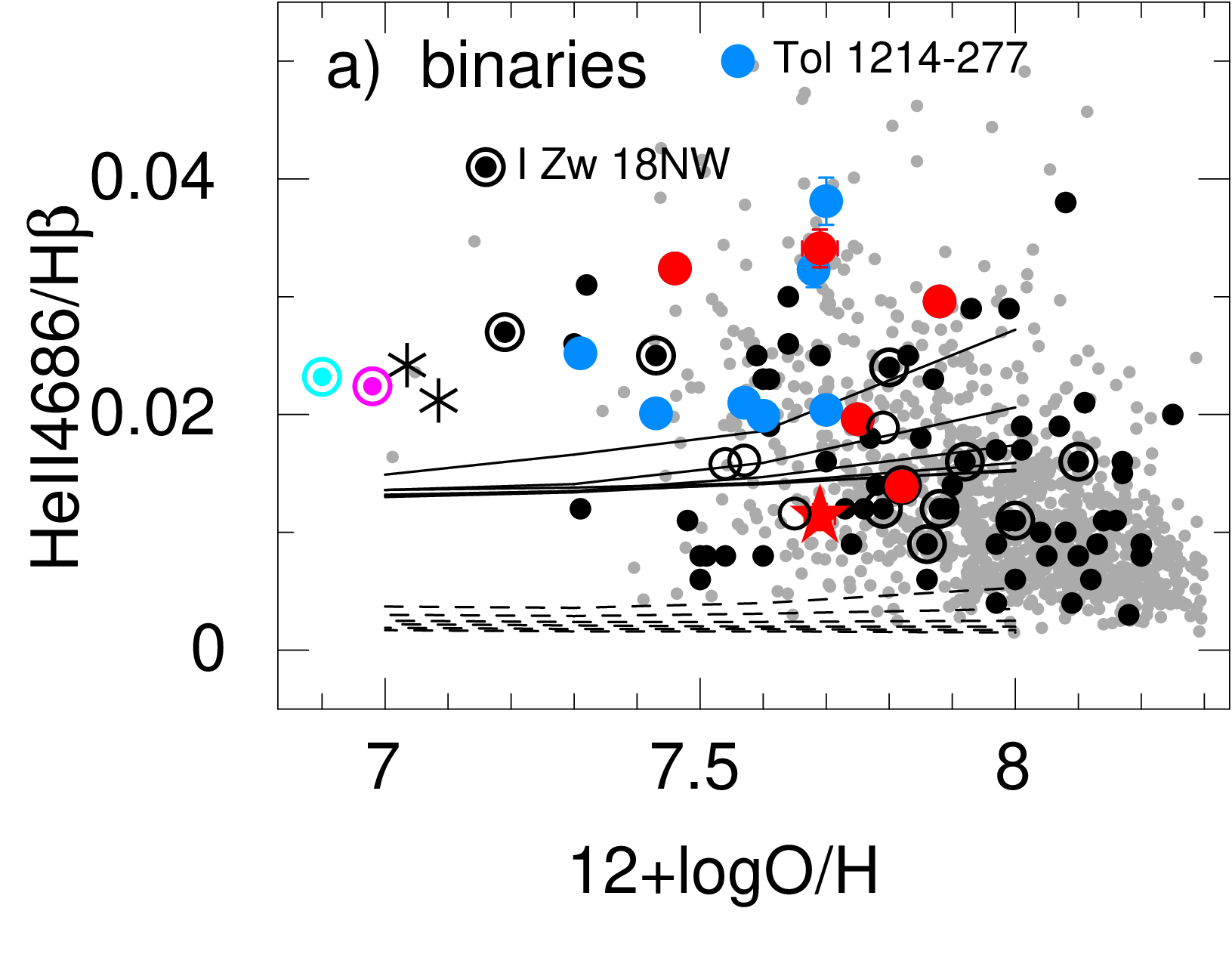}
\includegraphics[angle=-90,width=0.325\linewidth]{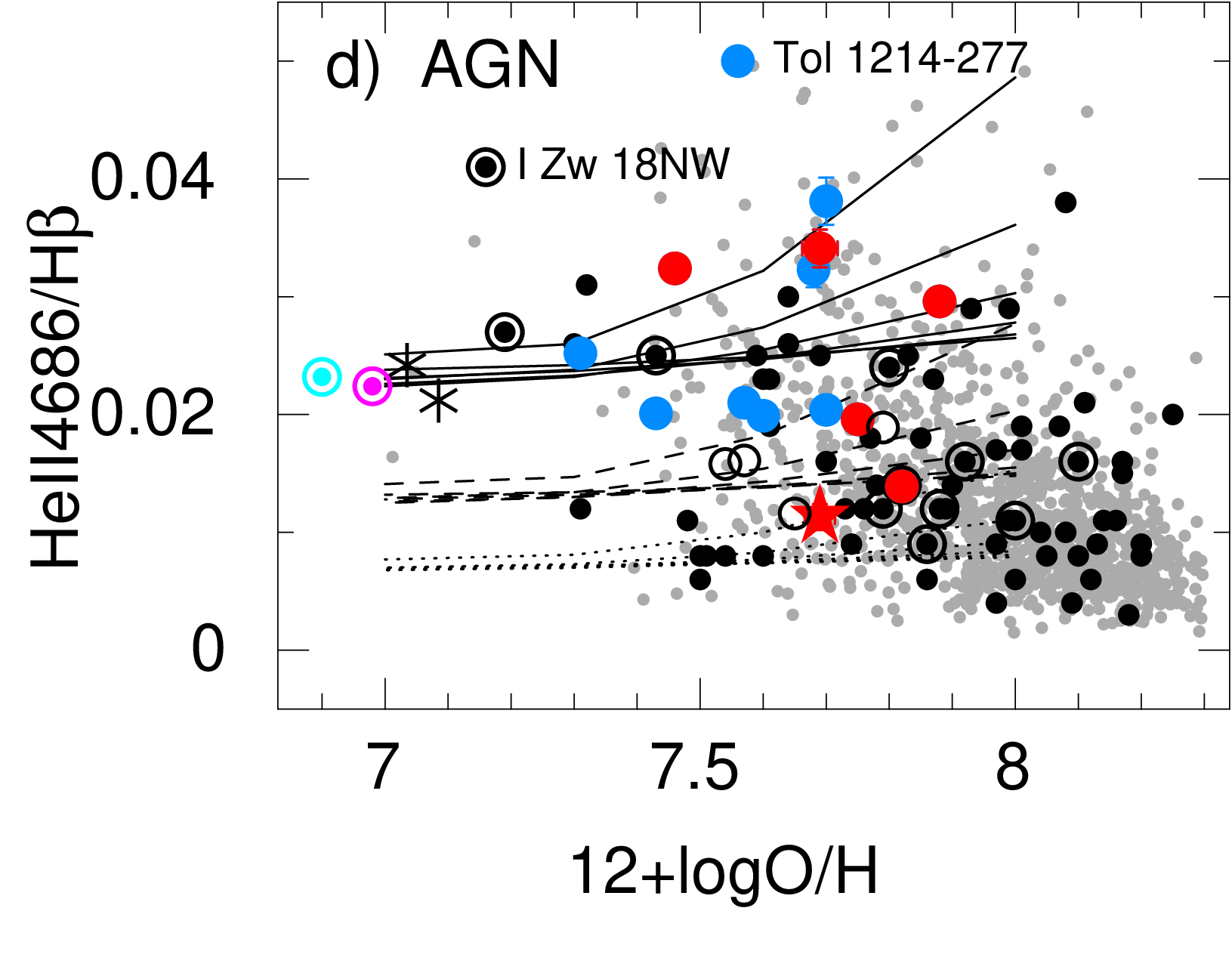}
\includegraphics[angle=-90,width=0.325\linewidth]{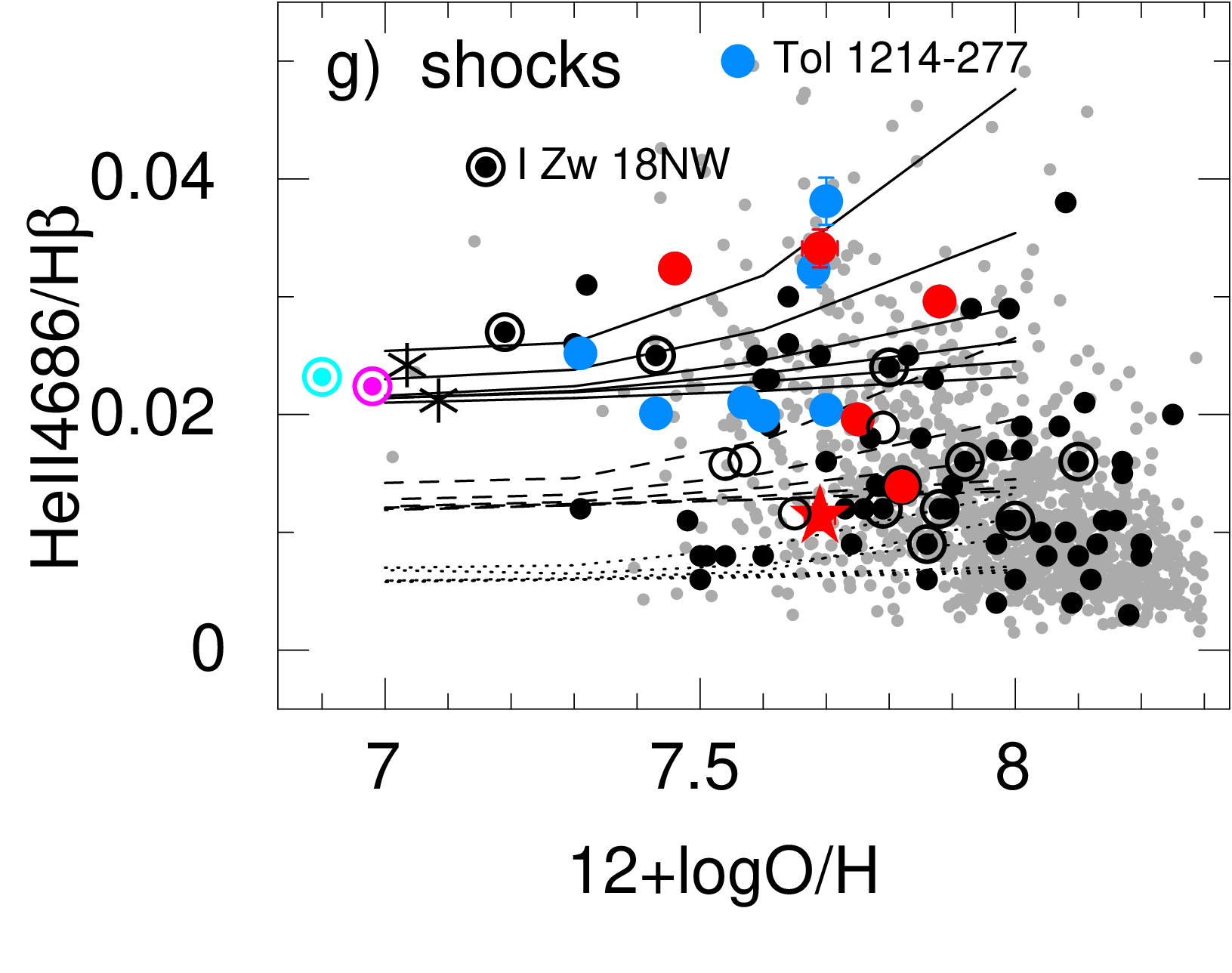}
}
\hbox{
\includegraphics[angle=-90,width=0.325\linewidth]{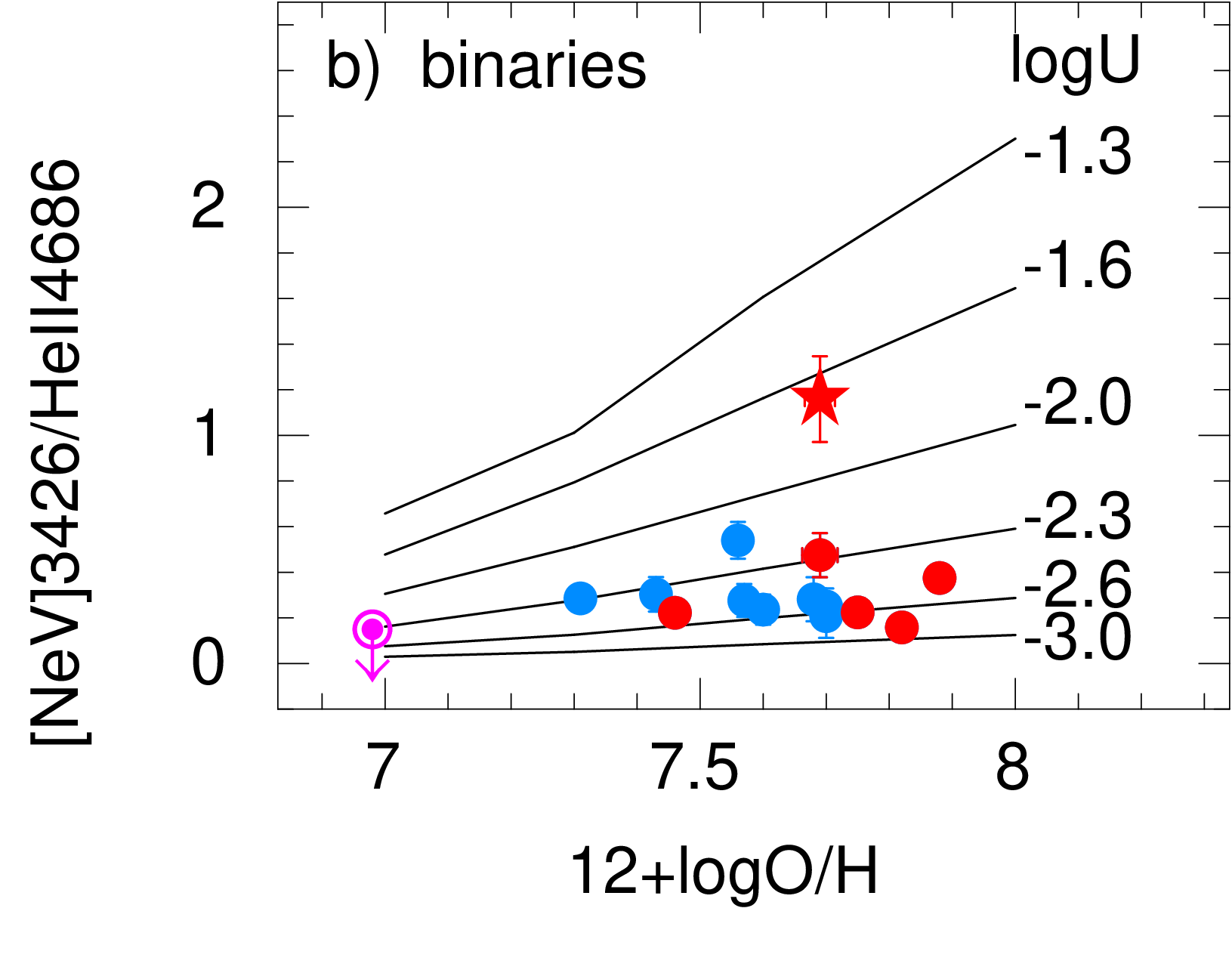}
\includegraphics[angle=-90,width=0.325\linewidth]{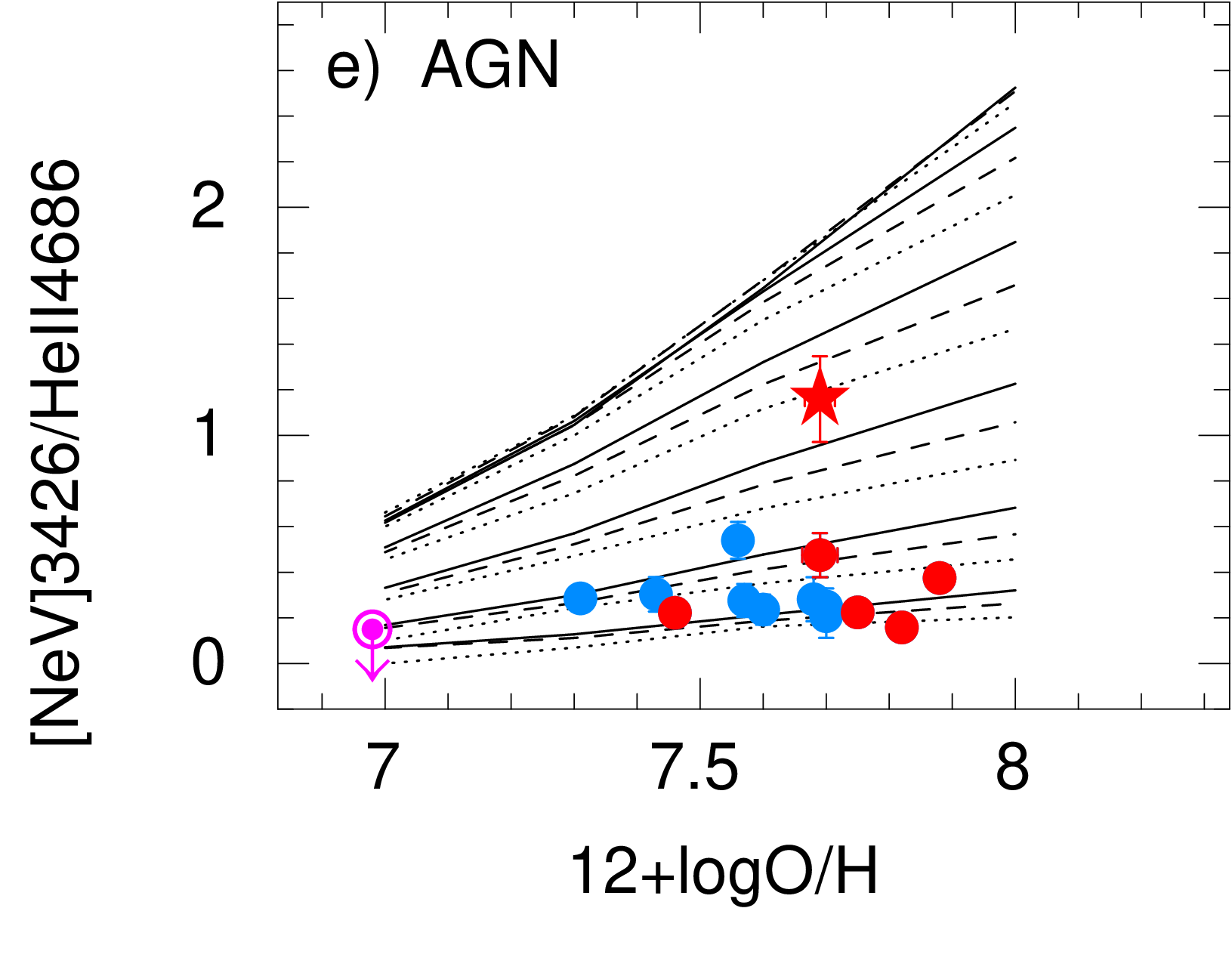}
\includegraphics[angle=-90,width=0.325\linewidth]{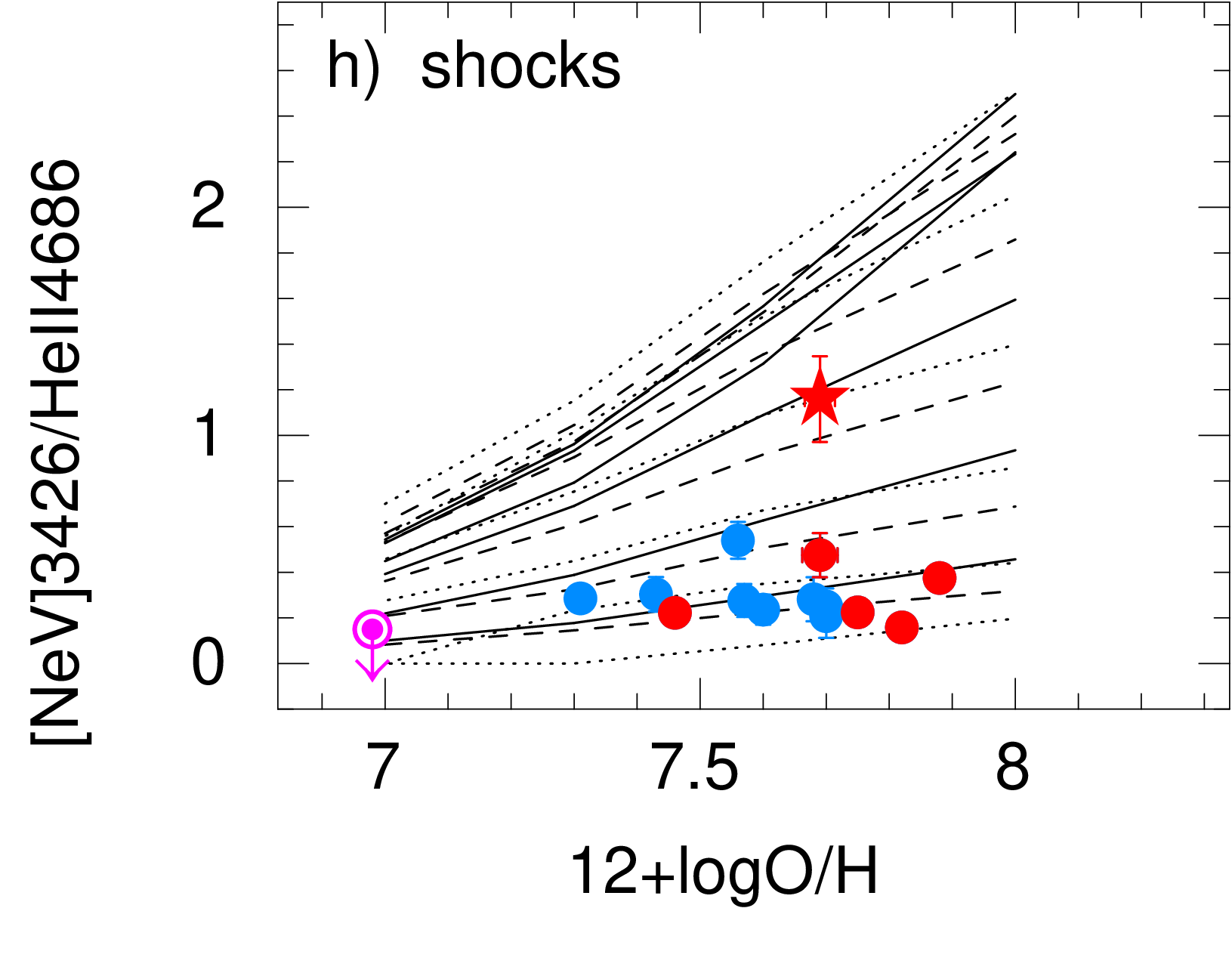}
}
\hbox{
\includegraphics[angle=-90,width=0.325\linewidth]{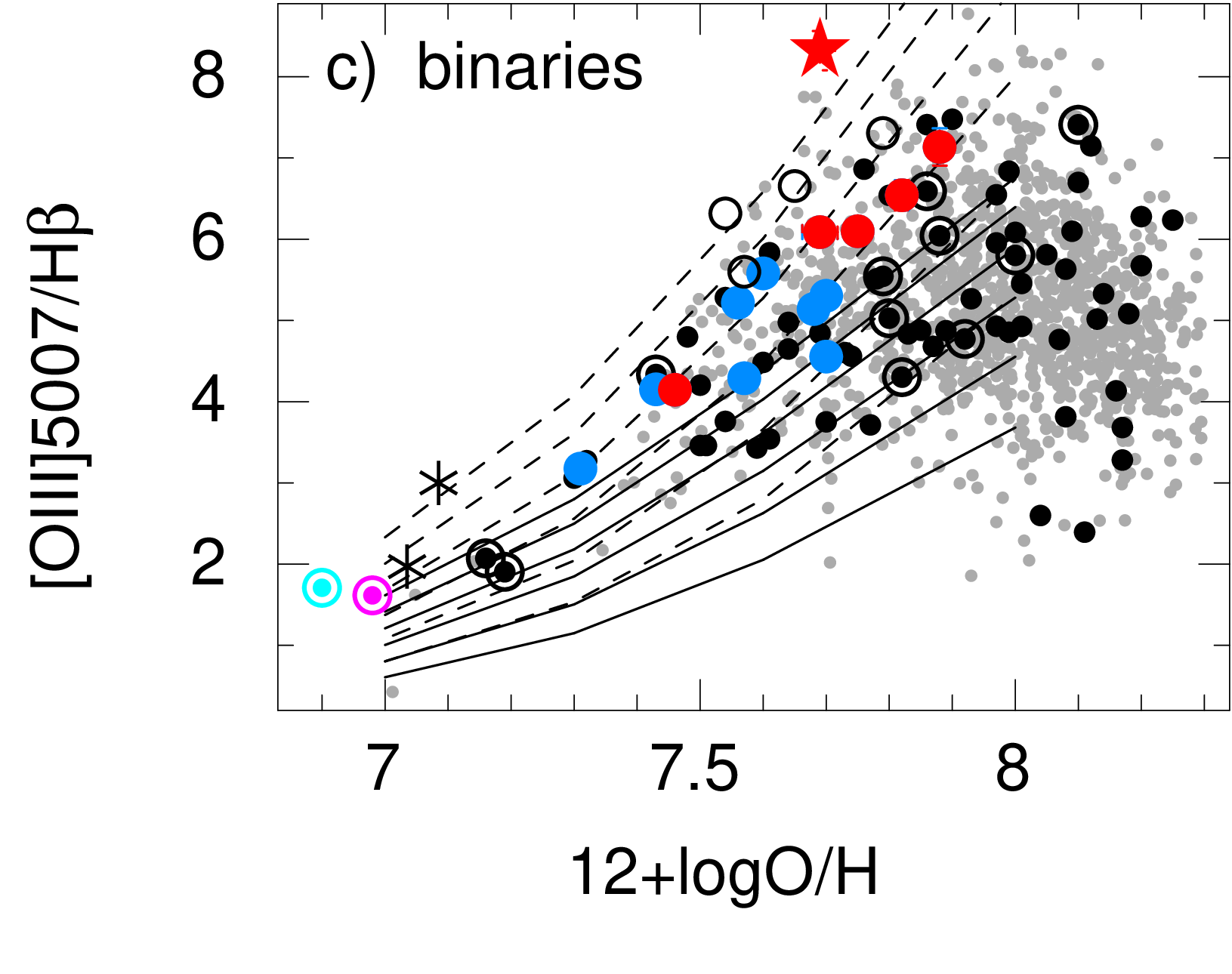}
\includegraphics[angle=-90,width=0.325\linewidth]{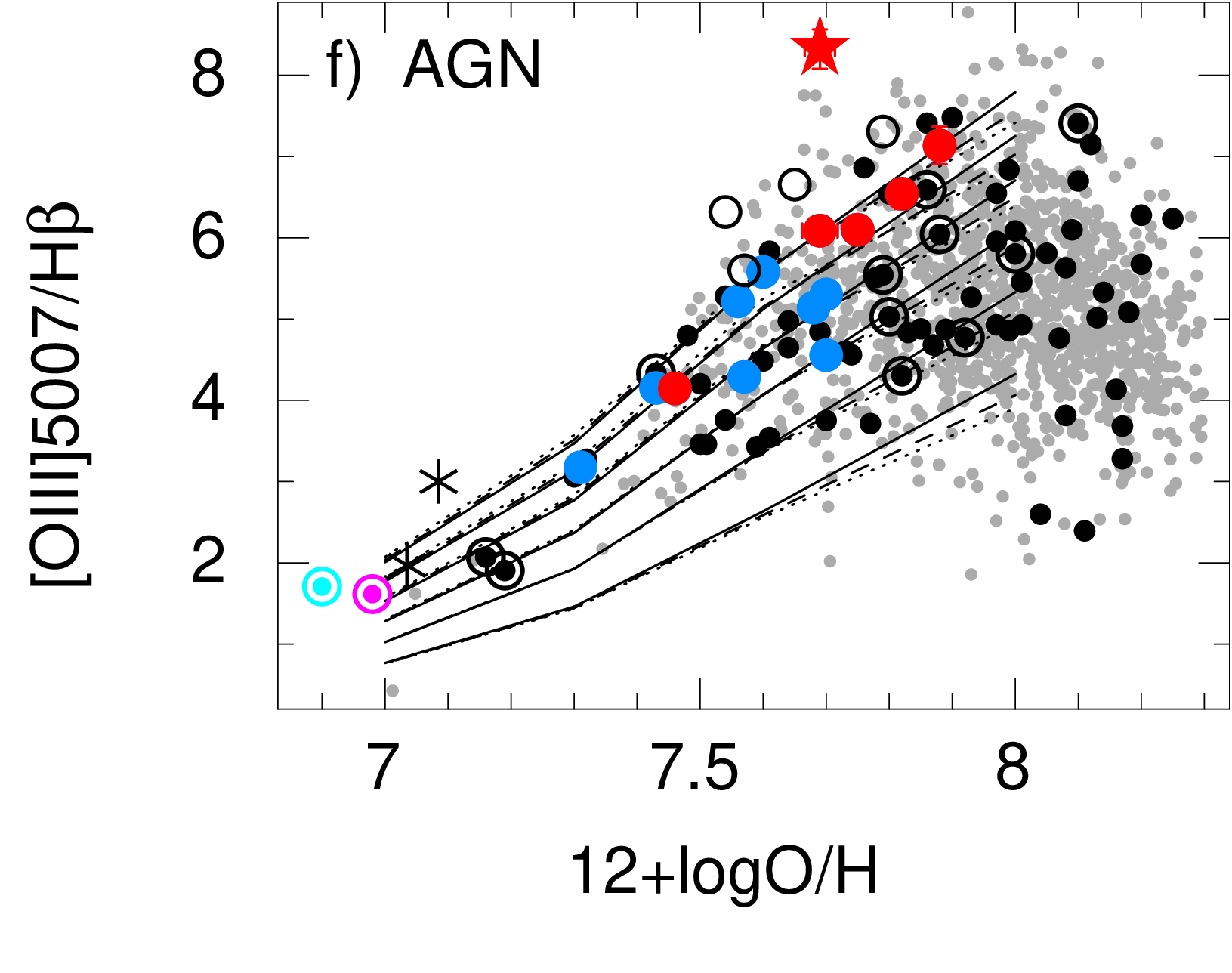}
\includegraphics[angle=-90,width=0.325\linewidth]{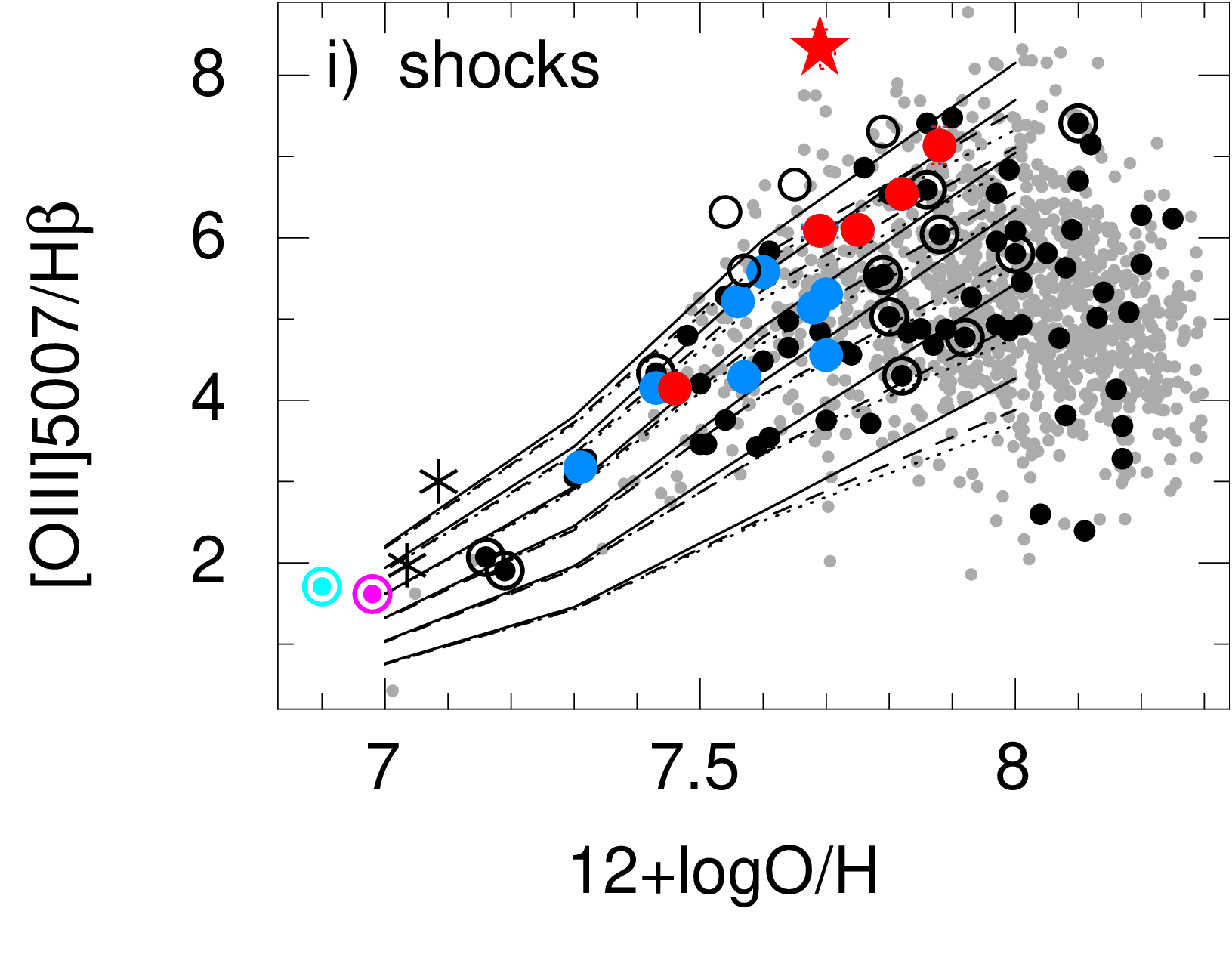}
}
\caption{{\bf a) - c)} Dependences of the He\,{\sc ii}\,$\lambda$4686/H$\beta$, 
[Ne\,{\sc v}]\,$\lambda$3426/He\,{\sc ii}\,$\lambda$4686 and 
[O\,{\sc iii}]\,$\lambda$5007/H$\beta$ emission-line
ratios on the nebular oxygen abundance 12\,+\,logO/H for {\sc bpass} models. 
Solid and dashed lines 
lines are for models with starburst ages of 20 Myr and 2 Myr, respectively.
Models are calculated with a stellar heavy element mass fraction of 10$^{-5}$ 
and various ionization parameters $U$, labelled in {\bf b)}. {\bf d) - f)} The 
same as in {\bf a) - c)}, but for {\sc starburst99} models including AGN and 
stellar ionizing radiation in proportions 5 and 95 per cent (dotted lines), 10 
and 90 per cent (dashed lines) and in proportions
20 and 80 per cent (solid lines) for various ionization parameters $U$.
{\bf h) - i)}  The same as in {\bf a) - c)}, but for models including 90 per 
cent of stellar ionization radiation of {\sc starburst99} models and 10 per cent
of shock ionizing radiation, with velocities
500\,km\,s$^{-1}$ (solid lines), 300\,km\,s$^{-1}$ (dashed lines) and 
200\,km\,s$^{-1}$ (dotted lines), for various
ionization parameters. In all panels, the galaxies from this paper are shown
with a red filled star (J1222$+$3602) and red filled circles (the five 
remaining galaxies). The galaxies with detected [Ne\,{\sc v}]\,$\lambda$3426 
emission from \citet{I04a,I12b} and \citet{TI05} are shown with blue filled 
circles. Two galaxies with the He\,{\sc ii}\,$\lambda$4686/H$\beta$ flux
ratio above 4 per cent, Tol\,1214$-$277 and I\,Zw\,18NW, are labelled in 
{\bf a), d)} and {\bf g)}. Meaning of other symbols in all panels is the same 
as in Fig.\,\ref{fig2}.}
\label{fig6}
\end{figure*}

We use the {\sc cloudy} v17.01 photoionization model calculations 
\citep{F98,F13}, adopting the ionizing radiation from {\sc bpass} v2.1 
population synthesis models, with the lowest heavy-element mass fraction 
$Z$ = 10$^{-5}$, corresponding to the hardest available ionizing radiation.
We also adopted the Salpeter IMF \citep{S55} with a slope $\alpha$=--2.35,
lower and upper mass limits $M_{\rm low}$ = 0.1 M$_\odot$ and 
$M_{\rm up}$ = 100 M$_\odot$, respectively. We note that adopting a top-heavy IMF 
with $\alpha$=--2.00 or an upper mass limit of 300\,M$_\odot$ will only slightly 
increase the calculated emission-line fluxes. Varying the ISM oxygen abundances 
12\,+\,logO/H in the range 7.0\,--\,8.0, we calculate the  
He\,{\sc ii}\,4686/H$\beta$, [Ne\,{\sc v}]\,$\lambda$3426/He\,{\sc ii}\,$\lambda$4686
and [O\,{\sc iii}]\,$\lambda$5007/H$\beta$ emission line ratios 
(Figs.\,\ref{fig6}a -- \ref{fig6}c) with the IMF parameters $\alpha$=--2.35, 
$M_{\rm low}$ = 0.1 M$_\odot$, $M_{\rm up}$ = 100\,M$_\odot$, for two starburst ages, 
$t$ = 2\,Myr (dashed lines) and $t$ = 20\,Myr (solid lines) and various ionization
parameters, labelled in Fig.\,\ref{fig6}b.

It is seen in Fig.\,\ref{fig6}a that only models with $t$ = 20 Myr
are able to produce He\,{\sc ii} $\lambda$4686/H$\beta$ ratios of $>$\,1 per cent,
but even then, the predicted values are still 
lower than those observed in most galaxies with detected 
[Ne\,{\sc v}]\,$\lambda$3426 emission (red and blue symbols) and in 
most metal-deficient SFGs (cyan and magenta encircled filled circles, and
asterisks). On other hand, they are similar to the observed intensities of 
the He\,{\sc ii}\,$\lambda$4686 emission line in HeBCD galaxies 
(black filled circles), compact SFGs from the SDSS DR16 (grey dots) and a 
sample of SFGs with extreme O$_{32}$ $>$ 20 (black open circles), which show a 
decrease of the He\,{\sc ii}\,$\lambda$4686/H$\beta$ flux ratio with increasing 
metallicity, confirming the finding by e.g. \citet{GIT00}. The 
He\,{\sc ii}\,$\lambda$4686/H$\beta$ ratio at larger 
ages ($>$\,20\,Myr) quickly decreases to zero due to a softening of the ionizing
radiation \citep{I19}. Correspondingly, it is unlikely to expect a detectable 
[Ne\,{\sc v}] emission at these ages.
We conclude that, with the higher observed values of the 
He\,{\sc ii}\,$\lambda$4686/H$\beta$ ratio, 
stellar ionizing radiation is unlikely to account for He\,{\sc ii} emission
in most SFGs with low oxygen abundances 12\,+\,logO/H\,$\la$\,7.7, 
including nearly all galaxies with detected [Ne~{\sc v}] emission 
(Fig.\,\ref{fig6}a).
In the same manner, {\sc cloudy} models in combination with {\sc bpass} v2.1 
population synthesis models including only single stars, and with 
{\sc starburst99} population synthesis models also predict a very low 
He\,{\sc ii}\,$\lambda$4686 line intensity, $\la$ 0.1 per cent that of H$\beta$.

In Fig.\,\ref{fig6}b, we show the dependence of the 
[Ne\,{\sc v}]\,$\lambda$3426/He\,{\sc ii}\,$\lambda$4686 flux ratio on oxygen
abundance 12\,+\,logO/H for the same SFGs as in Fig.\,\ref{fig6}a, but with detected 
[Ne\,{\sc v}]\,$\lambda$3426 emission.
The solid lines show {\sc cloudy} model predictions with ionizing radiation 
of the {\sc bpass} model with $Z$\,=\,10$^{-5}$ and 
$t$ = 20 Myr, for various ionization parameters $U$. 
We do not show models with $t$ = 2 Myr, because they do not produce any
[Ne\,{\sc v}]\,$\lambda$3426 emission. The increase of the modelled
[Ne\,{\sc v}]\,$\lambda$3426/He\,{\sc ii}\,$\lambda$4686 flux ratio with increasing 
12\,+\,logO/H, at a fixed $U$, is caused in part by the increased Ne abundance. 
At variance with the models, the observed ratios are nearly independent of 
12\,+\,logO/H, with log $U$ between --3.0 and --2.3. However, the assumption of an
extremely low stellar heavy element mass fraction, $Z$\,=\,10$^{-5}$, is not 
reasonable for SFGs with much higher observed nebular 12\,+\,logO/H (between 7.46 
and 7.88 or 1/17 and 1/7 solar, adopting 12\,+\,logO/H = 8.7 for the Sun).
The models with higher stellar heavy element mass fraction, 
$Z$\,=\,10$^{-3}$, corresponding to the observed nebular oxygen abundances, 
would result in several times lower fluxes of He~{\sc ii} and [Ne~{\sc v}].
In summary, we conclude that stellar radiation is unlikely to be the source of 
hard ionizing radiation responsible for the high ionization lines.

The relation between the [O\,{\sc iii}]\,$\lambda$5007/H$\beta$ emission line 
ratios and oxygen abundances are shown in Fig.\,\ref{fig6}c. It is seen that only
models with $t$\,=\,2\,Myr can reproduce the observations of the galaxies with
detected [Ne\,{\sc v}] $\lambda$3426 emission. Furthermore, the 
observed equivalent 
widths of the H$\beta$ emission line in the galaxies with detected 
[Ne\,{\sc v}]\,$\lambda$3426 emission are very high (Table\,\ref{taba1}) and are 
consistent with the $t$\,=\,2\,Myr instantaneous burst model, but not with 
the $t$\,=\,20\,Myr instantaneous burst model, 
characterised by EW(H$\beta$)\,$\sim$\,10\,--\,20\AA.
The highest EWs(H$\beta$) of $\sim$\,300\,--\,500\AA\ in some of our galaxies
can not also be explained by continuous star formation. 

These comparisons between observations and models  
support the overall conclusion that stellar ionizing radiation, calculated from 
both {\sc starburst99} and {\sc bpass} models, fails to reproduce 
simultaneously the observed emission line fluxes in galaxies with 
a detected [Ne\,{\sc v}]\,$\lambda$3426 
emission line. The instantaneous burst models with $t$\,=\,20\,Myr predict too
low line intensities of high-ionized 
(He\,{\sc ii}\,$\lambda$4686) and low-ionized ([O\,{\sc iii}]\,$\lambda$5007) 
species in these galaxies compared to observations (red and blue symbols in 
Fig.\,\ref{fig6}a -- \ref{fig6}c). 
However, the observed properties of most less 
extreme HeBCD galaxies (black filled and encircled filled circles), SFGs
with extreme O$_{32}$ $>$ 20 (black open circles) and compact 
SDSS DR16 galaxies (grey dots) with higher metallicities can be reproduced by 
the {\sc bpass} models with burst ages between 2 and 20 Myr.

\subsection{AGN} \label{agn}

Strong [Ne\,{\sc v}]\,$\lambda$3426 and He\,{\sc ii}\,$\lambda$4686 emission 
lines are commonly observed in high-excitation Sy2 galaxies, with typical 
[Ne\,{\sc v}]\,$\lambda$3426/He\,{\sc ii}\,$\lambda$4686 flux ratios of 
$\sim$\,2--3. These values can be obtained in an H\,{\sc ii} region with solar 
metallicity, ionized by a non-thermal source with a power-law distribution 
of ionizing radiation \citep[e.g. ][]{I04a}. 

The dominant source in our galaxies is softer 
stellar emission, with effective temperatures of the most massive main-sequence
stars $\la$ 55000\,K, which make it unable to excite detectable [Ne\,{\sc v}] and 
He\,{\sc ii} emission. Considerably higher effective temperatures, 
$\ga$ 100000\,K, are needed. These high temperatures can be found in
hot white dwarfs and early-type WR stars. However, luminosities of white dwarfs
are low, so that they are unable to produce large amount of ionizing
photons. Furthermore, emission lines in our SFGs are produced by young 
star-forming regions, whereas white dwarfs appear after 100 Myr. On the other 
hand, WR stellar populations are also unable to produce 
He\,{\sc ii}\,$\lambda$4686 emission line with the observed intensities. Therefore, 
an another additional source of hard ionizing radiation is needed. We consider 
here that additional source to be nonthermal power-law
radiation produced by an AGN.

\citet{I12b} have shown that to account for the observed strengths of both the
[Ne\,{\sc v}]\,$\lambda$3426 and He\,{\sc ii}\,$\lambda$4686 emission lines,
simultaneously with that of the [O\,{\sc iii}]\,$\lambda$5007 line in the galaxies with 
detected [Ne\,{\sc v}] emission, the 
number of the AGN ionizing photons should be $\sim$\,10 per cent of the 
number of softer stellar ionizing photons. Furthermore, the ISM oxygen 
abundances in our galaxies and the corresponding Ne abundances are 
about one order of magnitude lower than typical values for Sy2 galaxies. 
Therefore, if non-thermal sources of ionization are present in our SFGs, the 
[Ne\,{\sc v}]\,$\lambda$3426/He\,{\sc ii}\,$\lambda$4686 flux ratio is expected 
to be lower than in Sy2 galaxies because of their lower Ne abundance. This 
reduced ratio is indeed observed in SFGs with [Ne\,{\sc v}]\,$\lambda$3426 
emission. 

Figs.\,\ref{fig6}d\,--\,\ref{fig6}f display the same diagrams as in
Figs.\,\ref{fig6}a\,--\,\ref{fig6}c, but with population synthesis models 
combining the {\sc starburst99} stellar models with an AGN source of non-thermal
radiation, the latter contributing 5 per cent (dotted lines), 10 per cent 
(dashed lines) and 20 per cent 
(solid lines) of the total number of ionizing photons. The extreme-UV and
X-ray part of this AGN radiation, generating He\,{\sc ii} and [Ne\,{\sc v}] 
emission, is represented by a power-law \citep{F13}. Fig.\,\ref{fig6}d shows 
that the He\,{\sc ii}\,$\lambda$4686/H$\beta$ flux ratio in galaxies with 
detected [Ne\,{\sc v}]\,$\lambda$3426 emission can be reproduced 
by models mainly with 20 per cent of AGN contribution.  However, the ratios in some
galaxies can be modelled with 5 or 10 per cent of AGN contribution. 
There is one outlying galaxy with a detected [Ne\,{\sc v}] $\lambda$3426
emission line, Tol\,1214$-$277. It 
has a strong He\,{\sc ii}\,$\lambda$4686 emission line, with a flux equal to 
5 per cent that of H$\beta$, that can not be reproduced by any model.
Another object, I Zw 18NW, with a He\,{\sc ii}\,$\lambda$4686 emission 
line flux of $\sim$ 4 per cent that of the H$\beta$ emission line is located
in a region not covered by any model. However, no [Ne\,{\sc v}] emission is 
detected in I\,Zw\,18NW. On the other hand, the He\,{\sc ii}\,$\lambda$4686 
emission line is observed in most of the galaxies with a high O$_{32}$ from the 
HeBCD and SDSS samples (black open circles, black filled circles and
grey dots, respectively). It can be reproduced by models with AGN contributions
between 5 and 20 per cent.

The [Ne\,{\sc v}]\,$\lambda$3426/He\,{\sc ii}\,$\lambda$4686
and [O\,{\sc iii}]\,$\lambda$5007/H$\beta$ emission line ratios are shown in
Figs.\,\ref{fig6}e and \ref{fig6}f, respectively. It is seen that, with the 
exception of one galaxy, the models reproduce well the observations.
The exception is the dwarf AGN galaxy J1222$+$3602, where emission from
high-ionization species is formed in the narrow line region (NLR) of the AGN.
The observed [O\,{\sc iii}]\,$\lambda$5007/H$\beta$ emission line flux ratio 
in this galaxy is considerably 
higher than the range of values predicted by the models (Fig.\,\ref{fig6}f). 
In principle, increasing the fraction of ionizing photons due to the power-law 
source would increase the modelled ratio. However, this would unacceptably 
increase the predicted He\,{\sc ii}\,$\lambda$4686/H$\beta$ flux ratio, compared
to the observed one. What causes this difference between observations and models
is not clear.

All our galaxies were not detected in the X-ray range by the
  {\sl Roentgen Satellite} ({\sl ROSAT}). We also searched in the {\sl Chandra}
  source catalogue. Only J1222$+$3602 has been detected, supporting
  the conclusion that [Ne~{\sc v}]\,$\lambda$3426 and He~{\sc ii}\,$\lambda$4686
  emission is mainly powered in this galaxy by an AGN source, with a power-law
  ionizing
  emission extending to the X-ray range. All other galaxies are likely
  powered by the ionizing emission of stellar populations and radiative shocks.
  We note that
  even the bright blue compact dwarf (BCD) star-forming galaxy Tol~1214$-$277,
  with the highest
  [Ne~{\sc v}]\,$\lambda$3426/H$\beta$ flux ratio known \citep{I04a}, was not
  detected by either {\sl ROSAT} or {\sl Chandra}.

\subsection{Radiative shocks} \label{shock}

\citet{I12b} have suggested that interstellar radiative shocks with 
velocities $v_s$ of 300 -- 500 km s$^{-1}$ can also produce enough 
extreme UV radiation to account for the observed 
He\,{\sc ii}\,$\lambda$4686/H$\beta$ emission-line ratios 
of $\sim$ 2 -- 3 per cent, and for the 
[Ne\,{\sc v}]\,$\lambda$3426/He\,{\sc ii}\,$\lambda$4686 emission line ratios of
$\sim$ 0.2 -- 0.5 measured in the present sample of galaxies. The dependence of
this ratio on $v_s$ is complex, with both shock and 
precursor components having to be taken into account \citep[Fig.\,18 and Table\,6 in ][]{A08}. The fraction of
shock radiation in the H\,{\sc ii} region should be small compared to the 
fraction of softer stellar ionizing radiation in order to reproduce both the 
high- and lower-ionization emission lines. Shock radiation plays the main role in the 
production of He\,{\sc ii}\,$\lambda$4686 and [Ne\,{\sc v}]\,$\lambda$3426 emission,
whereas the softer stellar radiation is mainly responsible for the existence of lower-ionization
emission lines such as H\,{\sc i}, He\,{\sc i}, [O\,{\sc ii}], [O\,{\sc iii}] etc.
The dominance of stellar ionizing radiation is also required to reproduce 
the high observed O$_{32}$ ratios, because pure shock models predict very low O$_{32}$,
$\la$ 0.5. This is similar to the low O$_{32}$ predicted in pure AGN models.

To further investigate the role of radiative shocks in producing extreme-UV and
X-ray emission, we consider a set of {\sc cloudy} uniform and
spherically-symmetric composite models with ionizing radiation consisting of
two components. The first component is the radiation calculated with the 
{\sc starburst99} code, of a single stellar population with an age of 2 Myr, 
a production rate of ionizing photons above 1 ryd 
$Q_{\rm stellar}$ = 10$^{53}$ s$^{-1}$ (this corresponds to the average H$\beta$ 
luminosity of the studied galaxies) and various 
metallicities. The second component is the radiation from radiative shocks
calculated by \citet{A08}, coincident with the source of stellar radiation, 
and characterized by the three shock velocities of 200, 300 and 500 km s$^{-1}$.
We adopt the production rate of ionizing photons above 1 ryd, $Q_{\rm shock}$, to 
be 10 per cent of $Q_{\rm stellar}$, meaning that the fraction of the H$\beta$ 
luminosity due to shocks is set to be equal to 10 per cent of the H$\beta$ 
luminosity produced by stellar ionizing radiation. 

We calculate models varying the filling factor in a wide range so that the log 
of the ionization parameter $U$, averaged over the H\,{\sc ii} region volume,
is in the range $-$1.3 -- $-$3.0. The same range of log $U$ was adopted for 
models with pure stellar emission and for models with a mixture of stellar and 
non-thermal AGN emission (Sections \ref{stellar} and \ref{agn}). 

In Figs.\,\ref{fig6}g -- \ref{fig6}i, we show the dependences on oxygen 
abundance of the emission-line ratios discussed above, 
for composite models with shock velocities of 500 km s$^{-1}$ (solid lines),
300 km s$^{-1}$ (dashed lines) and 200 km s$^{-1}$ (dotted lines), and with 
$Q_{\rm shock}$/$Q_{\rm stellar}$ = 10 per cent.
In general, composite models with shock velocities of 500 km s$^{-1}$
reproduce well the observed emission-line ratios in galaxies with detected
[Ne\,{\sc v}]\,$\lambda$3426 emission (red and blue symbols), as is the case 
of models with power-law AGN ionizing spectra with 20 per cent of the
radiation produced by an AGN (Figs.\,\ref{fig6}d -- \ref{fig6}f). 
Correspondingly, the He\,{\sc ii}\,$\lambda$4686 emission-line intensities
in galaxies from the HeBCD and SDSS samples can better be reproduced by
models with shock velocities between 200 km s$^{-1}$ and 300 km s$^{-1}$. 
Alternatively, the observed He\,{\sc ii}\,$\lambda$4686 emission-line 
intensities can also be reproduced
by models in which the fraction of ionizing photons produced by a shock,
$Q_{\rm shock}$/$Q_{\rm stellar}$, is varied, instead of the shock velocity $v_s$.

On the other hand, given a fixed oxygen abundance, the 
[Ne\,{\sc v}]\,$\lambda$3426/He\,{\sc ii}\,$\lambda$4686 ratio
only weakly depends on variations of the fractions of the AGN or shock ionizing
radiation in the ranges considered in this paper, so that the
[Ne\,{\sc v}]\,$\lambda$3426 emission-line intensity can be estimated from the
He\,{\sc ii}\,$\lambda$4686 emission-line intensity and its detection 
in star-forming galaxies depends mainly on the signal-to-noise ratio of the 
spectrum. Practically, [Ne\,{\sc v}]\,$\lambda$3426 emission is detected in SFGs
with a He\,{\sc ii}\,$\lambda$4686/H$\beta$ line ratio $\ga$ 2 per cent
\citep[this paper, ][]{I04a,I12b}.

The reason for the similarity of the two types of models, with AGN and shock
emission, is that the ratio of the numbers of photons $Q$(He\,{\sc ii})
with energy $h\nu$ $>$ 54 eV ionizing He\,{\sc ii}, to 
$Q$(H\,{\sc i}) with energy 13.6 eV $<$ $h\nu$ $<$ 24 eV ionizing H\,{\sc i}, 
rapidly decreases with decreasing shock velocity $v_s$, in the range 
$v_s$ $\la$ 300 km s$^{-1}$. However, that ratio remains fairly constant at a 
high value, $Q$(He\,{\sc ii})/$Q$(H\,{\sc i}) $\sim$ 0.6 -- 0.7, in the velocity
range $v_s$ $\ga$ 500 km s$^{-1}$. The spectral energy distribution
for $h\nu$\,$>$\,24\,eV, including the photon energies above 97 eV responsible 
for [Ne\,{\sc v}] emission, can be represented by a power law with a slope 
$\alpha$ $\sim$\,$-$1, similar to the typical X-ray spectrum of an AGN 
\citep{F13}.

The common features which both AGN and shock models share are the following:
1) the\,He\,{\sc ii}\,$\lambda$4686/H$\beta$ emission-line ratios depend weakly on the
ionization parameter, 2) the [Ne\,{\sc v}]\,$\lambda$3426/He\,{\sc ii}\,$\lambda$4686 
emission line ratios can be reproduced only by models with low ionization
parameters and 3) the [O\,{\sc iii}]\,$\lambda$5007/H$\beta$ emission line ratios can 
be reproduced only by models with high ionization parameters. Thus, 
uniform H\,{\sc ii} region models, with a constant density, likely cannot 
account simultaneously for all observed emission-line ratios. More 
plausible are non-uniform models with low-density channels produced by outflows,
where He\,{\sc ii} and [Ne\,{\sc v}] are emitted via illumination by harder 
ionizing radiation of radiative shocks propagating through these channels,
whereas [O\,{\sc iii}] emission originates in denser regions exposed to 
softer stellar ionizing sources.

Finally, we note that the He\,{\sc ii}\,$\lambda$4686/H$\beta$ flux ratios
above 4 per cent in two BCDs, Tol\,1214$-$277 and I\,Zw\,18NW, are not reproduced 
by any model shown in Fig.\,\ref{fig6}. These unusually high values can, in 
principle, be explained if the He\,{\sc ii}\,$\lambda$4686 line is emitted in a 
density bounded H\,{\sc ii} region or in low-density holes produced by
outflows. This case is likely applicable to 
Tol\,1244$-$277 and to another SFG, J0240$-$0828, with [Ne\,{\sc v}] emission and
a He\,{\sc ii}\,$\lambda$4686/H$\beta$ flux ratio about 3 per cent. The strong 
Ly$\alpha$ emission line observed in the UV spectra of Tol\,1244$-$277 and
J0240$-$0828 \citep{TI97,J17} possibly indicates escape of ionizing 
radiation through low-density channels. In J0240$-$0828, the velocity separation
between the blue and red peaks of the Ly$\alpha$ emission-line profile
is small, $\sim$\,266\,km\,s$^{-1}$ \citep{J17}, whereas strong emission at
the center of the Ly$\alpha$ line is present in the spectrum of Tol\,1244$-$277
\citep{TI97}. The UV spectra of both objects indicate a low optical depth of
the neutral gas. However, the assumption of a density bounded H\,{\sc ii} 
region does not apply to I\,Zw\,18NW, because of the high
optical depth of neutral gas around this galaxy, as indicated by a strong 
absorption Ly$\alpha$ line \citep{K94}. 
Alternatively, \citet{Ke15,Ke21} have analysed the possible role of X-ray ionizing radiation 
from the high-mass X-ray binary (HMXB) in I\,Zw\,18NW, in generating strong He\,{\sc ii}\,$\lambda$4686 emission. 
They found that the X-ray luminosity of the HMXB is not large enough to reproduce 
the observed intensity of this line.
Thus, the presence of strong
He\,{\sc ii} emission and the absence of [Ne\,{\sc v}] emission in I\,Zw\,18NW
remains puzzling and can not be reproduced by any of the considered models.

\section{Conclusions}\label{sec:conclusions}

In this paper, we present Large Binocular Telescope (LBT) 
spectrophotometric observations of a sample of six compact star-forming galaxies (SFG)
with detected high-ionization He\,{\sc ii}\,$\lambda$4686 and
[Ne\,{\sc v}]\,$\lambda$3426 emission lines. All selected galaxies have been observed
in the optical range with the Multi-Object Dual Spectrograph (MODS) and four of them
have also been observed with the LBT Utility Camera
in the Infrared (LUCI) spectrograph, in the near-infrared $z$ and $J$ bands.
Our main results are as follows.

1. The He\,{\sc ii}\,$\lambda$4686 emission line (ionization potential 
$\sim$ 54 eV) with fluxes of $\sim$\,1 -- 3.5
per cent that of the H$\beta$ emission line, the [Ne\,{\sc v}]\,$\lambda$3426 
emission line (ionization potential $\sim$ 97 eV) with fluxes of 
$\sim$ 0.2 -- 1.6 per cent that of the H$\beta$ flux, and the [Fe\,{\sc v}]\,$\lambda$4227 emission line 
(ionization potential $\sim$ 54.8 eV) with fluxes of $\sim$ 0.3 -- 0.7 per cent that of the H$\beta$ flux
(excluding J1222+3602 which is likely a dwarf AGN) are detected in our sample of  
galaxies. The [Ne\,{\sc v}]\,$\lambda$3426/He\,{\sc ii}\,$\lambda$4686
flux ratio varies in the range from $\sim$\,0.2 to $\sim$\,0.4 in five galaxies,
and attaining the highest value of $\sim$\,1.2 in J1222$+$3602. These ratios 
are similar to those obtained for the eight galaxies with He\,{\sc ii}\,$\lambda$4686 and
[Ne\,{\sc v}]\,$\lambda$3426 emission discussed previously by \citet{I04a,I12b} and 
\citet{TI05}. However, the ratio is considerably higher in J1222$+$3602, likely 
due to the presence of an AGN and emission of both He\,{\sc ii} and 
[Ne\,{\sc v}] in its narrow-line region.

2. All sample galaxies are characterised by low oxygen abundances 
12\,+\,logO/H\,=\,7.46\,--\,7.88, very high equivalent widths 
EW(H$\beta$)\,$\sim$\,190\,--\,520\AA. Very dense H\,{\sc ii} regions with 
high electron number densities of $\sim$\,300\,--\,700\,cm$^{-3}$ are observed
in five out of six SFGs. These electron densities are derived from the 
[S\,{\sc ii}]\,$\lambda$6717/$\lambda$6731 and/or
He\,{\sc i}\,$\lambda$10831/$\lambda$6678 emission-line ratios. The 
signal-to-noise ratio of the MODS spectrum of J0344$-$0106 is too low to 
definitely derive $N_{\rm e}$(S\,{\sc ii}), and no LUCI observations are 
available for this galaxy to derive $N_{\rm e}$(He\,{\sc i}). All studied 
galaxies are also characterised by very high 
O$_{32}$ = [O\,{\sc iii}]\,$\lambda$5007/[O\,{\sc ii}]\,$\lambda$3727 
ratios, in the range 10 -- 30, indicating a high degree of ionization.

3. Five galaxies are located in the upper part of the SFG branch in the BPT
diagram, indicating ionization by stellar emission. Only H\,{\sc ii} regions in
J1222$+$3602 are likely, at least in part, powered by AGN ionizing radiation.
The [O\,{\sc iii}]\,$\lambda$5007/H$\beta$ ratio ranges from $\sim$ 4 in
the spectrum of the lowest-metallicity galaxy J1205$+$4551, to $\sim$ 8 in 
the spectrum of J1222$+$3602,
characteristic of high-excitation H\,{\sc ii} regions. Wolf-Rayet broad emission is
likely present in five galaxies. The signal-to-noise ratio of the J0344$-$0106
spectrum is too low to be sure of the presence of broad WR emission. 
Broad low-intensity components of the H$\alpha$ emission line are present in the spectra of all 
six galaxies. The flux ratios of the broad-to-narrow H$\alpha$ 
components are a few per cent in five galaxies, but goes up to $\sim$\,70 
per cent in J1222$+$3602.

4. We discuss possible sources of hard radiation producing He\,{\sc ii} and 
[Ne\,{\sc v}] emission. The main conclusion is that pure stellar ionizing 
radiation is unlikely to do the job.
Comparing the model predictions and observational data for a sample of fourteen
galaxies with simultaneously detected high-ionization lines 
He\,{\sc ii}\,$\lambda$4686 and 
[Ne\,{\sc v}]\,$\lambda$3426 (there are six new such galaxies in this paper),
we conclude that the most likely source of hard radiation is fast radiative 
shocks, with velocities of $\sim$\,500\,km\,s$^{-1}$, which are natural outcome
of massive star evolution through supernovae explosions in young star-forming 
regions. For the present sample, shock ionizing radiation in five galaxies and 
AGN emission in the sixth one, J1222$+$3602, can succesfully 
reproduce the observed fluxes of high-ionization lines. However, observed
emission-line ratios are also consistent with the presence an AGN source, 
contributing $\sim$\,10 per cent to the total luminosity of ionizing 
radiation, in all considered galaxies, in addition to the one in J1222$+$3602.

In both AGN and shock models,  the [Ne\,{\sc v}]\,$\lambda$3426 /He\,{\sc ii}\,$\lambda$4686 
emission line ratios can only be reproduced by models with low ionization
parameters while the [O\,{\sc iii}]\,$\lambda$5007/H$\beta$ emission line ratios can 
be reproduced only by models with high ionization parameters. Thus, most plausible are 
non-uniform models, where He\,{\sc ii} and [Ne\,{\sc v}] lines are 
emitted in low-density channels produced by outflows and illuminated by harder 
ionizing radiation from radiative shocks propagating through these channels,
whereas [O\,{\sc iii}] emission originates in denser regions exposed to 
softer stellar ionizing sources.

\section*{Acknowledgements}

Y.I. and N.G. acknowledge support from the National Academy of Sciences of 
Ukraine by its priority project No. 0120U100935 ``Fundamental properties of 
the matter in the relativistic collisions of nuclei and in the early Universe''.
Funding for the Sloan Digital Sky Survey IV has been provided by
the Alfred P. Sloan Foundation, the U.S. Department of Energy Office of
Science, and the Participating Institutions. SDSS-IV acknowledges
support and resources from the Center for High-Performance Computing at
the University of Utah. The SDSS web site is www.sdss.org.
SDSS-IV is managed by the Astrophysical Research Consortium for the 
Participating Institutions of the SDSS Collaboration. 
This research has made use of the NASA/IPAC Extragalactic Database (NED), which 
is operated by the Jet Propulsion Laboratory, California Institute of 
Technology, under contract with the National Aeronautics and Space 
Administration.

\section*{Data availability}

The data underlying this article will be shared on reasonable request to the 
corresponding author.

\input{ref.tex}
\appendix

\section{Emission-line fluxes and element abundances}

\begin{figure*}
\hbox{
\includegraphics[angle=-90,width=0.32\linewidth]{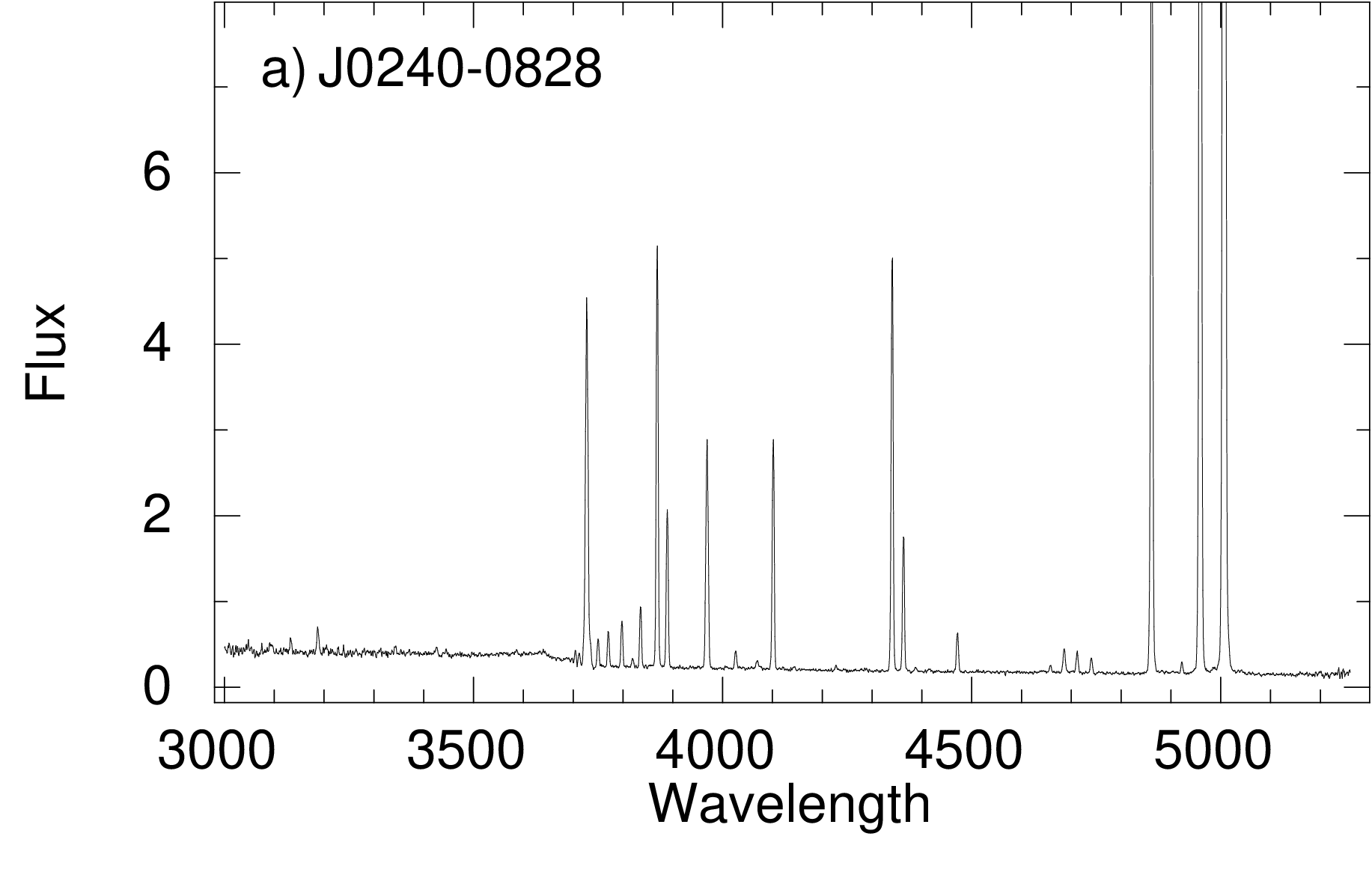}
\includegraphics[angle=-90,width=0.32\linewidth]{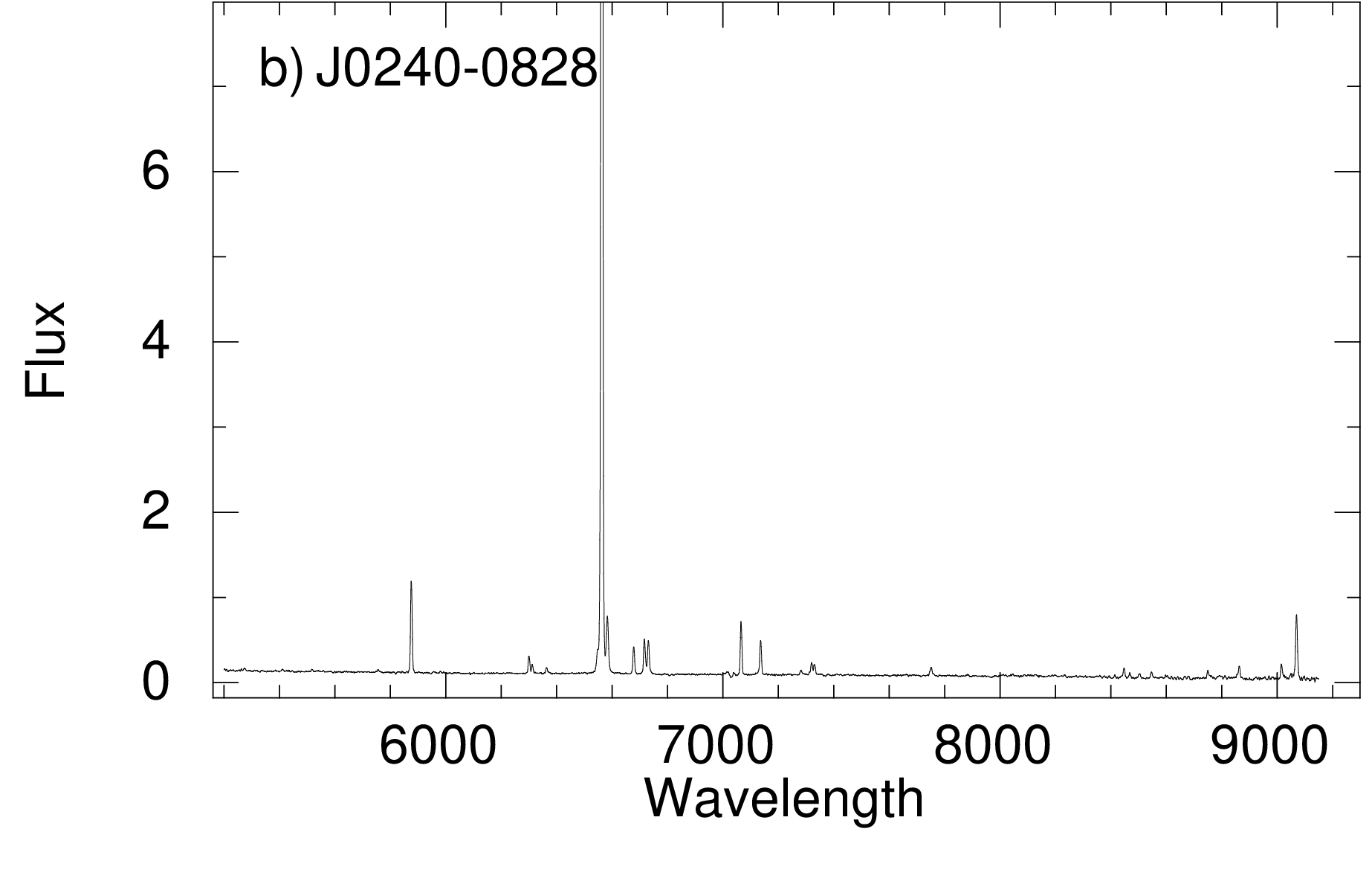}
\includegraphics[angle=-90,width=0.32\linewidth]{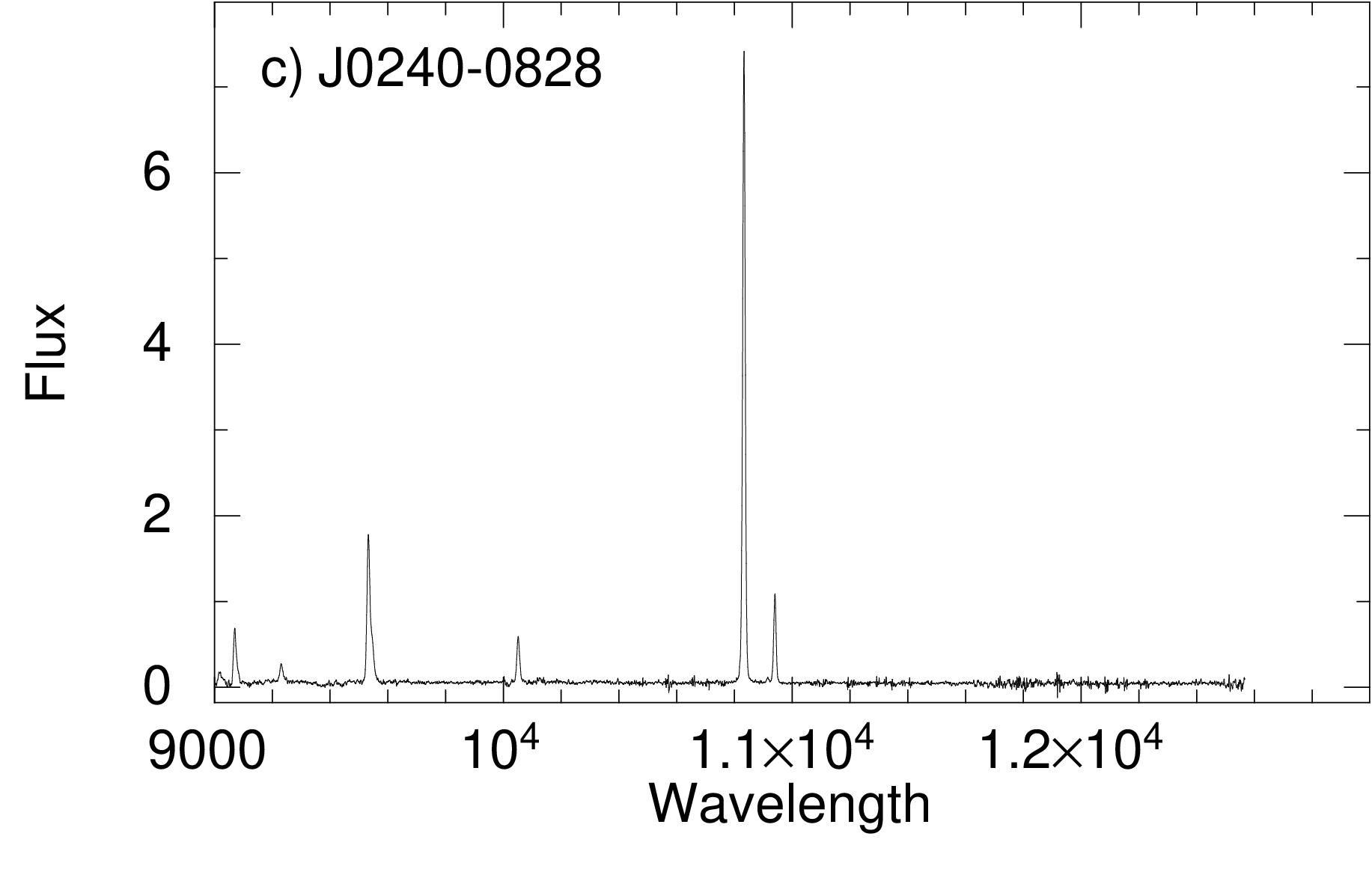}
}
\hbox{
\includegraphics[angle=-90,width=0.32\linewidth]{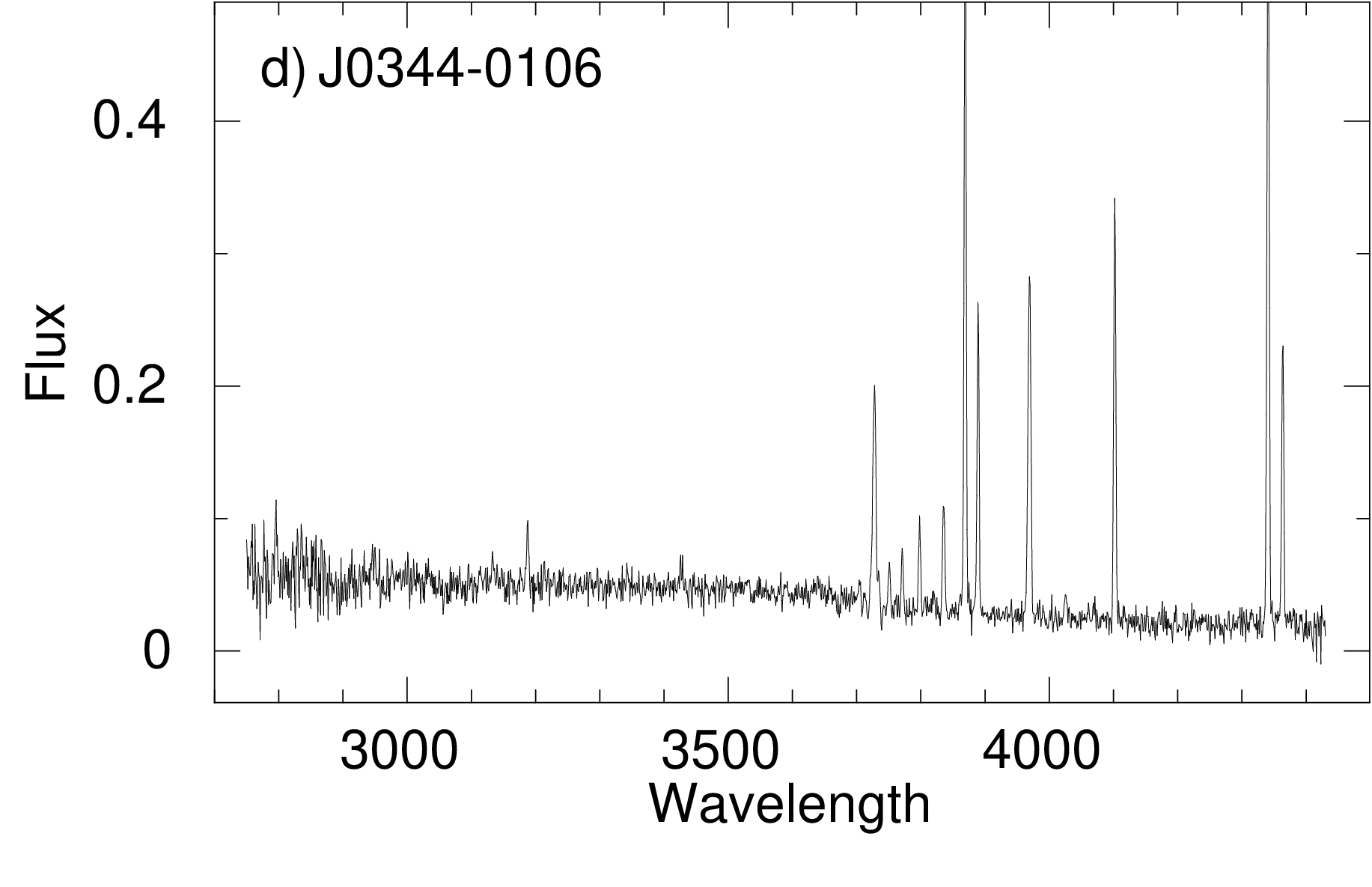}
\includegraphics[angle=-90,width=0.32\linewidth]{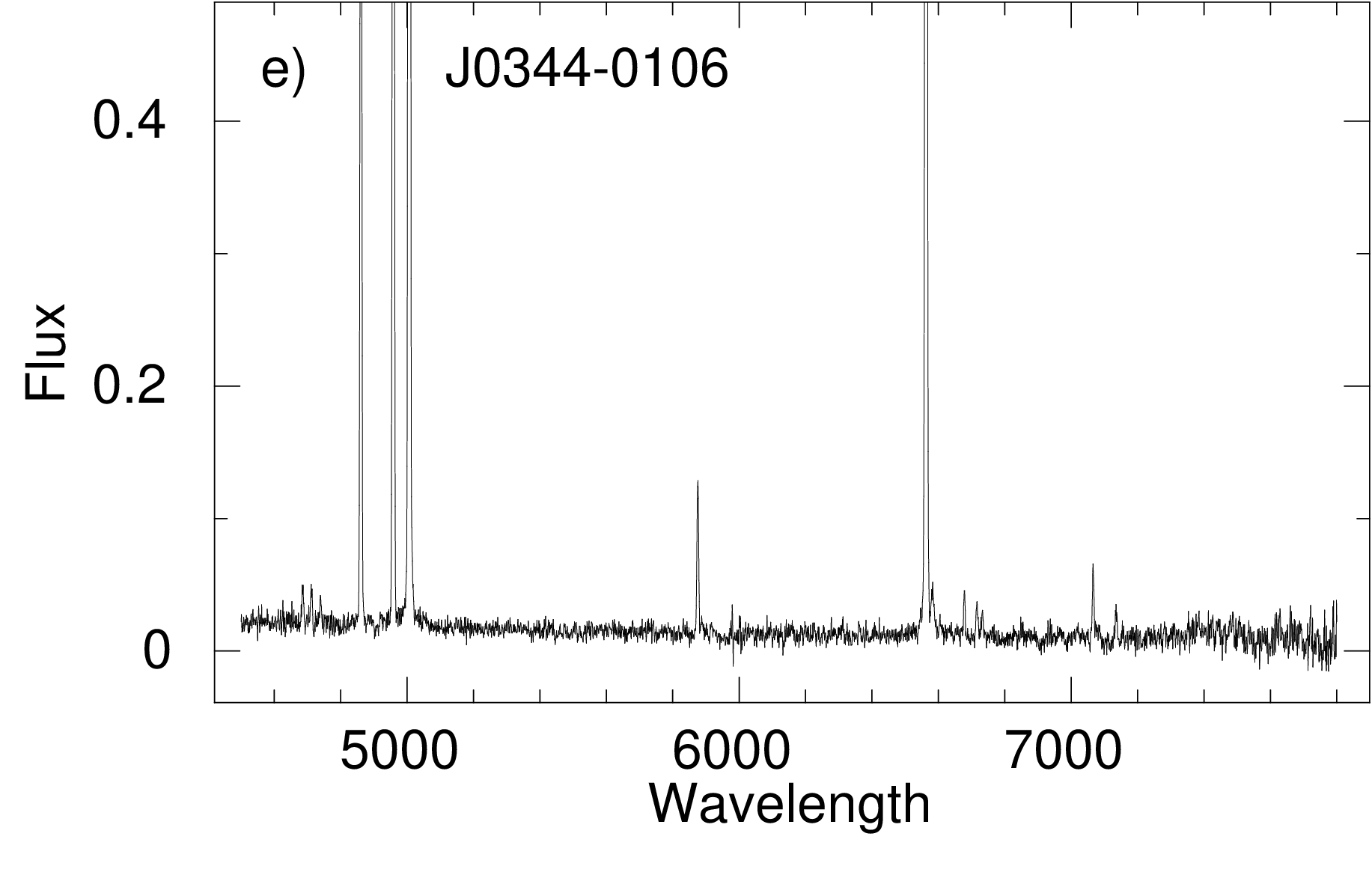}
}
\hbox{
\includegraphics[angle=-90,width=0.32\linewidth]{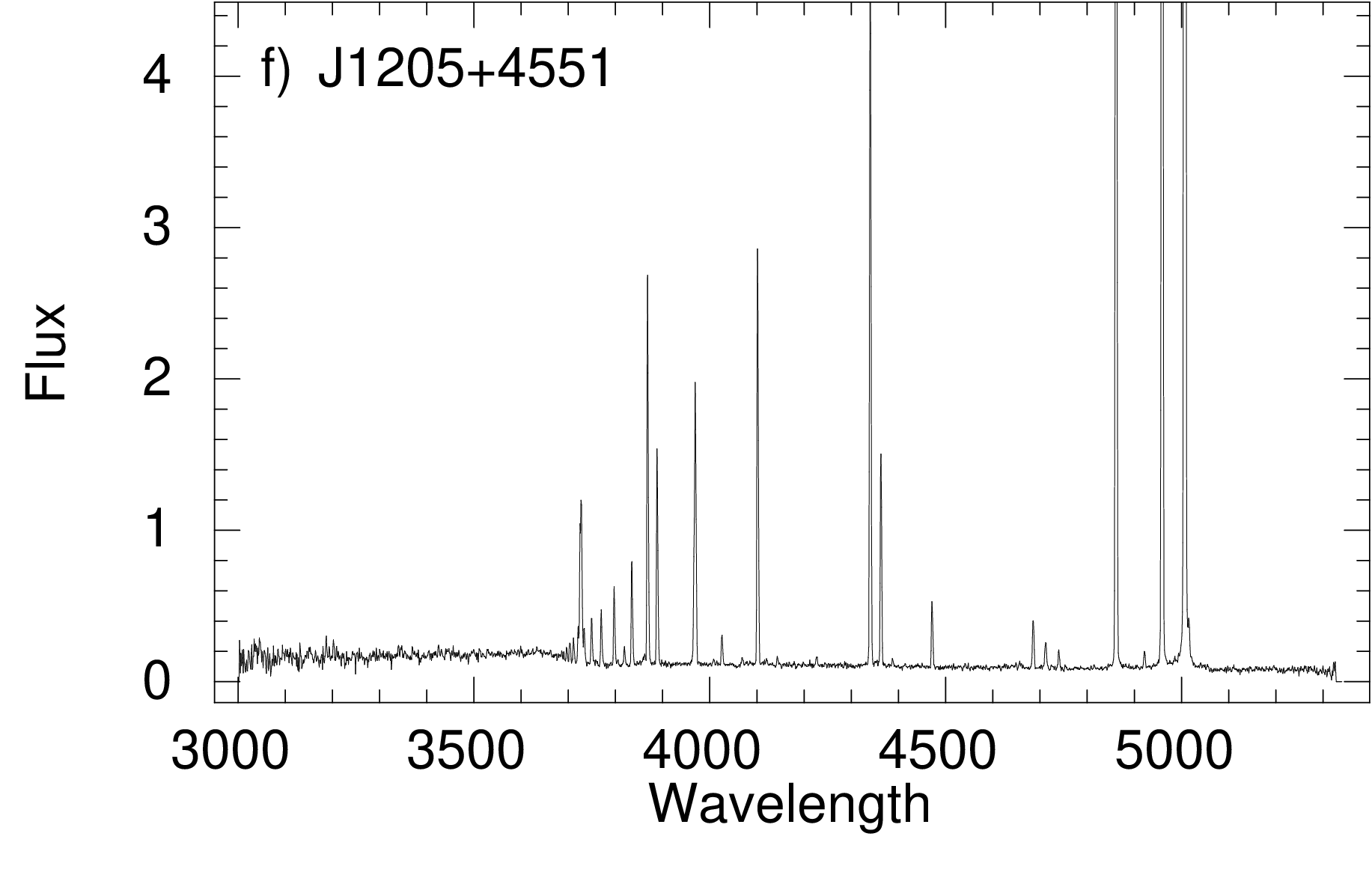}
\includegraphics[angle=-90,width=0.32\linewidth]{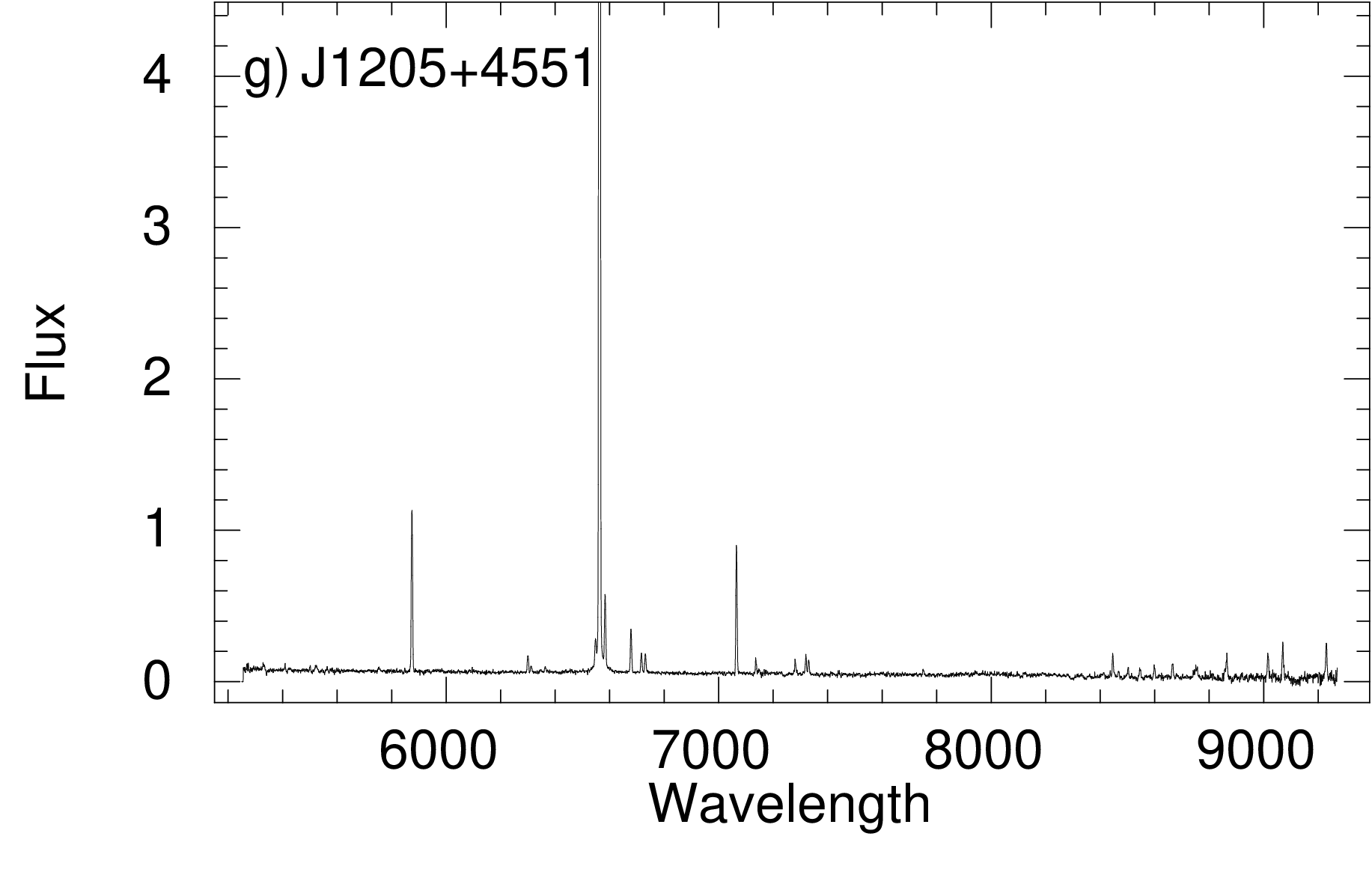}
\includegraphics[angle=-90,width=0.32\linewidth]{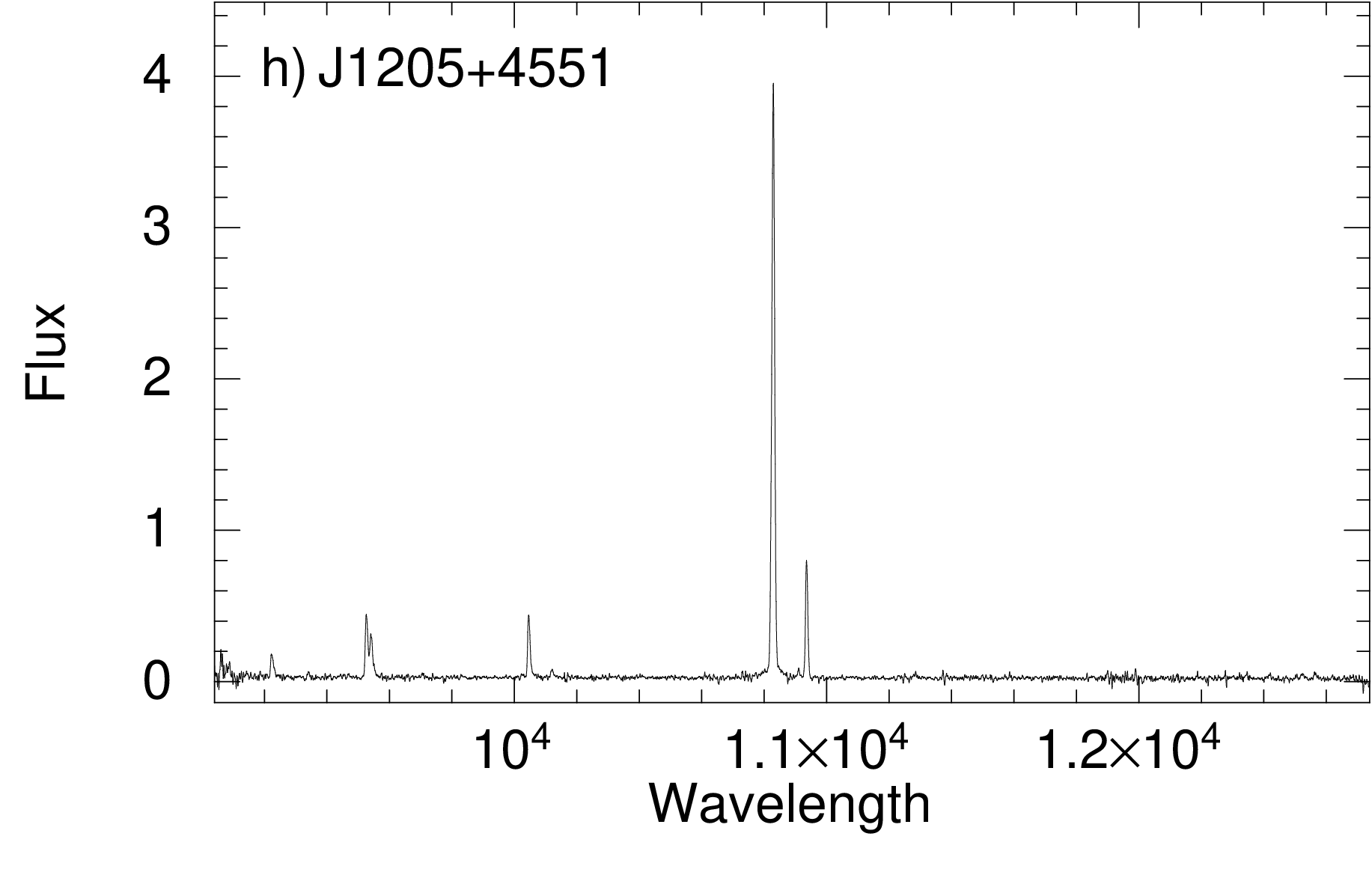}
}
\hbox{
\includegraphics[angle=-90,width=0.32\linewidth]{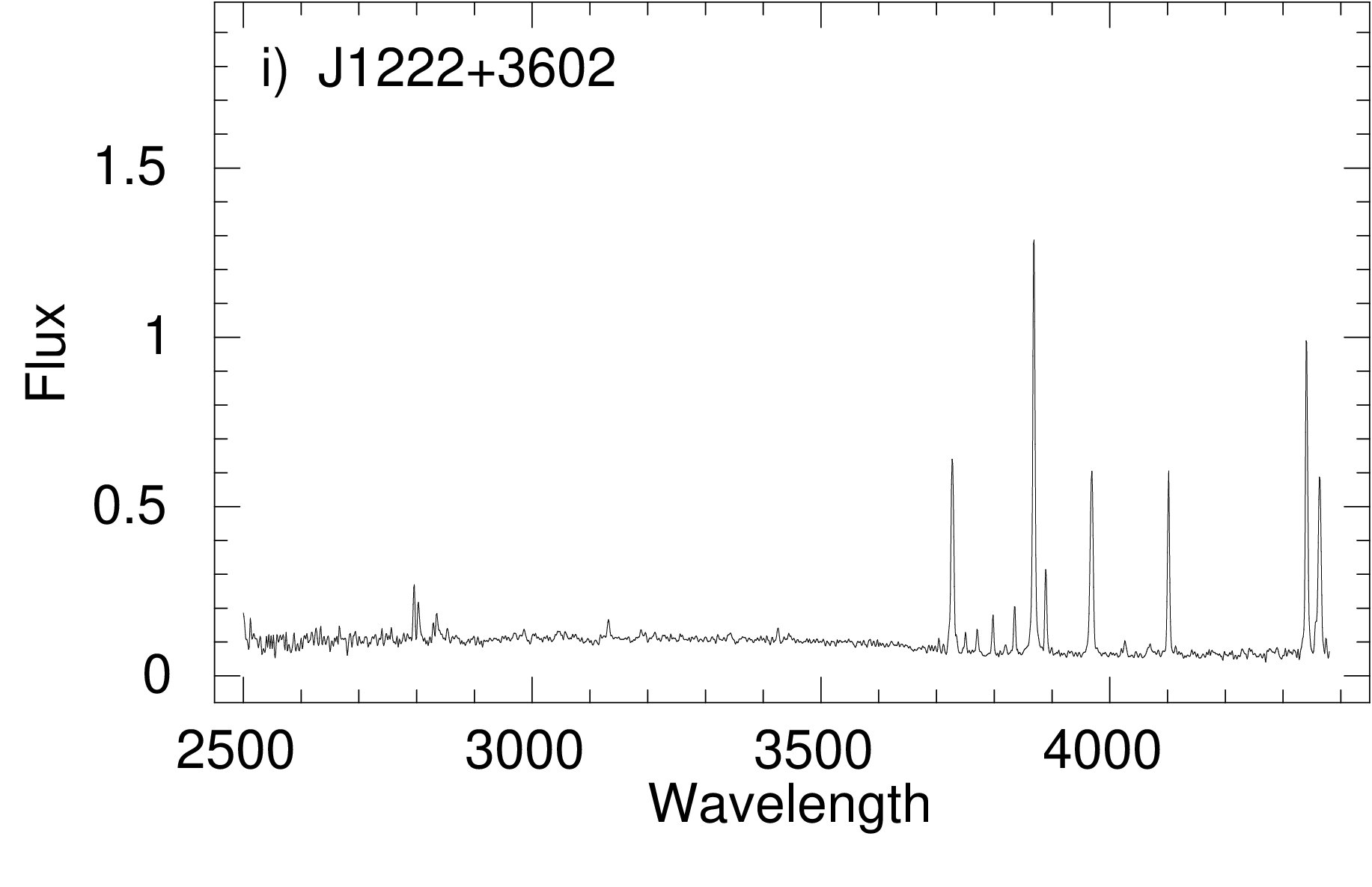}
\includegraphics[angle=-90,width=0.32\linewidth]{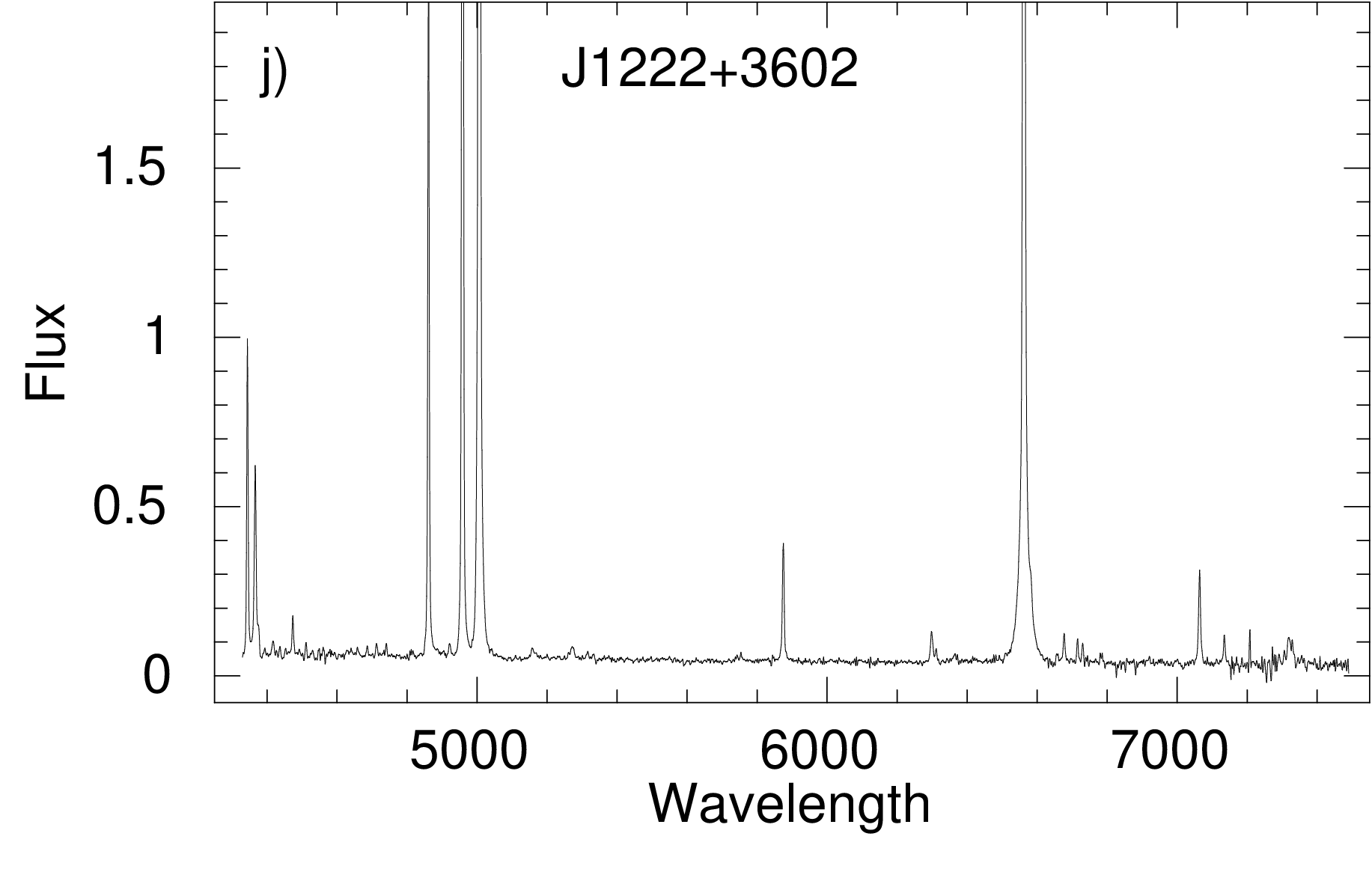}
}
\hbox{
\includegraphics[angle=-90,width=0.32\linewidth]{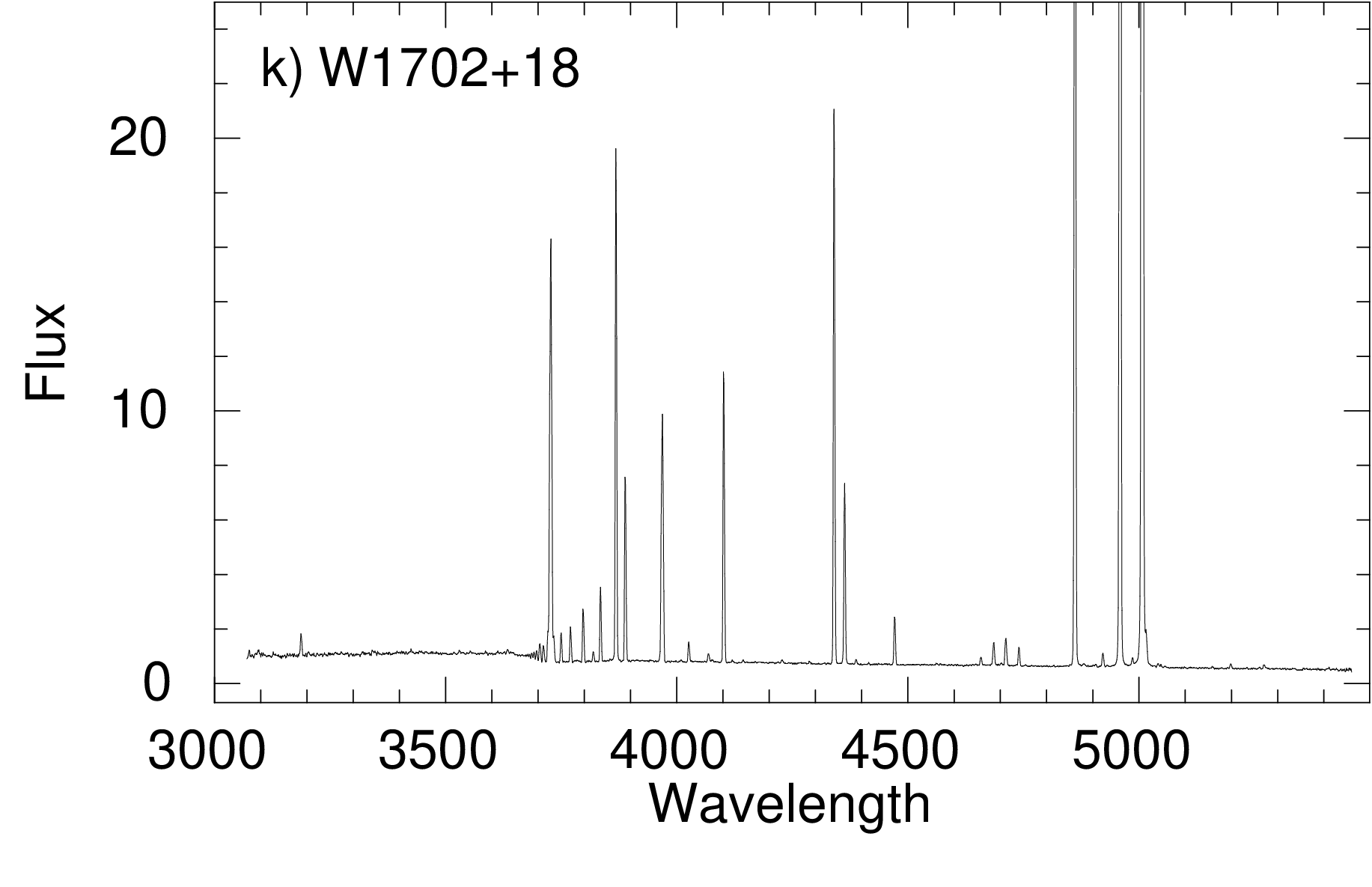}
\includegraphics[angle=-90,width=0.32\linewidth]{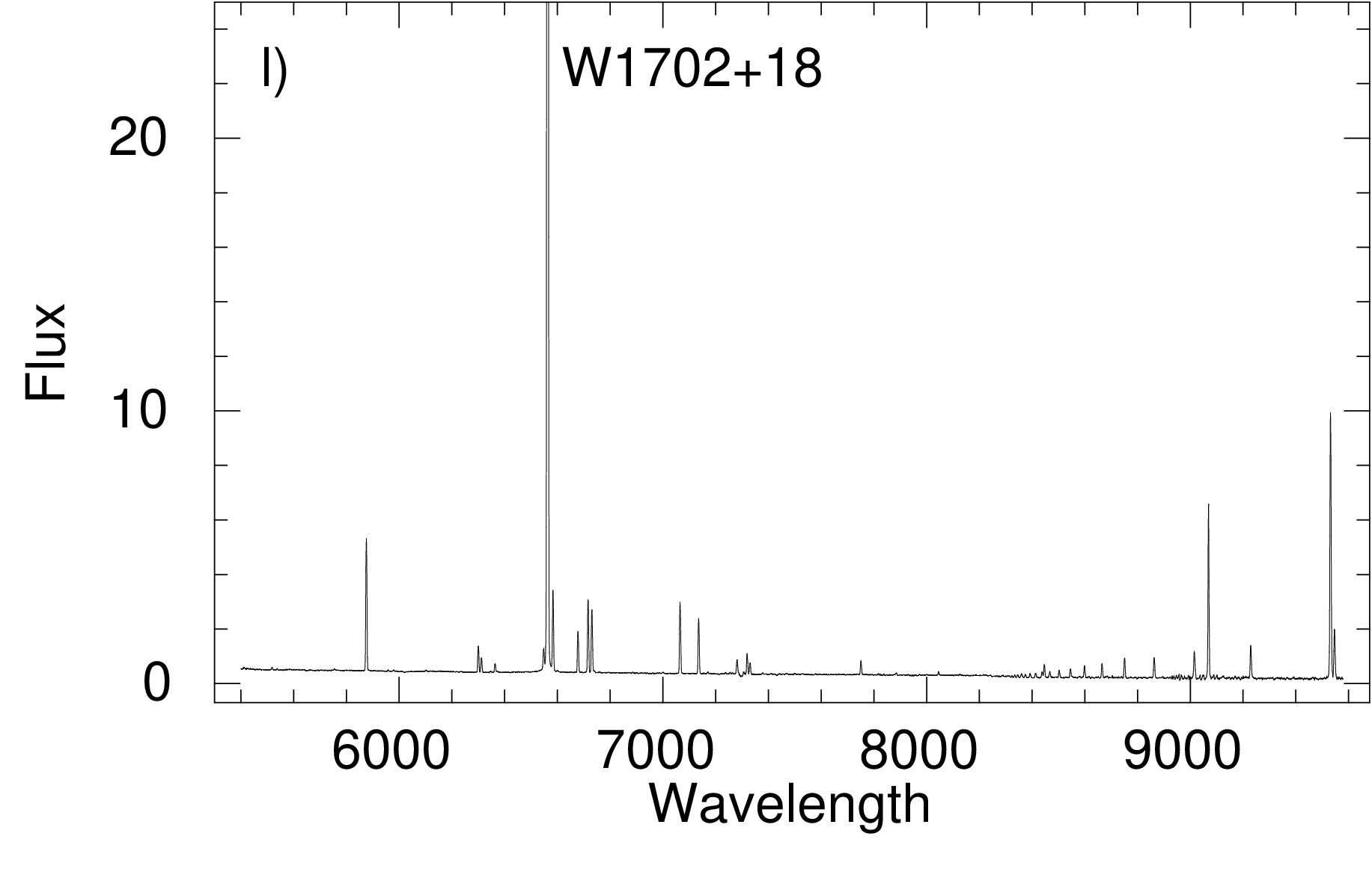}
\includegraphics[angle=-90,width=0.32\linewidth]{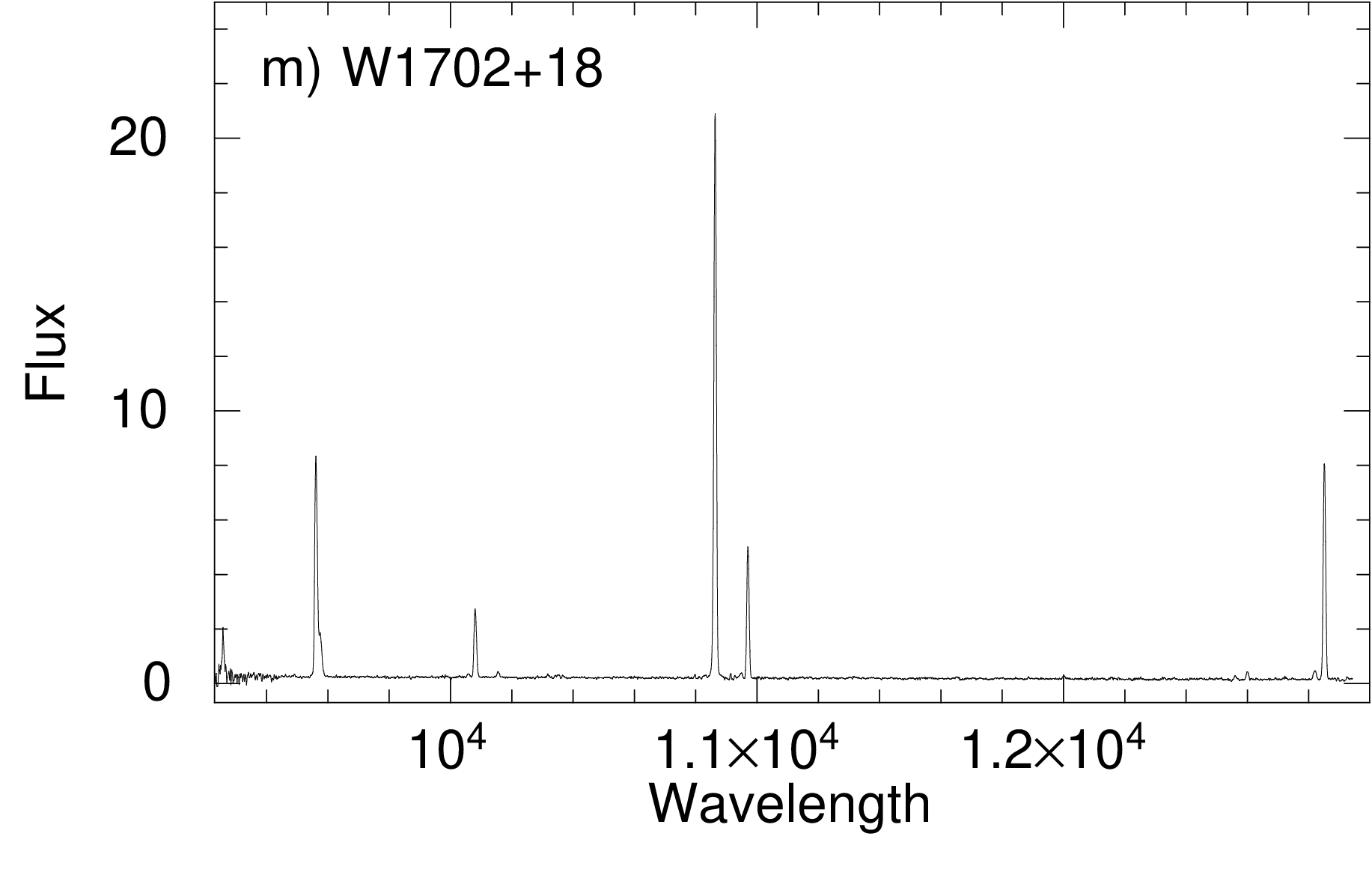}
}
\hbox{
\includegraphics[angle=-90,width=0.32\linewidth]{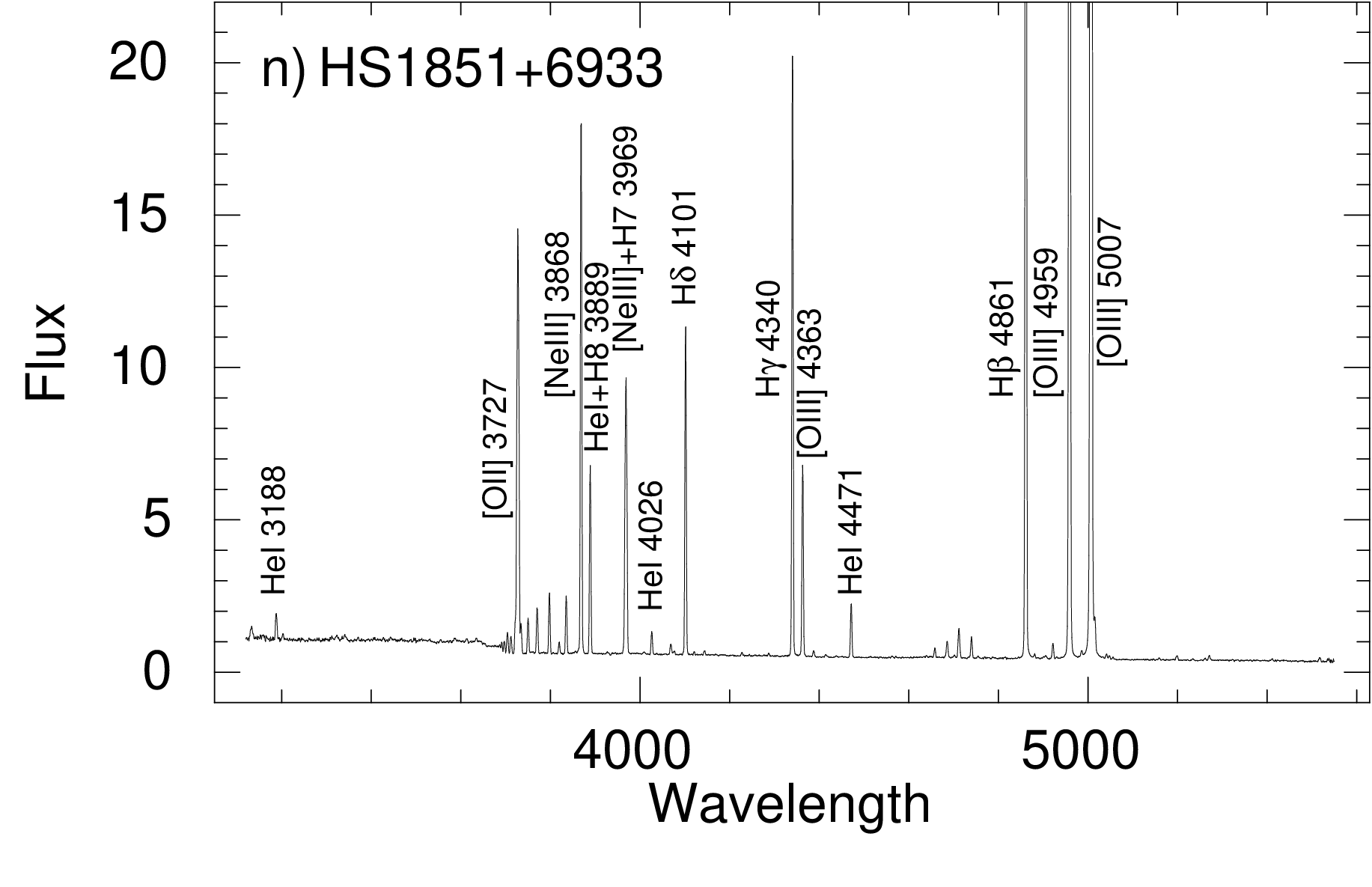}
\includegraphics[angle=-90,width=0.32\linewidth]{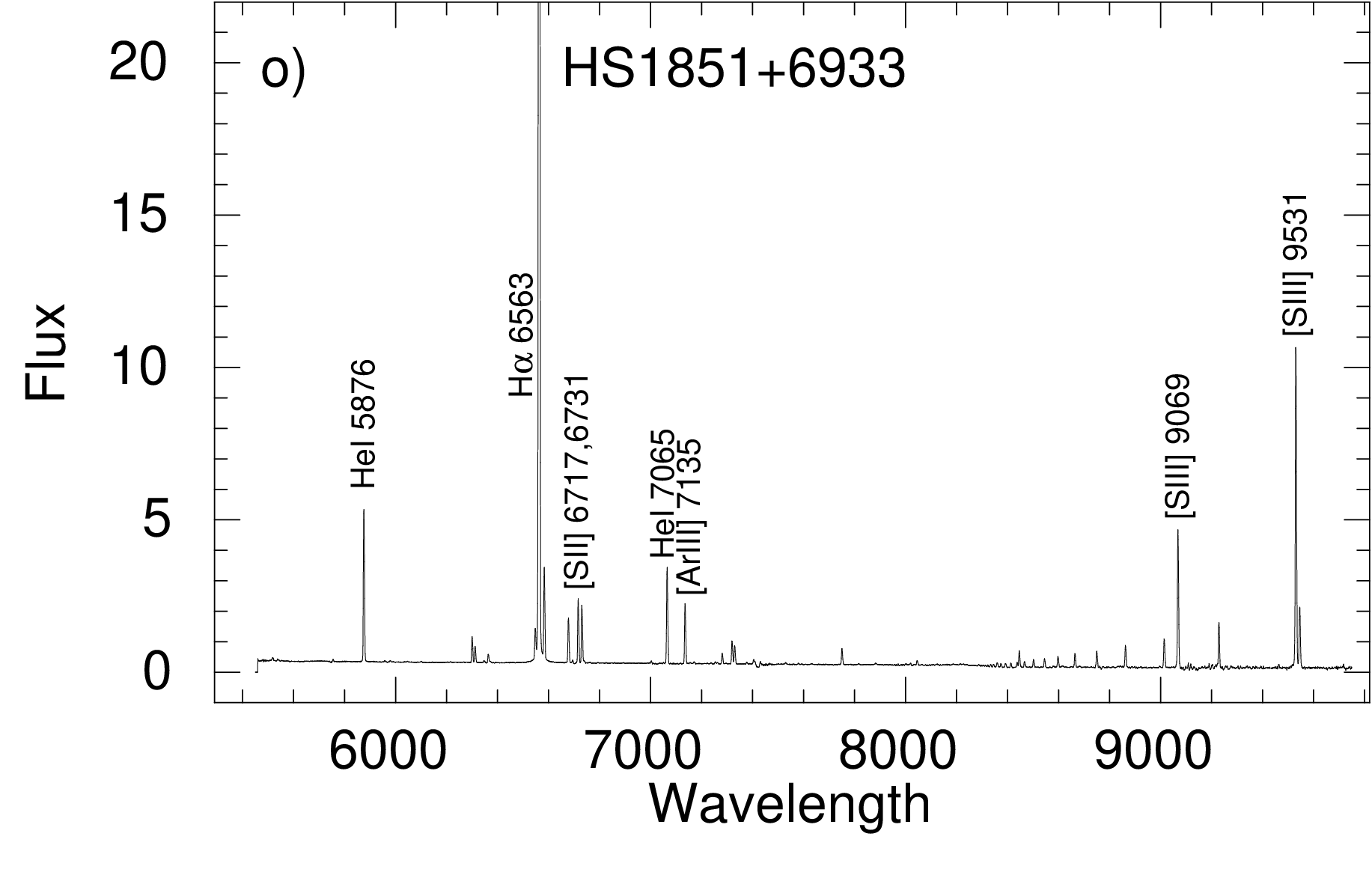}
\includegraphics[angle=-90,width=0.32\linewidth]{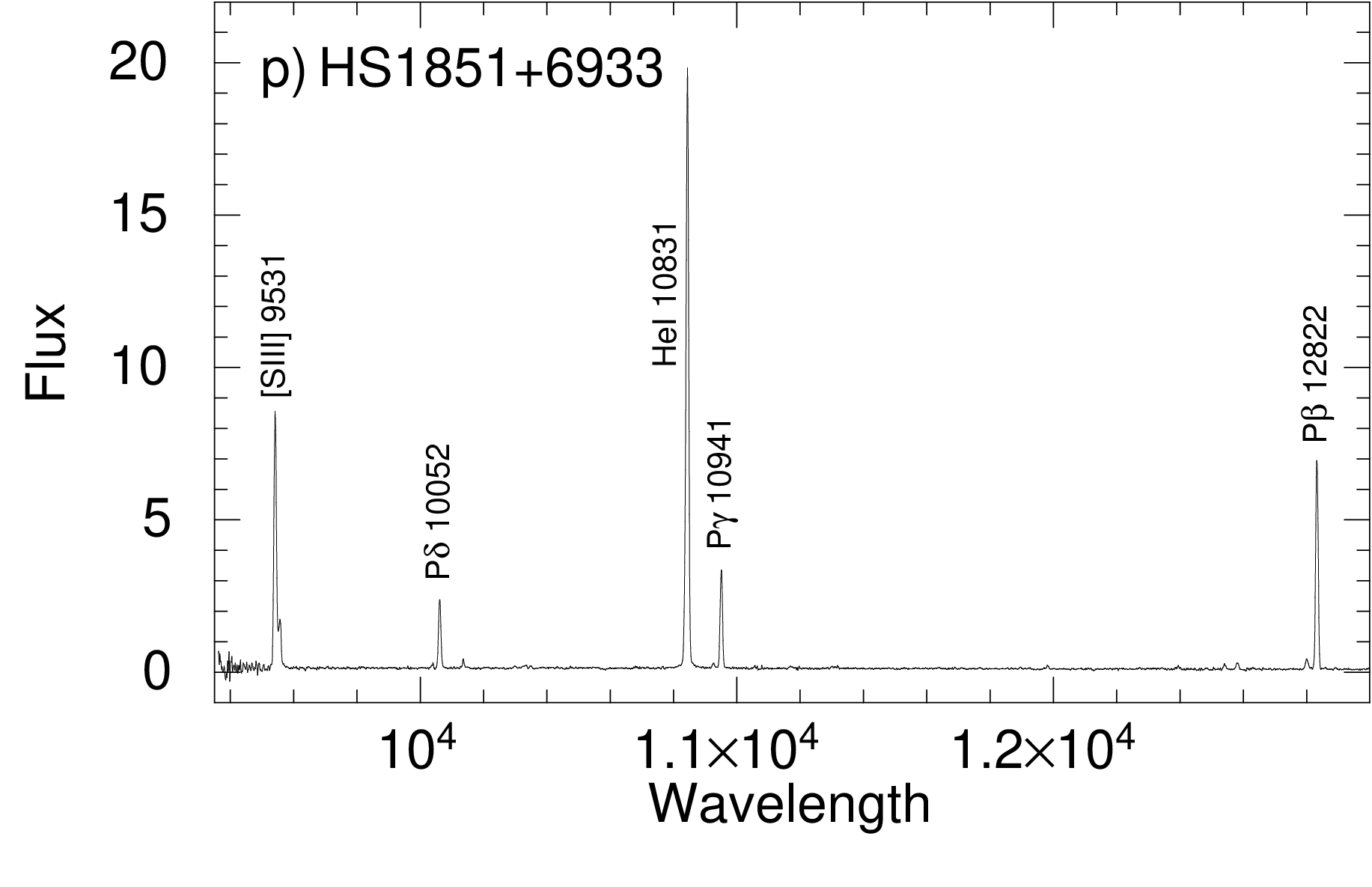}
}
\caption{Rest-frame LBT spectra of selected galaxies in the blue (left) 
red (centre) and infrared (right) wavelength regions. Emission lines are 
labelled in the spectra of HS\,1851$+$6933 in panels {\bf n) - p)}.}
\label{figa1}
\end{figure*}

\input{taba1_1.tex}

\input{taba2_1.tex}

\bsp

\label{lastpage}

\end{document}

%% file: tab1_1.tex
\begin{table*}
\caption{Observed characteristics of galaxies \label{tab1}}
\begin{tabular}{lcccccccccccccc} \hline
          &           &            &       &      &\multicolumn{2}{c}{SDSS$^{\rm a}$}&\multicolumn{2}{c}{{\sl GALEX}$^{\rm a}$}&\multicolumn{4}{c}{{\sl WISE}$^{\rm b}$} \\
Name      &R.A.(J2000)&Dec.(J2000) &O$_{32}$$^{\rm c}$&  $z$$^{\rm d}$ & $g$& $M_g$$^{\rm e}$&$FUV$&$NUV$&$W1$&$W2$&$W3$&$W4$  \\
\hline
J0240$-$0828 &02:40:52.21&$-$08:28:27.43&11&0.0822&18.98&$-$19.00&19.81&19.74&14.74&13.61& 9.96&6.98 \\
J0344$-$0106 &03:44:59.28&$-$01:06:16.89&30&0.2707&21.52&$-$19.30& ... &21.32&17.36&16.48& ... & ... \\
J1205$+$4551 &12:05:03.55&$+$45:51:50.94&23&0.0654&19.79&$-$17.62&21.04&20.58&15.14&13.55& 9.91&7.80 \\
J1222$+$3602 &12:22:45.70&$+$36:02:18.00&16&0.3011&21.18&$-$19.83&21.79&20.58&16.25&15.05&11.37&8.82 \\
W\,1702$+$18   &17:02:33.53&$+$18:03:06.44&10&0.0425&17.27&$-$19.39&19.14&19.11&14.29&12.04& 7.60&4.93 \\
HS\,1851$+$6933&18:51:17.16&$+$69:34:25.92&11&0.0253&17.00&$-$18.44& ... & ... & ... & ... & ... & ... \\
\hline
  \end{tabular}

\hbox{$^{\rm a}$AB magnitudes.}

\hbox{$^{\rm b}$Vega magnitudes.}

\hbox{$^{\rm c}$O$_{32}$ is the extinction-corrected [O~{\sc iii}]\,$\lambda$5007/[O~{\sc ii}]\,$\lambda$3727 flux ratio in the LBT spectrum.}

\hbox{$^{\rm d}$$z$ is the redshift derived from the LBT spectrum of HS\,1851$+$6933 and from SDSS spectra for the remaining galaxies.}

\hbox{$^{\rm e}$Absolute SDSS $g$-band magnitude corrected for Milky Way extinction and adopting the distance for a flat Universe}

\hbox{\,with $H_0$ = 69.6 km s$^{-1}$, $\Omega_{\rm m}$ = 0.286 and $\Omega_\Lambda$ = 0.714 \citep[NED, ][]{W06}.}
  \end{table*}

%% file: tab2_1.tex
\begin{table*}
\caption{Journal of observations \label{tab2}}
\begin{tabular}{llccccc} \hline
Name      &\multicolumn{1}{c}{Inst.}&Date&Exp. &\multicolumn{1}{c}{Slit}&Seeing    &Airmass  \\
          &                         &    &(s) &(arcsec)&(arcsec)&          \\ \hline
J0240$-$0828 &MODS1      &2013-12-07&3200&60$\times$1.2&1.8&1.33 \\
             &LUCI1+LUCI2&2019-01-04&1440+1440&240$\times$1.0\,~~&1.0&1.34 \\
J0344$-$0106 &MODS1      &2019-10-25&3600&60$\times$1.2&1.5&1.27 \\
J1205$+$4551 &MODS1      &2016-04-10&3200&60$\times$1.2&1.1&1.24 \\
             &LUCI1+LUCI2&2021-01-09&2400+2400&240$\times$1.0\,~~&1.0&1.03 \\
J1222$+$3602 &MODS1      &2015-05-18&3200&60$\times$1.2&0.8&1.01 \\
W\,1702$+$18   &MODS1      &2013-06-09&1800&60$\times$1.0&1.0&1.08 \\
             &LUCI2      &2016-06-16&1440&240$\times$1.0\,~~&0.8&1.03 \\
HS\,1851$+$6933&MODS1&2013-06-07&2400&60$\times$1.0&1.0&1.20 \\
             &LUCI2      &2016-06-16&1440&240$\times$1.0\,~~&1.0&1.25 \\
\hline
  \end{tabular}



  \end{table*}

%% file: taba1_1.tex
\begin{table*}
\caption{Extinction-corrected emission-line fluxes$^{\rm a}$ \label{taba1}}
\begin{tabular}{lrrrrrrr} \hline
Line &J0240$-$0828&J0344$-$0106&J1205$+$4551&J1222$+$3602$^{\rm b}$&W\,1702$+$18&HS\,1851$+$6933\\ \hline
3096.00 [Fe {\sc iv}]           &\multicolumn{1}{c}{...}&\multicolumn{1}{c}{...}&\multicolumn{1}{c}{...}&\multicolumn{1}{c}{...}&  0.81$\pm$0.30&\multicolumn{1}{c}{...}\\
3132.79 O {\sc iii}             &\multicolumn{1}{c}{...}&\multicolumn{1}{c}{...}&\multicolumn{1}{c}{...}&  2.70$\pm$0.25&\multicolumn{1}{c}{...}&  1.91$\pm$0.25\\
3187.74 He {\sc i}              &  4.09$\pm$0.30&  4.61$\pm$0.46&  1.21$\pm$0.17&  1.00$\pm$0.17&  2.81$\pm$0.22&  2.99$\pm$0.18\\
3203.10 He {\sc ii}             &\multicolumn{1}{c}{...}&\multicolumn{1}{c}{...}&  1.21$\pm$0.18&\multicolumn{1}{c}{...}&  0.54$\pm$0.15&  0.61$\pm$0.12\\
3322.54 [Fe {\sc ii}]           &\multicolumn{1}{c}{...}&\multicolumn{1}{c}{...}&\multicolumn{1}{c}{...}&\multicolumn{1}{c}{...}&  0.33$\pm$0.12&  0.80$\pm$0.14\\
3340.74 O {\sc iii}             &  0.97$\pm$0.12&\multicolumn{1}{c}{...}&  0.63$\pm$0.22&\multicolumn{1}{c}{...}&  0.44$\pm$0.09&  0.81$\pm$0.12\\
3370.00 S {\sc iii}             &\multicolumn{1}{c}{...}&\multicolumn{1}{c}{...}&\multicolumn{1}{c}{...}&\multicolumn{1}{c}{...}&\multicolumn{1}{c}{...}&  0.27$\pm$0.05\\
3405.74 O {\sc iii}             &\multicolumn{1}{c}{...}&\multicolumn{1}{c}{...}&  0.32$\pm$0.13&\multicolumn{1}{c}{...}&\multicolumn{1}{c}{...}&  0.23$\pm$0.07\\
3426.85 [Ne {\sc v}]            &  1.11$\pm$0.13&  1.62$\pm$0.32&  0.72$\pm$0.23&  1.31$\pm$0.17&  0.44$\pm$0.12&  0.22$\pm$0.06\\
3444.07 O {\sc iii}             &  0.74$\pm$0.11&\multicolumn{1}{c}{...}&\multicolumn{1}{c}{...}&  1.02$\pm$0.18&  0.19$\pm$0.07&  0.31$\pm$0.08\\
3465.94 He {\sc i}              &\multicolumn{1}{c}{...}&\multicolumn{1}{c}{...}&\multicolumn{1}{c}{...}&\multicolumn{1}{c}{...}&\multicolumn{1}{c}{...}&  0.27$\pm$0.08\\
3487.27 He {\sc i}              &\multicolumn{1}{c}{...}&\multicolumn{1}{c}{...}&\multicolumn{1}{c}{...}&\multicolumn{1}{c}{...}&\multicolumn{1}{c}{...}&  0.17$\pm$0.06\\
3498.66 He {\sc i}              &\multicolumn{1}{c}{...}&\multicolumn{1}{c}{...}&\multicolumn{1}{c}{...}&\multicolumn{1}{c}{...}&\multicolumn{1}{c}{...}&  0.25$\pm$0.08\\
3530.09 He {\sc i}              &\multicolumn{1}{c}{...}&\multicolumn{1}{c}{...}&  0.55$\pm$0.20&\multicolumn{1}{c}{...}&  0.42$\pm$0.07&  0.25$\pm$0.01\\
3554.39 He {\sc i}              &\multicolumn{1}{c}{...}&\multicolumn{1}{c}{...}&\multicolumn{1}{c}{...}&\multicolumn{1}{c}{...}&  0.26$\pm$0.06&  0.27$\pm$0.06\\
3587.28 He {\sc i}              &  0.61$\pm$0.09&\multicolumn{1}{c}{...}&\multicolumn{1}{c}{...}&\multicolumn{1}{c}{...}&  0.15$\pm$0.03&  0.37$\pm$0.05\\
3612.48 He {\sc i}              &\multicolumn{1}{c}{...}&\multicolumn{1}{c}{...}&\multicolumn{1}{c}{...}&\multicolumn{1}{c}{...}&  0.27$\pm$0.05&  0.38$\pm$0.05\\
3634.25 He {\sc i}              &\multicolumn{1}{c}{...}&\multicolumn{1}{c}{...}&  0.41$\pm$0.11&\multicolumn{1}{c}{...}&  0.41$\pm$0.04&  0.36$\pm$0.04\\
3676.37 H22                     &\multicolumn{1}{c}{...}&\multicolumn{1}{c}{...}&\multicolumn{1}{c}{...}&\multicolumn{1}{c}{...}&  0.26$\pm$0.03&\multicolumn{1}{c}{...}\\
3679.31 H21                     &\multicolumn{1}{c}{...}&\multicolumn{1}{c}{...}&\multicolumn{1}{c}{...}&\multicolumn{1}{c}{...}&  0.26$\pm$0.03&\multicolumn{1}{c}{...}\\
3682.81 H20                     &\multicolumn{1}{c}{...}&\multicolumn{1}{c}{...}&\multicolumn{1}{c}{...}&\multicolumn{1}{c}{...}&  0.29$\pm$0.03&\multicolumn{1}{c}{...}\\
3686.83 H19                     &\multicolumn{1}{c}{...}&\multicolumn{1}{c}{...}&  0.57$\pm$0.07&\multicolumn{1}{c}{...}&  0.44$\pm$0.03&\multicolumn{1}{c}{...}\\
3691.55 H18                     &\multicolumn{1}{c}{...}&\multicolumn{1}{c}{...}&  0.50$\pm$0.07&\multicolumn{1}{c}{...}&  0.65$\pm$0.04&  0.88$\pm$0.05\\
3697.15 H17                     &\multicolumn{1}{c}{...}&\multicolumn{1}{c}{...}&  0.82$\pm$0.07&\multicolumn{1}{c}{...}&  0.98$\pm$0.06&  1.01$\pm$0.05\\
3703.30 H16                     &  2.04$\pm$0.09&\multicolumn{1}{c}{...}&  1.32$\pm$0.08&\multicolumn{1}{c}{...}&  1.88$\pm$0.08&  1.89$\pm$0.07\\
3711.97 H15                     &  2.60$\pm$0.16&\multicolumn{1}{c}{...}&  2.85$\pm$0.21&\multicolumn{1}{c}{...}&  3.33$\pm$0.19&  2.30$\pm$0.11\\
3721.94 H14                     &\multicolumn{1}{c}{...}&\multicolumn{1}{c}{...}&  1.49$\pm$0.08&\multicolumn{1}{c}{...}&  1.41$\pm$0.06&  1.18$\pm$0.05\\
3727.00 [O {\sc ii}]            & 65.67$\pm$2.08& 20.17$\pm$0.74& 18.26$\pm$0.59& 36.42$\pm$1.21& 63.04$\pm$2.00& 59.79$\pm$1.79\\
3734.37 H13                     &\multicolumn{1}{c}{...}&\multicolumn{1}{c}{...}&  2.16$\pm$0.10&  2.04$\pm$0.18&  1.30$\pm$0.05&  1.92$\pm$0.07\\
3750.15 H12                     &  4.20$\pm$0.18&  4.71$\pm$0.44&  4.42$\pm$0.19&  4.83$\pm$0.42&  4.49$\pm$0.19&  3.78$\pm$0.14\\
3770.63 H11                     &  5.10$\pm$0.21&  4.97$\pm$0.45&  4.78$\pm$0.20&  5.59$\pm$0.40&  5.01$\pm$0.20&  4.83$\pm$0.17\\
3797.90 H10                     &  6.43$\pm$0.24&  6.58$\pm$0.46&  6.19$\pm$0.23&  6.85$\pm$0.39&  6.86$\pm$0.24&  5.97$\pm$0.20\\
3819.64 He {\sc i}              &  1.40$\pm$0.08&\multicolumn{1}{c}{...}&  1.19$\pm$0.07&  1.71$\pm$0.18&  0.93$\pm$0.05&  1.01$\pm$0.05\\
3835.39 H9                      &  8.31$\pm$0.29&  8.77$\pm$0.50&  8.15$\pm$0.28&  8.47$\pm$0.41&  8.72$\pm$0.29&  6.02$\pm$0.20\\
3868.76 [Ne {\sc iii}]          & 53.79$\pm$1.67& 41.08$\pm$1.38& 24.25$\pm$0.76& 61.92$\pm$2.00& 49.50$\pm$1.54& 47.64$\pm$1.48\\
3889.00 He {\sc i}+H8           & 20.59$\pm$0.65& 19.27$\pm$0.74& 15.13$\pm$0.49& 13.61$\pm$0.54& 20.16$\pm$0.64& 17.00$\pm$0.53\\
3926.50 He {\sc i}              &\multicolumn{1}{c}{...}&\multicolumn{1}{c}{...}&\multicolumn{1}{c}{...}&\multicolumn{1}{c}{...}&\multicolumn{1}{c}{...}&  0.22$\pm$0.03\\
3968.00 [Ne {\sc iii}]+H7       & 35.16$\pm$1.09& 28.44$\pm$1.00& 24.56$\pm$0.77& 37.65$\pm$1.24& 33.61$\pm$1.04& 34.06$\pm$1.04\\
4009.26 He {\sc i}              &\multicolumn{1}{c}{...}&\multicolumn{1}{c}{...}&  0.42$\pm$0.05&\multicolumn{1}{c}{...}&  0.17$\pm$0.03&  0.29$\pm$0.03\\
4026.19 He {\sc i}              &  2.30$\pm$0.10&  1.87$\pm$0.23&  1.97$\pm$0.08&  2.46$\pm$0.20&  1.81$\pm$0.07&  1.92$\pm$0.07\\
4068.60 [S {\sc ii}]            &  1.30$\pm$0.07&\multicolumn{1}{c}{...}&  0.65$\pm$0.06&  2.42$\pm$0.27&  1.00$\pm$0.06&  1.01$\pm$0.04\\
4076.35 [S {\sc ii}]            &\multicolumn{1}{c}{...}&\multicolumn{1}{c}{...}&  0.30$\pm$0.06&\multicolumn{1}{c}{...}&  0.28$\pm$0.04&  0.32$\pm$0.02\\
4101.74 H$\delta$               & 28.49$\pm$0.87& 25.47$\pm$0.89& 27.50$\pm$0.84& 26.28$\pm$0.89& 28.34$\pm$0.86& 27.78$\pm$0.84\\
4120.89 He {\sc i}              &\multicolumn{1}{c}{...}&\multicolumn{1}{c}{...}&  0.44$\pm$0.05&\multicolumn{1}{c}{...}&  0.28$\pm$0.03&  0.23$\pm$0.02\\
4143.81 He {\sc i}              &  0.39$\pm$0.04&\multicolumn{1}{c}{...}&  0.57$\pm$0.05&\multicolumn{1}{c}{...}&  0.29$\pm$0.03&  0.36$\pm$0.03\\
4227.20 [Fe {\sc v}]            &  0.72$\pm$0.06&  0.73$\pm$0.19&  0.73$\pm$0.05&\multicolumn{1}{c}{...}&  0.39$\pm$0.04&  0.32$\pm$0.03\\
4243.90 [Fe {\sc ii}]           &\multicolumn{1}{c}{...}&\multicolumn{1}{c}{...}&\multicolumn{1}{c}{...}&\multicolumn{1}{c}{...}&\multicolumn{1}{c}{...}&  0.09$\pm$0.02\\
4287.33 [Fe {\sc ii}]           &\multicolumn{1}{c}{...}&\multicolumn{1}{c}{...}&  0.49$\pm$0.05&\multicolumn{1}{c}{...}&  0.21$\pm$0.03&  0.19$\pm$0.02\\
4340.47 H$\gamma$               & 49.14$\pm$1.44& 45.93$\pm$1.46& 46.99$\pm$1.44& 48.25$\pm$1.49& 48.64$\pm$1.43& 48.65$\pm$1.42\\
4363.21 [O {\sc iii}]           & 15.41$\pm$0.45& 16.36$\pm$0.63& 12.87$\pm$0.38& 29.52$\pm$0.91& 15.41$\pm$0.45& 15.03$\pm$0.44\\
4387.93 He {\sc i}              &  0.59$\pm$0.05&\multicolumn{1}{c}{...}&  0.49$\pm$0.05&  1.16$\pm$0.13&  0.43$\pm$0.03&  0.51$\pm$0.03\\
4415.00 [Fe {\sc ii}]           &  0.40$\pm$0.05&\multicolumn{1}{c}{...}&\multicolumn{1}{c}{...}&  2.59$\pm$0.16&  0.15$\pm$0.03&  0.26$\pm$0.02\\
4437.55 He {\sc i}              &\multicolumn{1}{c}{...}&\multicolumn{1}{c}{...}&\multicolumn{1}{c}{...}&  1.53$\pm$0.13&  0.08$\pm$0.02&  0.15$\pm$0.02\\
4452.02 [Fe {\sc ii}]           &\multicolumn{1}{c}{...}&\multicolumn{1}{c}{...}&\multicolumn{1}{c}{...}&  1.28$\pm$0.11&\multicolumn{1}{c}{...}&  0.08$\pm$0.01\\
4471.48 He {\sc i}              &  4.62$\pm$0.15&\multicolumn{1}{c}{...}&  4.02$\pm$0.13&  4.68$\pm$0.21&  3.97$\pm$0.12&  4.18$\pm$0.12\\
4514.89 N {\sc iii}             &\multicolumn{1}{c}{...}&\multicolumn{1}{c}{...}&\multicolumn{1}{c}{...}&\multicolumn{1}{c}{...}&\multicolumn{1}{c}{...}&  0.10$\pm$0.03\\
4562.50 [Mg {\sc i}]            &\multicolumn{1}{c}{...}&\multicolumn{1}{c}{...}&\multicolumn{1}{c}{...}&  1.62$\pm$0.11&  0.15$\pm$0.02&  0.13$\pm$0.02\\
4571.10 Mg {\sc i}]             &\multicolumn{1}{c}{...}&\multicolumn{1}{c}{...}&\multicolumn{1}{c}{...}&\multicolumn{1}{c}{...}&  0.09$\pm$0.02&  0.08$\pm$0.01\\
4634.14 N {\sc iii}             &\multicolumn{1}{c}{...}&\multicolumn{1}{c}{...}&\multicolumn{1}{c}{...}&\multicolumn{1}{c}{...}&\multicolumn{1}{c}{...}&  0.06$\pm$0.02\\
4640.00 N {\sc iii}             &\multicolumn{1}{c}{...}&\multicolumn{1}{c}{...}&\multicolumn{1}{c}{...}&  0.66$\pm$0.09&  0.07$\pm$0.02&  0.14$\pm$0.02\\
4640.00 WR                      &  0.49$\pm$0.15&\multicolumn{1}{c}{...}&  0.87$\pm$0.18&  2.50$\pm$0.31&  0.20$\pm$0.05&  0.20$\pm$0.04\\
4658.10 [Fe {\sc iii}]          &  0.84$\pm$0.06&\multicolumn{1}{c}{...}&  0.58$\pm$0.05&  0.72$\pm$0.08&  0.65$\pm$0.04&  0.69$\pm$0.03\\
4685.94 He {\sc ii}             &  2.96$\pm$0.11&  3.41$\pm$0.16&  3.24$\pm$0.11&  1.13$\pm$0.11&  1.96$\pm$0.07&  1.39$\pm$0.05\\
\hline
  \end{tabular}
  \end{table*}

\begin{table*}
\contcaption{Extinction-corrected emission-line fluxes$^{\rm a}$ \label{tab2a}}
\begin{tabular}{lrrrrrrr} \hline
Line &J0240$-$0828&J0344$-$0106&J1205$+$4551&J1222$+$3602$^{\rm b}$&W\,1702$+$18&HS\,1851$+$6933\\ \hline
4701.56 [Fe {\sc iii}]          &\multicolumn{1}{c}{...}&\multicolumn{1}{c}{...}&\multicolumn{1}{c}{...}&\multicolumn{1}{c}{...}&  0.18$\pm$0.03&  0.20$\pm$0.02\\
4712.00 [Ar {\sc iv}]+He {\sc i}&  2.32$\pm$0.09&  3.16$\pm$0.15&  2.15$\pm$0.08&  1.61$\pm$0.12&  2.29$\pm$0.07&  2.37$\pm$0.07\\
4740.20 [Ar {\sc iv}]           &  1.70$\pm$0.07&  2.41$\pm$0.15&  1.20$\pm$0.06&  1.27$\pm$0.11&  1.37$\pm$0.05&  1.56$\pm$0.05\\
4754.72 [Fe {\sc iii}]          &\multicolumn{1}{c}{...}&\multicolumn{1}{c}{...}&  0.31$\pm$0.03&\multicolumn{1}{c}{...}&  0.13$\pm$0.02&  0.19$\pm$0.02\\
4769.60 [Fe {\sc iii}]          &\multicolumn{1}{c}{...}&\multicolumn{1}{c}{...}&\multicolumn{1}{c}{...}&\multicolumn{1}{c}{...}&\multicolumn{1}{c}{...}&  0.07$\pm$0.01\\
4788.09 N {\sc ii}              &\multicolumn{1}{c}{...}&\multicolumn{1}{c}{...}&\multicolumn{1}{c}{...}&\multicolumn{1}{c}{...}&\multicolumn{1}{c}{...}&  0.12$\pm$0.02\\
4814.47 [Fe {\sc ii}]           &\multicolumn{1}{c}{...}&\multicolumn{1}{c}{...}&  0.26$\pm$0.04&\multicolumn{1}{c}{...}&  0.09$\pm$0.02&  0.14$\pm$0.02\\
4861.33 H$\beta$                &100.00$\pm$2.85&100.00$\pm$2.91&100.00$\pm$2.86&100.00$\pm$2.92&100.00$\pm$2.86&100.00$\pm$2.84\\
4881.01 [Fe {\sc iii}]          &\multicolumn{1}{c}{...}&\multicolumn{1}{c}{...}&\multicolumn{1}{c}{...}&\multicolumn{1}{c}{...}&  0.28$\pm$0.03&  0.28$\pm$0.02\\
4901.11 [Fe {\sc iv}]           &\multicolumn{1}{c}{...}&\multicolumn{1}{c}{...}&\multicolumn{1}{c}{...}&\multicolumn{1}{c}{...}&  0.11$\pm$0.02&  0.11$\pm$0.01\\
4905.00 [Fe {\sc ii}]           &\multicolumn{1}{c}{...}&\multicolumn{1}{c}{...}&\multicolumn{1}{c}{...}&\multicolumn{1}{c}{...}&  0.13$\pm$0.02&  0.21$\pm$0.02\\
4921.93 He {\sc i}              &  1.06$\pm$0.06&\multicolumn{1}{c}{...}&  1.03$\pm$0.05&  1.94$\pm$0.14&  1.01$\pm$0.04&  1.04$\pm$0.04\\
4930.50 [Fe {\sc iii}]          &\multicolumn{1}{c}{...}&\multicolumn{1}{c}{...}&\multicolumn{1}{c}{...}&\multicolumn{1}{c}{...}&  0.09$\pm$0.01&  0.18$\pm$0.02\\
4958.92 [O {\sc iii}]           &239.06$\pm$6.81&199.09$\pm$5.76&139.15$\pm$3.98&281.07$\pm$8.16&205.07$\pm$5.87&218.41$\pm$6.21\\
4986.00 [Fe {\sc iii}]          &\multicolumn{1}{c}{...}&\multicolumn{1}{c}{...}&\multicolumn{1}{c}{...}&\multicolumn{1}{c}{...}&  0.62$\pm$0.03&  0.50$\pm$0.02\\
5006.80 [O {\sc iii}]           &713.61$\pm$20.3&608.79$\pm$17.6&414.63$\pm$11.9&832.21$\pm$24.2&609.85$\pm$17.0&654.13$\pm$18.6\\
5015.68 He {\sc i}              &\multicolumn{1}{c}{...}&\multicolumn{1}{c}{...}&  1.98$\pm$0.07&\multicolumn{1}{c}{...}&  2.03$\pm$0.07&  1.91$\pm$0.06\\
5041.03 Si {\sc ii}             &\multicolumn{1}{c}{...}&\multicolumn{1}{c}{...}&  0.39$\pm$0.04&\multicolumn{1}{c}{...}&  0.33$\pm$0.03&  0.37$\pm$0.02\\
5047.80 He {\sc i}              &\multicolumn{1}{c}{...}&\multicolumn{1}{c}{...}&  0.43$\pm$0.06&\multicolumn{1}{c}{...}&  0.18$\pm$0.02&  0.23$\pm$0.02\\
5056.05 Si {\sc ii}             &\multicolumn{1}{c}{...}&\multicolumn{1}{c}{...}&  0.36$\pm$0.04&\multicolumn{1}{c}{...}&  0.14$\pm$0.02&  0.20$\pm$0.02\\
5160.00 [Fe {\sc ii}]           &\multicolumn{1}{c}{...}&\multicolumn{1}{c}{...}&\multicolumn{1}{c}{...}&  2.15$\pm$0.12&  0.11$\pm$0.02&  0.14$\pm$0.02\\
5191.82 [Ar {\sc iii}]          &\multicolumn{1}{c}{...}&\multicolumn{1}{c}{...}&\multicolumn{1}{c}{...}&\multicolumn{1}{c}{...}&  0.06$\pm$0.02&\multicolumn{1}{c}{...}\\
5199.00 [N {\sc i}]             &\multicolumn{1}{c}{...}&\multicolumn{1}{c}{...}&\multicolumn{1}{c}{...}&\multicolumn{1}{c}{...}&  0.33$\pm$0.03&  0.37$\pm$0.02\\
5233.76 [Fe {\sc iv}]           &\multicolumn{1}{c}{...}&\multicolumn{1}{c}{...}&\multicolumn{1}{c}{...}&\multicolumn{1}{c}{...}&  0.08$\pm$0.02&  0.13$\pm$0.02\\
5261.62 [Fe {\sc ii}]           &  0.14$\pm$0.03&\multicolumn{1}{c}{...}&\multicolumn{1}{c}{...}&  0.99$\pm$0.08&  0.13$\pm$0.02&  0.18$\pm$0.02\\
5270.63 [Fe {\sc iii}]          &\multicolumn{1}{c}{...}&\multicolumn{1}{c}{...}&\multicolumn{1}{c}{...}&\multicolumn{1}{c}{...}&  0.35$\pm$0.03&  0.39$\pm$0.02\\
5323.27 [Cl {\sc iv}]           &\multicolumn{1}{c}{...}&\multicolumn{1}{c}{...}&\multicolumn{1}{c}{...}&\multicolumn{1}{c}{...}&\multicolumn{1}{c}{...}&  0.07$\pm$0.01\\
5411.52 He {\sc ii}             &  0.29$\pm$0.03&\multicolumn{1}{c}{...}&  0.32$\pm$0.03&\multicolumn{1}{c}{...}&  0.11$\pm$0.03&  0.11$\pm$0.02\\
5517.71 [Cl {\sc iii}]          &  0.30$\pm$0.02&\multicolumn{1}{c}{...}&\multicolumn{1}{c}{...}&\multicolumn{1}{c}{...}&  0.18$\pm$0.02&  0.30$\pm$0.02\\
5537.88 [Cl {\sc iii}]          &  0.10$\pm$0.02&\multicolumn{1}{c}{...}&\multicolumn{1}{c}{...}&\multicolumn{1}{c}{...}&  0.08$\pm$0.01&  0.20$\pm$0.02\\
5754.64 [N {\sc ii}]            &  0.30$\pm$0.02&\multicolumn{1}{c}{...}&  0.32$\pm$0.03&  0.98$\pm$0.10&  0.11$\pm$0.01&  0.14$\pm$0.01\\
5875.60 He {\sc i}              & 11.87$\pm$0.36& 10.90$\pm$0.37& 11.77$\pm$0.35& 17.47$\pm$0.58& 10.46$\pm$0.31& 11.60$\pm$0.34\\
5957.55 Si {\sc ii}             &  0.20$\pm$0.02&\multicolumn{1}{c}{...}&\multicolumn{1}{c}{...}&\multicolumn{1}{c}{...}&  0.08$\pm$0.01&  0.10$\pm$0.01\\
5978.91 Si {\sc ii}             &  0.21$\pm$0.02&\multicolumn{1}{c}{...}&\multicolumn{1}{c}{...}&\multicolumn{1}{c}{...}&  0.11$\pm$0.01&  0.12$\pm$0.01\\
6046.00 O {\sc i}               &\multicolumn{1}{c}{...}&\multicolumn{1}{c}{...}&\multicolumn{1}{c}{...}&\multicolumn{1}{c}{...}&  0.03$\pm$0.01&  0.05$\pm$0.01\\
6087.00 [Fe {\sc vii}]          &\multicolumn{1}{c}{...}&\multicolumn{1}{c}{...}&  0.11$\pm$0.02&\multicolumn{1}{c}{...}&\multicolumn{1}{c}{...}&  0.03$\pm$0.01\\
6102.00 He {\sc ii}             &  0.07$\pm$0.01&\multicolumn{1}{c}{...}&\multicolumn{1}{c}{...}&\multicolumn{1}{c}{...}&  0.09$\pm$0.01&  0.08$\pm$0.01\\
6300.30 [O {\sc i}]             &  2.16$\pm$0.07&\multicolumn{1}{c}{...}&  1.21$\pm$0.04&  5.58$\pm$0.23&  1.90$\pm$0.06&  1.81$\pm$0.06\\
6312.10 [S {\sc iii}]           &  1.01$\pm$0.04&\multicolumn{1}{c}{...}&  0.50$\pm$0.03&  2.26$\pm$0.12&  1.07$\pm$0.04&  1.13$\pm$0.04\\
6347.08 Si {\sc ii}             &\multicolumn{1}{c}{...}&\multicolumn{1}{c}{...}&  0.13$\pm$0.02&\multicolumn{1}{c}{...}&  0.08$\pm$0.01&  0.13$\pm$0.01\\
6363.80 [O {\sc i}]             &  0.75$\pm$0.04&\multicolumn{1}{c}{...}&  0.36$\pm$0.03&  1.48$\pm$0.12&  0.63$\pm$0.02&  0.61$\pm$0.02\\
6371.36 Si {\sc ii}             &\multicolumn{1}{c}{...}&\multicolumn{1}{c}{...}&\multicolumn{1}{c}{...}&\multicolumn{1}{c}{...}&  0.11$\pm$0.01&\multicolumn{1}{c}{...}\\
6548.10 [N {\sc ii}]            &\multicolumn{1}{c}{...}&  1.24$\pm$0.09&  1.65$\pm$0.06&\multicolumn{1}{c}{...}&  1.37$\pm$0.04&  1.95$\pm$0.06\\
6562.80 H$\alpha$               &282.31$\pm$8.70&277.88$\pm$8.70&278.70$\pm$8.62&360.21$\pm$11.3&282.48$\pm$8.75&282.07$\pm$8.69\\
6583.40 [N {\sc ii}]            &  8.09$\pm$0.25&  3.43$\pm$0.15&  4.61$\pm$0.15&  3.14$\pm$0.20&  5.37$\pm$0.17&  6.15$\pm$0.19\\
6678.10 He {\sc i}              &  3.46$\pm$0.11&  2.77$\pm$0.15&  2.91$\pm$0.10&  4.15$\pm$0.19&  2.88$\pm$0.09&  3.07$\pm$0.10\\
6716.40 [S {\sc ii}]            &  4.74$\pm$0.15&  2.72$\pm$0.16&  1.34$\pm$0.05&  2.51$\pm$0.15&  5.06$\pm$0.16&  4.35$\pm$0.14\\
6730.80 [S {\sc ii}]            &  4.44$\pm$0.14&  1.95$\pm$0.14&  1.31$\pm$0.05&  2.52$\pm$0.14&  4.36$\pm$0.14&  4.01$\pm$0.13\\
6739.80 [Fe {\sc iv}]           &\multicolumn{1}{c}{...}&\multicolumn{1}{c}{...}&\multicolumn{1}{c}{...}&\multicolumn{1}{c}{...}&  0.09$\pm$0.01&  0.16$\pm$0.01\\
6933.91 He {\sc i}              &\multicolumn{1}{c}{...}&\multicolumn{1}{c}{...}&\multicolumn{1}{c}{...}&\multicolumn{1}{c}{...}&  0.06$\pm$0.01&\multicolumn{1}{c}{...}\\
7002.00 O {\sc i}               &\multicolumn{1}{c}{...}&\multicolumn{1}{c}{...}&\multicolumn{1}{c}{...}&\multicolumn{1}{c}{...}&\multicolumn{1}{c}{...}&  0.16$\pm$0.01\\
7065.30 He {\sc i}              &  5.68$\pm$0.19&  4.78$\pm$0.21&  7.87$\pm$0.25& 15.89$\pm$0.58&  4.66$\pm$0.15&  6.36$\pm$0.20\\
7135.80 [Ar {\sc iii}]          &  3.94$\pm$0.13&  2.16$\pm$0.14&  1.03$\pm$0.05&  4.05$\pm$0.24&  3.62$\pm$0.12&  3.85$\pm$0.12\\
7155.08 [Fe {\sc ii}]           &\multicolumn{1}{c}{...}&\multicolumn{1}{c}{...}&\multicolumn{1}{c}{...}&\multicolumn{1}{c}{...}&  0.03$\pm$0.01&  0.09$\pm$0.01\\
7170.62 [Ar {\sc iv}]           &\multicolumn{1}{c}{...}&\multicolumn{1}{c}{...}&\multicolumn{1}{c}{...}&\multicolumn{1}{c}{...}&  0.10$\pm$0.01&  0.13$\pm$0.01\\
7237.00 [Ar {\sc iv}]           &\multicolumn{1}{c}{...}&\multicolumn{1}{c}{...}&\multicolumn{1}{c}{...}&\multicolumn{1}{c}{...}&  0.08$\pm$0.01&  0.06$\pm$0.01\\
7254.00 O {\sc i}               &\multicolumn{1}{c}{...}&\multicolumn{1}{c}{...}&\multicolumn{1}{c}{...}&\multicolumn{1}{c}{...}&  0.11$\pm$0.01&  0.14$\pm$0.01\\
7262.76 [Ar {\sc iv}]           &\multicolumn{1}{c}{...}&\multicolumn{1}{c}{...}&\multicolumn{1}{c}{...}&\multicolumn{1}{c}{...}&  0.06$\pm$0.01&  0.08$\pm$0.01\\
7281.20 He {\sc i}              &  0.72$\pm$0.03&\multicolumn{1}{c}{...}&  0.89$\pm$0.03&\multicolumn{1}{c}{...}&  1.00$\pm$0.06&  0.68$\pm$0.02\\
7298.00 He {\sc i}              &\multicolumn{1}{c}{...}&\multicolumn{1}{c}{...}&\multicolumn{1}{c}{...}&\multicolumn{1}{c}{...}&\multicolumn{1}{c}{...}&  0.04$\pm$0.01\\
7319.90 [O {\sc ii}]            &  1.57$\pm$0.06&\multicolumn{1}{c}{...}&  1.13$\pm$0.04&  5.50$\pm$0.36&  1.19$\pm$0.04&  1.51$\pm$0.05\\
7330.20 [O {\sc ii}]            &  1.20$\pm$0.05&\multicolumn{1}{c}{...}&  0.87$\pm$0.04&  3.63$\pm$0.34&  0.75$\pm$0.03&  1.17$\pm$0.04\\
7377.83 [Ni {\sc ii}]           &\multicolumn{1}{c}{...}&\multicolumn{1}{c}{...}&\multicolumn{1}{c}{...}&\multicolumn{1}{c}{...}&  0.11$\pm$0.01&  0.12$\pm$0.01\\
\hline
  \end{tabular}
  \end{table*}

\begin{table*}
\contcaption{Extinction-corrected emission-line fluxes$^{\rm a}$ \label{tab2b}}
\begin{tabular}{lrrrrrrr} \hline
Line &J0240$-$0828&J0344$-$0106&J1205$+$4551&J1222$+$3602$^{\rm b}$&W\,1702$+$18&HS\,1851$+$6933\\ \hline
7468.31 N {\sc i}               &\multicolumn{1}{c}{...}&\multicolumn{1}{c}{...}&\multicolumn{1}{c}{...}&\multicolumn{1}{c}{...}&  0.06$\pm$0.01&\multicolumn{1}{c}{...}\\
7499.85 He {\sc i}              &\multicolumn{1}{c}{...}&\multicolumn{1}{c}{...}&\multicolumn{1}{c}{...}&\multicolumn{1}{c}{...}&  0.06$\pm$0.01&\multicolumn{1}{c}{...}\\
7751.12 [Ar {\sc iii}]          &  1.02$\pm$0.04&\multicolumn{1}{c}{...}&  0.33$\pm$0.02&\multicolumn{1}{c}{...}&  0.84$\pm$0.03&  1.00$\pm$0.04\\
7816.80 He {\sc ii}             &\multicolumn{1}{c}{...}&\multicolumn{1}{c}{...}&\multicolumn{1}{c}{...}&\multicolumn{1}{c}{...}&  0.06$\pm$0.01&  0.06$\pm$0.01\\
8045.63 [Cl {\sc iv}]           &  0.14$\pm$0.02&\multicolumn{1}{c}{...}&\multicolumn{1}{c}{...}&\multicolumn{1}{c}{...}&  0.19$\pm$0.01&  0.27$\pm$0.01\\
8298.83 P28                     &\multicolumn{1}{c}{...}&\multicolumn{1}{c}{...}&\multicolumn{1}{c}{...}&\multicolumn{1}{c}{...}&\multicolumn{1}{c}{...}&  0.08$\pm$0.01\\
8306.22 P27                     &\multicolumn{1}{c}{...}&\multicolumn{1}{c}{...}&\multicolumn{1}{c}{...}&\multicolumn{1}{c}{...}&\multicolumn{1}{c}{...}&  0.09$\pm$0.01\\
8314.26 P26                     &\multicolumn{1}{c}{...}&\multicolumn{1}{c}{...}&\multicolumn{1}{c}{...}&\multicolumn{1}{c}{...}&  0.05$\pm$0.01&  0.09$\pm$0.01\\
8323.43 P25                     &\multicolumn{1}{c}{...}&\multicolumn{1}{c}{...}&\multicolumn{1}{c}{...}&\multicolumn{1}{c}{...}&  0.06$\pm$0.01&  0.09$\pm$0.01\\
8333.78 P24                     &\multicolumn{1}{c}{...}&\multicolumn{1}{c}{...}&  0.16$\pm$0.02&\multicolumn{1}{c}{...}&  0.11$\pm$0.01&  0.14$\pm$0.01\\
8345.44 P23                     &\multicolumn{1}{c}{...}&\multicolumn{1}{c}{...}&\multicolumn{1}{c}{...}&\multicolumn{1}{c}{...}&  0.13$\pm$0.01&  0.17$\pm$0.01\\
8359.01 P22                     &  0.16$\pm$0.02&\multicolumn{1}{c}{...}&\multicolumn{1}{c}{...}&\multicolumn{1}{c}{...}&  0.24$\pm$0.01&  0.29$\pm$0.01\\
8374.48 P21                     &  0.18$\pm$0.02&\multicolumn{1}{c}{...}&\multicolumn{1}{c}{...}&\multicolumn{1}{c}{...}&  0.19$\pm$0.01&  0.19$\pm$0.01\\
8392.40 P20                     &  0.17$\pm$0.02&\multicolumn{1}{c}{...}&\multicolumn{1}{c}{...}&\multicolumn{1}{c}{...}&  0.23$\pm$0.02&  0.23$\pm$0.01\\
8413.32 P19                     &  0.26$\pm$0.02&\multicolumn{1}{c}{...}&  0.22$\pm$0.02&\multicolumn{1}{c}{...}&  0.28$\pm$0.02&  0.23$\pm$0.01\\
8437.96 P18                     &\multicolumn{1}{c}{...}&\multicolumn{1}{c}{...}&  0.21$\pm$0.03&\multicolumn{1}{c}{...}&  0.33$\pm$0.02&  0.27$\pm$0.01\\
8446.34 O {\sc i}               &  0.96$\pm$0.04&\multicolumn{1}{c}{...}&  1.41$\pm$0.06&\multicolumn{1}{c}{...}&  0.71$\pm$0.03&  0.92$\pm$0.03\\
8467.26 P17                     &  0.45$\pm$0.03&\multicolumn{1}{c}{...}&  0.33$\pm$0.03&\multicolumn{1}{c}{...}&  0.31$\pm$0.02&  0.31$\pm$0.01\\
8502.49 P16                     &  0.52$\pm$0.03&\multicolumn{1}{c}{...}&  0.63$\pm$0.04&\multicolumn{1}{c}{...}&  0.39$\pm$0.02&  0.45$\pm$0.02\\
8545.38 P15                     &  0.65$\pm$0.04&\multicolumn{1}{c}{...}&  0.65$\pm$0.04&\multicolumn{1}{c}{...}&  0.53$\pm$0.02&  0.48$\pm$0.02\\
8598.39 P14                     &  0.35$\pm$0.03&\multicolumn{1}{c}{...}&  0.68$\pm$0.04&\multicolumn{1}{c}{...}&  0.64$\pm$0.03&  0.66$\pm$0.03\\
8665.02 P13                     &\multicolumn{1}{c}{...}&\multicolumn{1}{c}{...}&  0.90$\pm$0.05&\multicolumn{1}{c}{...}&  0.76$\pm$0.03&  0.76$\pm$0.03\\
8683.40 N {\sc i}               &\multicolumn{1}{c}{...}&\multicolumn{1}{c}{...}&\multicolumn{1}{c}{...}&\multicolumn{1}{c}{...}&  0.12$\pm$0.01&  0.12$\pm$0.01\\
8750.47 P12                     &  0.73$\pm$0.04&\multicolumn{1}{c}{...}&\multicolumn{1}{c}{...}&\multicolumn{1}{c}{...}&  1.05$\pm$0.04&  0.99$\pm$0.04\\
8862.79 P11                     &  1.13$\pm$0.06&\multicolumn{1}{c}{...}&  1.20$\pm$0.07&\multicolumn{1}{c}{...}&  1.13$\pm$0.05&  1.27$\pm$0.05\\
8996.99 He {\sc i}              &\multicolumn{1}{c}{...}&\multicolumn{1}{c}{...}&\multicolumn{1}{c}{...}&\multicolumn{1}{c}{...}&\multicolumn{1}{c}{...}&  0.09$\pm$0.01\\
9014.91 P10                     &  1.44$\pm$0.08&\multicolumn{1}{c}{...}&  1.25$\pm$0.06&\multicolumn{1}{c}{...}&  1.64$\pm$0.07&  1.68$\pm$0.06\\
9069.00 [S {\sc iii}]           &  6.70$\pm$0.26&\multicolumn{1}{c}{...}&  2.05$\pm$0.10&\multicolumn{1}{c}{...}&  8.02$\pm$0.38&  7.94$\pm$0.29\\
9229.02 P9                      &  1.68$\pm$0.10&\multicolumn{1}{c}{...}&  2.06$\pm$0.10&\multicolumn{1}{c}{...}&  1.77$\pm$0.08&  2.48$\pm$0.10\\
9463.62 He {\sc i}              &\multicolumn{1}{c}{...}&\multicolumn{1}{c}{...}&\multicolumn{1}{c}{...}&\multicolumn{1}{c}{...}&  0.07$\pm$0.03&  0.18$\pm$0.02\\
9530.60 [S {\sc iii}]           & 23.02$\pm$0.86&\multicolumn{1}{c}{...}&  4.60$\pm$0.18&\multicolumn{1}{c}{...}& 14.72$\pm$0.56& 18.42$\pm$0.69\\
9545.98 P8                      &\multicolumn{1}{c}{...}&\multicolumn{1}{c}{...}&  3.51$\pm$0.14&\multicolumn{1}{c}{...}&  2.74$\pm$0.12&  3.53$\pm$0.14\\
10030.00 He~{\sc i}    &\multicolumn{1}{c}{...}&\multicolumn{1}{c}{...}&\multicolumn{1}{c}{...}&\multicolumn{1}{c}{...}&  0.31$\pm$0.01&  0.34$\pm$0.04  \\
10052.15 P$\delta$     &\multicolumn{1}{c}{...}&\multicolumn{1}{c}{...}&  5.75$\pm$0.17&\multicolumn{1}{c}{...}&  6.25$\pm$0.17&  5.89$\pm$0.17 \\
10290.00 [S~{\sc ii}]  &\multicolumn{1}{c}{...}&\multicolumn{1}{c}{...}&\multicolumn{1}{c}{...}&\multicolumn{1}{c}{...}&  0.33$\pm$0.03&  0.19$\pm$0.02  \\
10320.00 [S~{\sc ii}]  &\multicolumn{1}{c}{...}&\multicolumn{1}{c}{...}&\multicolumn{1}{c}{...}&\multicolumn{1}{c}{...}&  0.43$\pm$0.03&\multicolumn{1}{c}{...} \\
10831.00 He~{\sc i}    & 84.21$\pm$2.49&\multicolumn{1}{c}{...}& 62.96$\pm$1.86&\multicolumn{1}{c}{...}& 50.59$\pm$1.34& 55.82$\pm$1.63  \\
10910.00 He~{\sc i}    &\multicolumn{1}{c}{...}&\multicolumn{1}{c}{...}&\multicolumn{1}{c}{...}&\multicolumn{1}{c}{...}&\multicolumn{1}{c}{...}&  0.38$\pm$0.04  \\
10941.12 P$\gamma$     & 10.84$\pm$0.32&\multicolumn{1}{c}{...}&  9.66$\pm$0.29&\multicolumn{1}{c}{...}& 10.73$\pm$0.29&  7.94$\pm$0.24 \\
11630.00 He~{\sc ii}   &\multicolumn{1}{c}{...}&\multicolumn{1}{c}{...}&\multicolumn{1}{c}{...}&\multicolumn{1}{c}{...}&\multicolumn{1}{c}{...}&  0.11$\pm$0.01  \\
11970.00 He~{\sc i}    &\multicolumn{1}{c}{...}&\multicolumn{1}{c}{...}&\multicolumn{1}{c}{...}&\multicolumn{1}{c}{...}&  0.31$\pm$0.02&  0.28$\pm$0.02  \\
12530.00 He~{\sc i}    &\multicolumn{1}{c}{...}&\multicolumn{1}{c}{...}&\multicolumn{1}{c}{...}&\multicolumn{1}{c}{...}&  0.30$\pm$0.03&  0.41$\pm$0.03  \\
12570.00 [Fe~{\sc ii}] &\multicolumn{1}{c}{...}&\multicolumn{1}{c}{...}&\multicolumn{1}{c}{...}&\multicolumn{1}{c}{...}&  0.61$\pm$0.04&  0.59$\pm$0.03  \\
12790.00 He~{\sc i}    &\multicolumn{1}{c}{...}&\multicolumn{1}{c}{...}&\multicolumn{1}{c}{...}&\multicolumn{1}{c}{...}&  0.83$\pm$0.04&  0.96$\pm$0.04  \\
12821.62 P$\beta$      &\multicolumn{1}{c}{...}&\multicolumn{1}{c}{...}&\multicolumn{1}{c}{...}&\multicolumn{1}{c}{...}& 17.29$\pm$0.48& 16.66$\pm$0.51 \\ \\
$C$(H$\beta$)$^{\rm c}$         &\multicolumn{1}{c}{0.330$\pm$0.037}&\multicolumn{1}{c}{0.150$\pm$0.037}&\multicolumn{1}{c}{0.310$\pm$0.037}&\multicolumn{1}{c}{0.105$\pm$0.037}&\multicolumn{1}{c}{0.440$\pm$0.037}&\multicolumn{1}{c}{0.410$\pm$0.037}\\
$F$(H$\beta$)$^{\rm d}$         &\multicolumn{1}{c}{56.3$\pm$0.2}&\multicolumn{1}{c}{5.86$\pm$0.2}&\multicolumn{1}{c}{ 44.4$\pm$0.1}&\multicolumn{1}{c}{ 10.5$\pm$0.1}&\multicolumn{1}{c}{196.0$\pm$0.2}&\multicolumn{1}{c}{183.3$\pm$0.2}\\
EW(H$\beta$)$^{\rm e}$          &\multicolumn{1}{c}{345.1$\pm$1.0}&\multicolumn{1}{c}{311.8$\pm$1.2}&\multicolumn{1}{c}{519.4$\pm$0.9}&\multicolumn{1}{c}{190.4$\pm$0.6}&\multicolumn{1}{c}{309.8$\pm$1.0}&\multicolumn{1}{c}{412.6$\pm$1.0}\\
EW(abs)$^{\rm e}$               &\multicolumn{1}{c}{1.2$\pm$0.1}&\multicolumn{1}{c}{3.0$\pm$0.3}&\multicolumn{1}{c}{3.6$\pm$0.2}&\multicolumn{1}{c}{3.6$\pm$0.2}&\multicolumn{1}{c}{3.0$\pm$0.1}&\multicolumn{1}{c}{1.3$\pm$0.1}\\
\hline
  \end{tabular}

\hbox{$^{\rm a}$Fluxes are in units 100$\times$$I(\lambda)$/$I$(H$\beta$).} 

\hbox{$^{\rm b}$Corrected with extinction coefficient derived excluding 
the H$\alpha$ emission line.} 

\hbox{$^{\rm c}$Extinction coefficient, derived from the observed hydrogen 
Balmer decrement.}

\hbox{$^{\rm d}$Observed flux in units of 10$^{-16}$ erg s$^{-1}$ cm$^{-2}$.}

\hbox{$^{\rm e}$Equivalent width in \AA.}

  \end{table*}

%% file: taba2_1.tex
\begin{table*}
\caption{Electron temperatures, electron number densities, ionic and elemental abundances \label{taba2}}
\begin{tabular}{lcccccc} \hline
Property                             &J0240$-$0828&J0344$-$0106&J1205$+$4551&J1222$+$3602&W\,1702$+$18&HS\,1851$+$6933          \\ \hline
$T_{\rm e}$(O {\sc iii}), K          &16000$\pm$300&17700$\pm$400&19000$\pm$400&20600$\pm$500&17000$\pm$300&16300$\pm$300       \\
$T_{\rm e}$(O {\sc ii}), K           &14500$\pm$200&15200$\pm$300&15500$\pm$300&15600$\pm$300&15000$\pm$300&14600$\pm$300       \\
$T_{\rm e}$(S {\sc iii}), K          &14600$\pm$200&16600$\pm$400&17900$\pm$300&18100$\pm$400&15900$\pm$300&15100$\pm$300      \\
$N_{\rm e}$(S {\sc ii}), cm$^{-3}$    &485$\pm$103& 28$\pm$78 &616$\pm$142&676$\pm$227&319$\pm$86&449$\pm$98         \\ 
$N_{\rm e}$(He {\sc i}), cm$^{-3}$    &674$\pm$150&    ...    &512$\pm$130&    ...    &295$\pm$70&684$\pm$160         \\ 
\\
O$^+$/H$^+$$\times$10$^5$            &0.697$\pm$0.039&0.177$\pm$0.012&0.160$\pm$0.010&0.312$\pm$0.020&0.597$\pm$0.034&0.583$\pm$0.032  \\
O$^{2+}$/H$^+$$\times$10$^5$          &6.745$\pm$0.341&4.519$\pm$0.278&2.631$\pm$0.144&4.496$\pm$0.262&4.949$\pm$0.256&5.909$\pm$0.299 \\ 
O$^{3+}$/H$^+$$\times$10$^6$          &1.965$\pm$0.132&1.769$\pm$0.164&0.943$\pm$0.066&0.393$\pm$0.048&1.161$\pm$0.077&0.903$\pm$0.060 \\
O/H$\times$10$^5$                   &7.639$\pm$0.344&4.873$\pm$0.278&2.886$\pm$0.145&4.848$\pm$0.263&5.662$\pm$0.259&6.582$\pm$0.300 \\
12+log(O/H)                         &7.883$\pm$0.020&7.688$\pm$0.028&7.460$\pm$0.022&7.686$\pm$0.024&7.753$\pm$0.020&7.818$\pm$0.020     \\ \\
N$^{+}$/H$^+$$\times$10$^6$          &0.640$\pm$0.025&0.247$\pm$0.013&0.320$\pm$0.014&0.215$\pm$0.013&0.399$\pm$0.016&0.477$\pm$0.018 \\
ICF(N)                              &9.721 &16.058 &23.863&13.787 &8.631 &10.074  \\
N/H$\times$10$^6$                   &6.222$\pm$0.272&5.893$\pm$0.351&5.139$\pm$0.245&2.960$\pm$0.208&3.442$\pm$0.150&4.807$\pm$0.208 \\
log(N/O)                            &$-$1.089$\pm$0.027~~~&$-$0.917$\pm$0.036~~~&$-$0.749$\pm$0.030~~~&$-$1.214$\pm$0.039~~~&$-$1.216$\pm$0.027~~~&$-$1.137$\pm$0.027~~~\\ \\
Ne$^{2+}$/H$^+$$\times$10$^5$        &1.207$\pm$0.067&0.704$\pm$0.045&0.345$\pm$0.020&0.734$\pm$0.043&0.931$\pm$0.052&1.015$\pm$0.056 \\
ICF(Ne)                             &1.041 &1.025&1.037 &1.025 &1.050 &1.034 \\
Ne/H$\times$10$^5$                  &1.256$\pm$0.075&0.721$\pm$0.049&0.358$\pm$0.021&0.752$\pm$0.046&0.977$\pm$0.059&1.050$\pm$0.062 \\
log(Ne/O)                           &$-$0.784$\pm$0.033~~~&$-$0.830$\pm$0.039~~~&$-$0.907$\pm$0.034~~~&$-$0.809$\pm$0.036~~~&$-$0.763$\pm$0.033~~~&$-$0.797$\pm$0.032~~~\\ \\
S$^{+}$/H$^+$$\times$10$^6$          &0.099$\pm$0.004&     ...       &0.026$\pm$0.001&0.048$\pm$0.003&0.094$\pm$0.003&0.088$\pm$0.003 \\
S$^{2+}$/H$^+$$\times$10$^6$         &0.566$\pm$0.033&     ...       &0.152$\pm$0.011&0.677$\pm$0.049&0.462$\pm$0.025&0.571$\pm$0.030 \\
ICF(S)                              &1.777& ... &2.639&1.908&1.543&1.737 \\
S/H$\times$10$^6$                   &1.181$\pm$0.059&      ...       &0.470$\pm$0.029&1.385$\pm$0.094&0.858$\pm$0.039&1.144$\pm$0.053 \\
log(S/O)                            &$-$1.811$\pm$0.029~~~&    ...   &$-$1.788$\pm$0.034~~~&$-$1.544$\pm$0.038~~~&$-$1.820$\pm$0.028~~~&$-$1.760$\pm$0.028~~~\\ \\
Cl$^{2+}$/H$^+$$\times$10$^8$        &1.117$\pm$0.100&     ...       &       ...    &       ...    &0.584$\pm$0.050&1.270$\pm$0.088 \\
ICF(Cl)                             &1.504& ... & ... & ... &1.489&1.510 \\
Cl/H$\times$10$^8$                  &1.680$\pm$0.150&      ...       &       ...    &       ...    &0.869$\pm$0.074&1.918$\pm$0.133 \\
log(Cl/O)                           &$-$3.658$\pm$0.044~~~&     ...   &       ...    &       ...    &$-$3.814$\pm$0.042~~~&$-$3.536$\pm$0.036~~~\\ \\
Ar$^{2+}$/H$^+$$\times$10$^7$        &1.670$\pm$0.066&0.727$\pm$0.049 &0.309$\pm$0.018&1.203$\pm$0.075&1.320$\pm$0.051&1.543$\pm$0.060 \\
Ar$^{3+}$/H$^+$$\times$10$^7$        &1.475$\pm$0.081&1.637$\pm$0.125 &0.683$\pm$0.040&0.610$\pm$0.058&1.016$\pm$0.050&1.288$\pm$0.061 \\
ICF(Ar)                             &1.372&2.223&1.771&1.630&1.329&1.401 \\
Ar/H$\times$10$^7$                  &2.291$\pm$0.144&1.616$\pm$0.298&0.546$\pm$0.078&1.961$\pm$0.155&1.753$\pm$0.095&2.162$\pm$0.120 \\
log(Ar/O)                           &$-$2.523$\pm$0.034~~~&$-$2.479$\pm$0.084&$-$2.723$\pm$0.066~~~&$-$2.393$\pm$0.042~~~&$-$2.509$\pm$0.031~~~&$-$2.484$\pm$0.031~~~\\ \\
$[$Fe~{\sc iii}$]$ 4658: \\
Fe$^{2+}$/H$^+$$\times$10$^6$        &0.173$\pm$0.013& ...  &0.102$\pm$0.010&0.124$\pm$0.015&0.124$\pm$0.008&0.139$\pm$0.008 \\
ICF(Fe)                             &14.187& ... &24.636&20.767&12.555&14.789 \\
Fe/H$\times$10$^6$                  &2.451$\pm$0.189& ... &2.516$\pm$0.247&2.580$\pm$0.307&1.558$\pm$0.103&2.057$\pm$0.115 \\
log(Fe/O)                           &$-$1.494$\pm$0.039~~~& ... &$-$1.059$\pm$0.048~~~&$-$1.274$\pm$0.057~~~&$-$1.560$\pm$0.035~~~&$-$1.505$\pm$0.031~~~\\ \\
$[$Fe~{\sc iii}$]$ 4986: \\
Fe$^{2+}$/H$^+$$\times$10$^6$        & ... & ... & ... & ... &0.119$\pm$0.008&0.102$\pm$0.006  \\
ICF(Fe)                             & ... & ... & ... & ... &12.555&14.789 \\
Fe/H$\times$10$^6$                  & ... & ... & ... & ... &1.490$\pm$0.096&1.505$\pm$0.087 \\
log(Fe/O)                           & ... & ... & ... & ... &$-$1.580$\pm$0.034~~~&$-$1.641$\pm$0.032~~~\\ \\
\hline
  \end{tabular}
  \end{table*}